%% file: thesis.tex
\documentstyle[12pt,uwthesis,epsf,rotate,twoside]{thesisme}
\def\bq{\begin{equation}}
\def\eq{\end{equation}}
\def\ba{\begin{eqnarray}}
\def\ea{\end{eqnarray}}
\def\begr{\begin{array}}
\def\endr{\end{array}}
\newcommand{\sla}[1]{/\!\!\!#1}

\begin{document}
 \title{Intermediate-Mass Higgs Searches in Weak Boson Fusion}
 \author{David Landry Rainwater}
 \principaladviser{Dieter Zeppenfeld}
 \dept{Physics} 
 \date{1999} 

\beforepreface
 \prefacesection{Abstract}
 \input{body/abstract}
 \prefacesection{Acknowledgments}
 \input{body/ackn}
\afterpreface

\chapter{Introduction}
 \label{ch:intro}
 \input{body/intro}

\chapter{The Weak Boson Fusion Signature}
 \label{ch:WBF}
 \input{body/WBF}

\chapter{The Search for $H\to\gamma\gamma$}
 \label{ch:gammagamma}
 \input{body/gammagamma}

\chapter{The Search for $H\to W^{(*)}W^{(*)}$}
 \label{ch:WW}
 \input{body/WW}

\chapter{The Search for $H\to\tau^+\tau^-$}
 \label{ch:tautau}
 \input{body/tautau}

\chapter{Conclusions}
 \label{ch:concl}
 \input{body/concl}

\appendix

\chapter{Important Parameters}
 \label{app:parms}
 \input{body/parms}

\chapter{Detector Resolution}
 \label{app:smear}
 \input{body/smear}

\chapter{Minijet Approximations}
 \label{app:minijet}
 \input{body/minijet}

\chapter{Tau Decay and Reconstruction}
 \label{app:taudecay}
 \input{body/taudecay}

\chapter{Monte Carlo Programs}
 \label{app:programs}
 \input{body/programs}

\chapter{Implications for the MSSM}
 \label{app:mssm}
 \input{body/mssm}


\end{document}

%% file: body/abstract.tex

Weak boson fusion is a copious source of intermediate mass Higgs bosons at the 
LHC. The additional very energetic forward jets in these events are powerful 
background suppression tools. I analyze the decays $H\to\gamma\gamma$ and 
$H\to W^{(*)}W^{(*)} \to e^{\pm} \mu^{\mp} \sla{p_T}$, with the latter a 
potential discovery channel, and the decay 
$H\to\tau^+\tau^- \to \ell^\pm h^\mp \sla{p}_T$ as a method for achieving the 
first direct measurement of a Higgs-fermion coupling.

I perform parton level analyses of the signal and dominant backgrounds for each 
decay mode, and demonstrate kinematic cuts and other important tools neccessary 
to achieve an S/B~$> 1/1$ rate in all cases. I also perform cross section 
calculations with additional gluon emission which provide an estimate of a 
minijet veto probability.

I show that a $5\sigma$ $H\to\gamma\gamma$ observation can be made for 
110~GeV~$< M_H <$~150~GeV with modest luminosity, order 40-50~fb$^{-1}$ at low 
machine luminosity, overlapping the region explored by the CERN LEP and Fermilab 
Tevatron. For 130~GeV~$< M_H <$~200~GeV, I show that $H\to W^{(*)}W^{(*)}$ can 
achieve a $5\sigma$ observation with S/B much greater than 1/1 with extremely 
low luminosity, about 2-10~fb$^{-1}$ over almost the entire range. This is the 
most promising search channel in the 130-200~GeV mass range. It overlaps the 
$H\to\gamma\gamma$ region and nicely complements the $H\to W^{(*)}W^{(*)}$ 
measurement that can be made with very low luminosity in inclusive $gg\to H$ 
production. I further show that a Higgs-fermion coupling can be directly 
measured via the $H\to\tau\tau$ decay with only about 60~fb$^{-1}$ (assuming low 
luminosity running).


%% file: body/ackn.tex
I am most indebted to my advisor, Dieter Zeppenfeld, for both his teaching of 
physics and the extraordinary patience and kindness it surely took to withstand 
the onslaught of questions to which I subjected him, especially the repeated 
ones and especially early on. I am grateful he agreed to take me on as an 
apprentice, and only hope I can one day try to hold the same standards of 
mentoring for my students.

I thank the rest of the Phenomenology Institute professors, Vernon Barger, 
Francis Halzen, Tao Han and Martin Olsson, for their support and the friendly 
atmosphere they maintain in these halls. I did well to choose Wisconsin for my 
studies.

I will long remember the camaraderie shared with my fellow high-energy students 
and colleagues John Beacom, Tim Kinnel, Nikolas Kauer and Bill Long, as well as 
with our wonderful librarian Kerry Kresse, in everything from rag sessions 
about physics, computers, or people and politics, to a passion for flying. 
I thank my fellow students, most notably Steve Kadlecek, Rob Haslinger, 
Chris O'Dell, Brian Schwartz and Greg Jaczko, for the late-night Jackson 
problem-solving sessions that helped me make it this far, and the great parties 
after every ordeal we endured.

I owe thanks to my colleagues outside of the Phenomenology Institute with 
whom I was lucky to have worked while still a graduate student, 
Kaoru Hagiwara, Tim Stelzer and Rob Szalapski. 
Their advice and encouragement has been invaluable.

I thank my former teachers Keith Vemmer and Mark Schuermann, who first sparked 
my interest in science, and my undergraduate physics professors Cliff Tompson, 
David Cowan, Meera Chadrasekhar and the late Justin Huang, for planting my 
feet firmly on the path to a career in physics.

Finally, I cannot forget my parents Gary and Becky, whom I thank for 
supporting me in pursuing a goal that I think cannot be reached without an 
endless supply of encouragement.

%% file: body/intro.tex

\section{Where Are We?}

The year 1999 finds the particle physics community on the cusp of understanding 
the fundamental particles and forces of nature. We have, over the past three 
decades, formulated a theory governing the electromagnetic, weak and strong 
forces and their constituents, called the Standard Model, which now predicts 
with uncanny accuracy the results of most particle physics experiments, and 
there is a considerable volume of data with which to compare~\cite{PDG}. During 
this time numerous observations have been made that suggested new physics 
beyond the Standard Model, or that it was wrong, but all have essentially 
vanished as statistical fluctuations or experimental flaws.

Yet while the community continually rejoices in the predictive success of the 
Standard Model, it also faces the somber realization that the theory is 
unsatisfactory, that it is incomplete at a fundamental level. Aside from the 
deeper questions of the large number of input parameters required to define the 
theory (of order 20)~\cite{th-input}, the intimation of unification of forces 
at energy scales higher than can be probed~\cite{unification}, and perhaps even 
the necessity of such principles as renormalizability, there are the more 
immediate problems of electroweak symmetry breaking (ESB) and mass generation, 
on which this dissertation focuses. 

The Standard Model hinges on an explanation of these observations at a time when 
the community is on the verge of having the technical capability of exploring 
some of these issues with conclusion. It remains to be seen whether or not the 
Higgs mechanism, the (currently) most viable explanation of ESB, is correct. It 
is certain that this area has not yet been fully explored phenomenologically, 
and this dissertation outlines strategies for experimental searches designed to 
lay the issue to rest with efficiency and certainty.

In this Chapter, I introduce the physics of the electroweak sector of the 
Standard Model which describes all presently known interactions and forces other 
than gravity, explaining its structure and the mathematical limitations in 
formulating the theory with only the particles observed in nature. I describe 
how the observed ESB of the theory can be obtained via addition of a simple 
Higgs sector, and show the Standard Model couplings that result. I explain the 
origin of mass in the theory, for both known particles and the theoretically 
introduced but unobserved Higgs scalar, and why a light Higgs mass is expected 
from radiative corrections to the masses of the weak gauge bosons in light of 
the current status of electroweak (EW) precision fits to data. As a prelude to 
introduction of the minijet veto tool in Chapter~\ref{ch:WBF}, I discuss the 
basics of QCD, the strong force responsible for the interactions of colored 
particles. Finally, to introduce my calculations for a Higgs search in weak 
boson fusion I give a brief description of parton-level Monte Carlo calculation 
of cross sections, and an introduction to {\sc madgraph}~\cite{Madgraph}, a tool 
I use for generating matrix elements in {\sc fortran}. 


\section{The Standard Model}
\label{sec:SM}

The Standard Model (SM) has evolved from a field theoretical description of only 
electromagnetism (Quantum Electro-Dynamics, or QED)~\cite{QED}, to a much richer 
theory encompassing also the weak force, first observed in nuclear beta decay, 
and the strong force, which binds together nucleons and their constituents, 
quarks, which are discussed further in Section~\ref{sec:QCD}. All known 
particles in nature (other than the graviton) are described by the SM. Two 
classes of fermions, leptons and quarks, make up matter. Bosonic particles carry 
the three forces which act between them: the massless photon, $\gamma$, for 
electromagnetism; the neutral $Z$ and charged $W^\pm$, both of which are 
massive, for the weak force; and the massless gluons, $g$, for the strong force.

I begin by describing the group theoretic structure of the SM - the symmetries 
$SU(3)_c \otimes SU(2)_L \otimes U(1)_Y$, where $c$ refers to the color group of 
QCD, $L$ refers to a left-handed group structure, and $Y$ is the hypercharge 
group~\cite{SM,HalMar}. To this gauge structure one adds a scalar $SU(2)$ 
doublet of hypercharge $+1$, called the Higgs doublet, which breaks 
$SU(2)_L \otimes U(1)_Y \longrightarrow U(1)_{EM}$~\cite{HalMar,PesSch}: massive 
weak gauge bosons $W^\pm,Z$ plus the electromagnetism of massless photons. The 
left-handed nature of the $SU(2)$ group comes from direct observation of 
left-handed doublets of matter fields in nature, the quarks and leptons, along 
with right-handed singlets:
\bq
\label{eq:matter}
\left[
\begin{array} {ccc}
\left(
\begin{array}{c} u \\ d \end{array}
\right)_L
& d_R & u_R \\
\left(
\begin{array}{c} \nu_e \\ e \end{array}
\right)_L
& e_R & 
\end{array}
\right]
\left[
\begin{array} {ccc}
\left(
\begin{array}{c} c \\ s \end{array}
\right)_L
& s_R & c_R \\
\left(
\begin{array}{c} \nu_\mu \\ \mu \end{array}
\right)_L
& \mu_R & 
\end{array}
\right]
\left[
\begin{array} {ccc}
\left(
\begin{array}{c} t \\ b \end{array}
\right)_L
& b_R & t_R \\
\left(
\begin{array}{c} \nu_\tau \\ \tau \end{array}
\right)_L
& \tau_R & 
\end{array}
\right]
\eq
While all matter fields transform non-trivially under $U(1)_Y$, only the 
left-handed doublets transform non-trivially under $SU(2)_L$, and only quarks 
and gluons carry the conserved charges of QCD. Note that while we have observed 
all~\footnote{The tau neutrino, $\nu_\tau$, has as yet not been observed 
directly, but is believed to exist based on indirect evidence, such as the total 
decay rate $\Gamma(Z\to\nu\bar\nu)$.} the particles of the three complete 
generations shown above, only the first generation is found in normal matter: 
particles of the second and third generations are more massive and unstable, 
decaying in all cases to particles of the first generation. The SM fails to 
predict the masses of all matter fields.

Each symmetry group has its own generators of transformation, and thus its own 
force-carrying bosonic particles, in the adjoint representation. For the color 
symmetry of QCD there is an octet of bosons, for $SU(2)_L$ there is a triplet, 
and for $U(1)_Y$ there can be only one. One may write a Lagrangian for the SM 
gauge and matter fields as~\cite{CORE}
\bq
\label{eq:Lagr1}
{\cal L}_{SM} \; = \; 
{\cal L}_G + {\cal L}_f + {\cal L}_{GF} + {\cal L}_{FP} 
\, ,
\eq
where ${\cal L}_G$ is the kinetic term for gauge fields, ${\cal L}_f$ is the 
kinetic term for fermions, which also (via minimal coupling) describes the 
interactions of fermions with gauge bosons, and ${\cal L}_{GF}$ and 
${\cal L}_{FP}$ are gauge-fixing pieces I do not address here. A fundamental 
principle in the construction of this Lagrangian is that of gauge invariance - 
the only theories of vector bosons that can be renormalized are gauge theories. 
Thus, the gauge principle is a basis for the predictive power of the SM.

The first problem with this Lagrangian is that by construction all gauge and 
matter fields must be massless. Inserting mass terms by hand leads to disaster, 
as gauge boson mass terms are not gauge invariant, and matter field mass terms 
will not respect the $SU(2)_L \otimes U(1)_Y$ symmetry. However, in nature we 
observe all matter fields to have a mass, as well as the weak bosons $W^\pm,Z$.

The second problem is that of bad behavior at high energy for massive $W$ and 
$Z$ bosons. This same feature was present in the Fermi theory of four-fermion 
interactions, the predecessor to the SM~\cite{Willenbrock}. Weak processes, 
first observed in nuclear beta decay, were described via the coupling $G_F$ and 
an expansion in powers of the center-of-mass (CoM) energy, $E^2$, for the 
scattering amplitude. $G_F$ was measured in experiment, and the theory worked 
very well, but was doomed to ultimate failure as it predicted violation of 
unitarity for $E \gtrsim 600$~GeV. To regulate this behavior, a modification to 
the theory was proposed, whereby the interaction of four fermions is replaced 
by a force-carrying massive boson, a charged $W^\pm$, which predicts an 
amplitude of the form
\bq
\label{eq:EW_amp}
A \sim g^2 {E^2 \over E^2 - M^2_W} \, ,
\eq
where $g$ is the coupling, related to the Fermi coupling by 
$G_F = {g^2 \over 4\sqrt{2}M^2_W}$. By inspection we see that the low-energy 
($E \ll M_W$) behavior asymptotically approaches that described by the Fermi 
theory, amplitude proportional to $E^2$, while at high energy, the amplitude 
approaches a constant, thus avoiding unitarity violation. High-energy data 
supported this description to incredible accuracy, and eventually the massive 
$W^\pm$ bosons were directly observed along with a massive neutral $Z$ boson, 
which confirmed an $SU(2)_L \otimes U(1)_Y$ structure as integral to the SM.

This theoretical fix to Fermi theory merely put off the problem, however: 
calculation of SM weak boson scattering at high energies, {\em e.g.} 
$W^+W^- \to W^+W^-$, reveals that unitarity is again violated at CoM energies 
of order $\sim 1$~TeV. The theoretical success of the Higgs mechanism lies in 
supplying not only masses for gauge bosons and matter fields, but also in 
regulating this bad behavior in weak boson scattering. Including Higgs exchange 
in $W^+W^- \to W^+W^-$ scattering is exactly what restores unitarity.


\section{Electroweak Symmetry Breaking}
\label{sec:ESB}

To obtain massive gauge bosons the $SU(2)_L \otimes U(1)_Y$ group structure must 
be broken. We must obtain after the breaking three massive gauge bosons with the 
observed mass ratio, and a massless photon, reflecting a leftover, unbroken 
$U(1)_{EM}$. Massless vector fields have two degrees of freedom, whereas massive 
fields have three, so one must introduce additional degrees of freedom that 
account for the third helicity state of the bosons that become massive. 
Following mostly the notation of Ref.~\cite{HalMar}, this can be done most 
simply by introducing an additional spin-0 $SU(2)_L$ complex doublet of 
hypercharge $Y = +1$, written as
\bq
\label{eq:phi}
\Phi = 
\left(
\begr {c} \phi^+ \\ \phi^\circ \endr
\right)
= \, \sqrt{1\over 2} 
\left(
\begr {c} w_1 + iw_2 \\ \phi + iw_3 \endr
\right)
\, ,
\eq
The electric charges of the components are known via the relation 
$Q = T_3 + {Y\over 2}$, where $T_3$ is the third component of the $SU(2)_L$ 
quantum number, $\pm {1\over 2}$, and $Y$ is the hypercharge of the doublet. 
This doublet ``spontaneously'' breaks the symmetry when it acquires a non-zero 
vacuum expectation value. It is inserted into the theory via the additional 
Lagrangian term 
\ba
\label{eq:Lagr2}
{\cal L}_H \, = \, (D_\mu \Phi)^\dagger (D^\mu \Phi) - V(\Phi) \, , \\
V(\Phi) \, = \, \mu^2 (\Phi^\dagger\Phi) + \lambda (\Phi^\dagger\Phi)^2 \, ,
\ea
with minimal coupling to the gauge fields via the covariant derivative
\bq
\label{eq:cov_dev}
D_\mu \, = \, \partial_\mu + ig{\tau^i \over 2}W^i_\mu + ig'{Y\over 2}B_\mu
\quad (i = 1,2,3) \, ,
\eq
where $W^i_\mu$ are the massless $SU(2)_L$ gauge fields and $B_\mu$ is the 
massless hypercharge gauge field. This potential is the simplest that can be 
written which gives a non-zero expectation value for the field and is bounded 
from below, provided that $\mu^2 < 0$ and $\lambda > 0$. It has a global 
minimum at 
\bq
\label{eq:phi_sq_min}
{1\over 2} (w^2_1 + w^2_2 + \phi^2 + w^2_3) \, = \, -{\mu^2\over 2 \lambda} 
\, .
\eq

One may choose which component of $\Phi$ lives at this minimum, and for 
convenience choose it to be $\phi$:
\bq
\label{eq:phi_vev}
w_1 = w_2 = w_3 = 0 \; , \; \phi^2 = -{\mu^2\over\lambda} \;
\equiv \; v^2 \: .
\eq
This is the vacuum expectation value (vev) of the Higgs field. $\Phi$ can now 
be written as small perturbations about this minimum,
\bq
\label{eq:rewrite1}
\Phi \, = \, \exp\left({i\tau^i\theta^i \over v}\right) \sqrt{1\over 2}
\left(
\begr {c} 0 \\ v + H \endr
\right)
\, .
\eq
Since $\Phi$ and the component fields respect the $SU(2)_L$ symmetry, one may 
make a transformation to the Unitary gauge and write the simpler expression
\bq
\label{eq:rewrite2}
\Phi \, = \, \sqrt{1\over 2}
\left(
\begr {c} 0 \\ v + H \endr
\right)
\, .
\eq

Insertion of this resulting expression for $\Phi$ into the kinetic term in 
${\cal L}_H$, we find the Lagrangian terms
\bq
\label{eq:bos_mass}
{\cal L}_H = ... 
+ {1\over 8} (v+H)^2 g^2 \left[ W^1_\mu W^{1\mu} + W^2_\mu W^{2\mu} \right]
+ {1\over 8} (v+H)^2 (g' B_\mu - g W^3_\mu)(g' B^\mu - g W^{3\mu})
\, .
\eq
Defining the observed fields as $W^\pm = {1\over\sqrt{2}}(W^1 \mp W^2)$, we 
immediately find a mass term (${1\over 4} v^2 g^2 W^+_\mu W^{-\mu}$), a 
trilinear coupling term (${1\over 2} v g^2 H W^+_\mu W^{-\mu}$), and a quartic 
coupling term (${1\over 4} g^2 HH W^+_\mu W^{-\mu}$). The $W$ boson mass may be 
immediately read off to be $M_W = {1\over 2}vg$.

The $v^2$ part of the second term's expansion may be expressed as a mass mixing 
matrix
\bq
\label{eq:AZ_mass_matrix}
{1\over 8} v^2 (W^3_\mu, B_\mu)
\left(
\begr {cc} g^2 & -gg' \\ -gg' & g'^2 \endr
\right)
\left(
\begr {c} W^{3\mu} \\ B^\mu \endr
\right)
\, ,
\eq
which has eigenvalues 0 and ${v\over 2}\sqrt{g^2+g'^2} \;$, the observed masses 
of the photon and $Z$ boson. Their fields are similarly defined in terms of 
$W^3_\mu$ and $B_\mu$ via a mixing angle, called the Weinberg angle, given by 
the relation
\bq
\label{eq:sintw}
\sin\,\theta_W \: = \: {g' \over \sqrt{g^2+g'^2}} \: .
\eq
This leads to the relation for electric charge,
\bq
\label{eq:e}
e \: = \: g \, \sin\,\theta_W \: = \: g' \, \cos\,\theta_W \: .
\eq
There are similar terms for $H Z_\mu Z^\mu$ and $HH Z_\mu Z^\mu$ couplings. 
Note that there are no trilinear $H A_\mu A^\mu$ coupling or $HH A_\mu A^\mu$ 
quartic couplings! Thus, the $H\to\gamma\gamma$ decay is a rare process 
induced by loop diagrams.

The Higgs mechanism also predicts a relation between the $W$ and $Z$ masses, 
which is a crucial test of the SM:
\bq
\label{eq:rho}
\rho \: \equiv \: {M^2_W \over M^2_Z \, \cos^2 \,\theta_W} \: = \: 1 \: .
\eq
Another crucial test of the SM is the requirement to observe the trilinear $HVV$ 
couplings - only models with scalars that acquire a vev can produce this 
feature. Thus, observation of a trilinear $\phi VV$ coupling would identify the 
Higgs as the agent responsible for ESB.

It is worth noting that the four additional fields added via $\Phi$ have been 
rearranged such that three of the fields have disappeared while three degrees of 
freedom have shown up as the longitudinal components of the newly-massive weak 
gauge bosons, and the fourth field has become a real, interacting particle.
To see how the Higgs field itself acquires a mass, Eq.~(\ref{eq:rewrite2}) must 
be inserted into the potential $V(\Phi)$ of ${\cal L}_H$ to obtain additional 
terms in the Lagrangian:
\bq
\label{eq:Hmass}
{1\over 4} \lambda ( v H + H^2 ) ^2
\eq
which give Higgs self-interactions ($HHH$ and $HHHH$ couplings) as well as a 
Higgs mass, $m_H = v\sqrt{2\lambda}$. It is very important to realize that the 
only places in the theory where $\lambda$ appears are in the Higgs 
self-couplings and the Higgs mass term. Thus, experiment cannot determine the 
Higgs mass directly without observing the Higgs or a process involving the Higgs 
propagator. Other methods can, however, give us clues as to its mass, as 
explained in the next Section.

Fermion masses, while distinct from the issue of ESB, can also be generated 
from a Higgs $SU(2)$ doublet via Lagrangian terms of the form 
$Y^u_f \bar{F}_L \Phi_c f^u_R$ and $Y^d_f \bar{F}_L \Phi f^d_R$, and their 
Hermitian conjugates, where $F_L$ are left-handed fermion doublets, $f_R$ are 
right-handed fermion singlets, and $Y_f$ are the Yukawa couplings for up-type 
and down-type fermions, respectively. Taking the first-generation leptons as an 
example, one may write the Lagrangian terms
\bq
\begr {ccccc}
\label{eq:fermion}
{\cal L}_f 
& = & ... & - Y_e \bar{E}_L \Phi e_R & - Y_e \bar{e}_R \Phi^\dagger E_L \\
& = & ... & - Y_e ( \bar\nu_e , \bar{e} )_L
              {1\over\sqrt{2}} \left( \begr {c} 0 \\ v+H \endr \right) e_R
          & - Y_e \bar{e}_R {1\over\sqrt{2}} ( 0 , v+H ) 
              \left( \begr {c} \nu_e \\ e \endr \right)_L \\
& = & ... & - {Y_e \over\sqrt{2}} (v+H) \bar{e}_L e_R 
          & - {Y_e \over\sqrt{2}} (v+H) \bar{e}_R e_L \, ,
\endr
\eq
which yield a mass for the electron, ${1\over\sqrt{2}} Y_e v$, and an 
electron-Higgs coupling, naturally proportional to the mass, $m_e \over v$. 
Each fermion has a different Yukawa coupling $Y_f$, and for the quarks the 
values are different for $Y^u_q$ and $Y^d_q$. The Standard Model does not 
explain why the Yukawa couplings exhibit the seemingly unrelated values that 
they do; here, the mechanism merely provides mathematical consistency for 
construction of the theory. It is quite unsatisfactory in explaining {\it why} 
the fermions have their observed masses.

It is also possible to introduce two Higgs doublets to the theory. Typically, 
this is done such that each Higgs acquires a vev. By construction this must give 
the same ESB and gauge boson masses, but one doublet's vev will give masses to 
up-type fermions and the other's to down-type fermions. While there has been 
much research on general two-Higgs doublet models, this scenario is more 
commonly explored in the context of the minimal supersymmetric Standard Model 
(MSSM). I discuss this model briefly in Appendix~\ref{app:mssm}, and also 
extend the SM results of Chapter~\ref{ch:tautau} to the MSSM case.


\section{The Higgs Mass}
\label{sec:Hmass}

While the Standard Model does not predict the Higgs mass, there are theoretical 
constraints on the allowed mass range. First and foremost is the unitarity 
limit, obtained by examining weak boson scattering at high energies. Unitarity 
is maintained if the Higgs mass is less than about a TeV~\cite{unitarity},
\bq
\label{eq:TeVlimit}
m_H \lesssim 1 {\: \rm TeV \quad (unitarity)} \, .
\eq
A closer examination of the $HHHH$ coupling suggests a stricter 
limit~\cite{Quigg,Djouadi}. This coupling runs according to the relation
\bq
\label{eq:lambda_run}
{1\over\lambda(\mu)} \, = \, 
{1\over\lambda(\Lambda)} + {3\over 2\pi^2} \, 
{\rm log} \left( {\Lambda\over\mu} \right) \, ,
\eq
where $\mu$ is some low scale at which we measure $\lambda(\mu)$, and $\Lambda$ 
is a higher scale. The first requirement is that $\lambda(\Lambda)$ never be 
negative, an unphysical value. This leads, after some algebra, to an upper bound 
on the Higgs mass,
\bq
\label{eq:uplim}
M_H\leq\sqrt{{8v^2\pi^2\over 3\, {\rm log}\left({\Lambda\over M_H}\right)}}\, .
\eq
This is known as the ``triviality'' condition, because if the SM Higgs sector is 
to be valid for arbitrarily large energies (effectively $\Lambda\to\infty$ with 
$\lambda(\Lambda)$ finite) then $\lambda(M_H)\to 0$, making it a trivial theory. 
For example, for $\Lambda = 10^{19}$~GeV, one finds the limit 
$M_H \lesssim 175$~GeV. For small values of $\Lambda$ there is still a limit, 
although in this case the running of $\lambda(\mu)$ enters a nonperturbative 
region and the above equation becomes significantly modified; lattice 
simulations of the small-$\Lambda$ limit estimate that 
$m_H \lesssim 630 \; {\rm GeV} = \Lambda$~\cite{Djouadi}, much less restrictive 
than for $\Lambda\to\infty$.

The requirement of vacuum stability places instead a lower bound on the Higgs 
mass~\cite{Quigg}. Here, one-loop corrections to the Higgs potential, which must 
have an absolute minimum at 
$\phi = \left( \begr {c} 0 \\ {v\over\sqrt{2}} \endr \right)$, yield the 
relation
\bq
\label{eq:lolim}
M^2_H \: \geq \: {3\sqrt{2}G_F\over 4\pi^2} (2M^4_W + M^4_Z - m^4_t) \: 
                 {\rm log}\left({\Lambda\over v}\right)
      \: \gtrsim \, -4500 \: {\rm log}\left({\Lambda\over v}\right) 
      \: {\rm GeV^2} \, .
\eq
That this bound is negative and therefore not a real constraint is a result of 
the fact that $m_t \gg M_Z$. Two-loop corrections, however, are sizeable and 
yield a positive-definite lower bound.

An important feature to note of the above relations is that for small $\Lambda$, 
there is essentially no bound on $M_H$. Thus, while there are theoretical 
arguments for expecting $\Lambda$ to be large, {\em e.g.} the GUT or Planck 
scale, one cannot take these expressions as real restrictions since there is no 
experimental confirmation of a desert.

Instead, the most important hints of what the Higgs mass may be come from 
indirect measurements in present-day experiments - precision fits to electroweak 
observables. Both the Higgs and the top quark appear in loop radiative 
corrections to the gauge boson propagators. Now that the top quark has been 
observed and its mass measured, precision fits can place upper and lower 
confidence limits on the Higgs mass. Unfortunately, while these radiative 
corrections are a function of $m^2_t$, and therefore very sensitive to the top 
quark mass, they are a function of only the logarithm of the Higgs mass, 
log($m_H$), making the sensitivity very weak. Nevertheless, these fits give us 
an important clue: the Higgs mass should be low, on the order of 100~GeV, as 
opposed to several hundred GeV or close to the unitarity limit. Recent (1999) 
fits give the best-fit value for the Higgs mass to be~\cite{Quigg}
\bq
\label{eq:bestfit}
m_H = 107^{+67}_{-45} \: {\rm GeV} \, ,
\eq
with a $95\%$~C.L. upper limit of $m_H \lesssim 255$~GeV. While the uncertainty 
on this is quite large, the point is that indirect data suggests the Higgs has 
a fairly low mass, at the low end of the ``intermediate'' range (100-200 GeV). 
This should not be surprising, as from the historical example of Fermi theory 
one would expect the regulating physics to step in at a scale much lower than 
the unitarity limit; the weak bosons have masses $m_W = 80.35$~GeV and 
$m_Z = 91.19$~GeV, much lower than the upper bound of 600~GeV imposed by 
unitarity.

Direct searches have established an ever-increasing lower limit on the Higgs 
mass as machine energies have risen. The current bound, established by the LEP 
collider at CERN in 1998, places the $95\%$~C.L. limit~\cite{LEP99}
\bq
\label{eq:LEPlimit}
m_H > 99 \: {\rm GeV} \, .
\eq
In its next (and last) two runs through 2000, LEP will either discover a Higgs 
up to a mass of about 105~GeV, or place a $95\%$~C.L. exclusion limit of about 
110~GeV.


\section{QCD: Theory and Calculation}
\label{sec:QCD}

Quantum Chromo-Dynamics, or QCD, accounts for the remaining sector of the SM. 
It is an unbroken local gauge symmetry, described by the non-Abelian $SU(3)_c$ 
gauge group, where $c$ refers to color~\cite{QCD}. The non-Abelian aspect means 
the generators of $SU(3)$ transformation do not commute. This leads to some 
interesting properties which will be relevant for an investigation of the 
minijet veto tool and calculations I perform for QCD processes. The non-Abelian 
nature leads to gauge bosons that carry the conserved charges. Thus, the eight 
massless gluons of $SU(3)_c$ QCD are also color-charged particles. In contrast, 
the massless gauge boson of QED, the photon, carries no electric charge. Since 
interactions are governed by conserved charges, photons may not have 
self-interactions, but gluons must.

The value of the conserved charge in a gauge theory is modified by higher-order 
corrections to the vertex, as well as vacuum polarizations in the gauge boson 
propagator. In QED, these can only be fermion loops, which lead to a running of 
the coupling constant, 
\bq
\label{eq:QED_run}
\alpha(Q^2) \, = \, {\alpha(\mu^2) \over 
                     1 - {\alpha(\mu^2) \over 3\pi}{\rm log}
                     \left( {Q^2 \over \mu^2} \right) }
\, ,
\eq
where $Q^2$ is the energy scale being probed and $\mu^2$ is some reference scale 
where $\alpha(\mu^2)$ is finite and known. It is easy to see that at low energy, 
the QED coupling is essentially constant, but at very high energy it becomes 
large - calculations in this regime become nonperturbative. It turns out that 
this regime is well beyond the reach of experiment.

In QCD, the opposite happens. Because gluons self-couple, gluon loops in the 
vacuum polarization propagator corrections must also be calculated. This leads 
to a running coupling of the form 
\bq
\label{eq:QCD_run}
\alpha_s(Q^2) \, = \, {\alpha_s(\mu^2) \over 1 + {\alpha_s(\mu^2) \over 12\pi}
                       (33-2n_f){\rm log}\left( {Q^2 \over \mu^2} \right) }
\, .
\eq
Here, the coupling blows up as $Q^2$ becomes small, but approaches zero as 
$Q^2 \to \infty$; this feature, which is general of non-Abelian gauge theories, 
is known as ``asymptotic freedom''. Because of asymptotic freedom, QCD 
calculations instead become nonperturbative at low energy, which are referred 
to as the soft regime or soft effects. Our minijet veto tool naturally requires 
soft jets, which may be nonperturbative in the experimental phase space regions 
of interest. I explain how this can be dealt with in the next Chapter and in 
Appendix~\ref{app:minijet}.

The self-coupling of gluons has an additional effect on colored particles: lines 
of flux between color charges attract each other and so become constricted, such 
that the lines are confined to a narrow ``flux tube'' connecting the two 
physical particles. In QCD, the energy density per unit length in a flux tube is 
approximately constant, leading to a color field potential that grows 
proportional to the separation between the colored particles, $V(r) \sim r$, and 
ultimately results in confinement. That is, colored particles must exist only in 
``colorless'' bound states, such as color/anti-color mesons or red/green/blue 
(or anti-red/anti-green/anti-blue) baryons, such as the proton and neutron. High 
energy collisions may separate colored particles; if the momentum transfer is 
sufficient, we may calculate this hard scattering process perturbatively. 
Additional quark/gluon splittings may then occur, called parton showering 
because the relatively large probability of this occurring typically results in 
many more final state particles than the $2\to n$ hard process under 
consideration. Showering, too, may be calculated perturbatively, down to 
momentum transfer values of order $Q^2 \approx 1\; {\rm GeV}^2$. Below this 
scale, the colored partons separate from each other sufficiently that one must 
consider the long-range behavior of QCD, which is inherently non-perturbative. 
The kinetic energy of the separating quarks and gluons is converted into 
potential energy in the flux tubes maintaining the color connections. At large 
enough distances, when the energy in the flux tube is sufficient, a pair of 
colored particles is created out of the vacuum, which then separate themselves. 
This continues until not enough energy is left in a flux tube to create new 
quark/anti-quark pairs and the system condenses to a collection of colorless 
bound states. This process is known as hadronization.

Showering and hadronization lead to important consequences for collider physics. 
For collisions involving either initial- or final-state colored particles, for 
example $u\bar{u}\to d'\bar{d'}$, these processes lead not to the final state 
$d\bar{d}$ with only two observed particle tracks, but to two clusters of 
hadrons, which may be grouped together into jets. Showering and hadronization 
are well-understood phenomenologically, but for practical purposes can be 
simulated only via tuned Monte Carlo programs~\cite{Sjostrand} (Monte Carlo 
refers to the random nature of the numbers used in generating phase space 
points). Various programs such as {\sc phytia}~\cite{Pythia}, 
{\sc herwig}~\cite{Herwig}, and {\sc isajet}~\cite{Isajet}, exist that take into 
account successively the hard scattering, parton showering, and hadronization. 
They can also simulate showering for collider processes involving more than two 
final-state particles in the hard scattering process. For the background 
processes I examine in the following Chapters, however, these programs would 
first require additional tuning because the acceptance requirements imposed on 
the final states do not necessarily satisfy the collinearity assumptions of the 
programs' approximations. For example, {\sc pythia} and {\sc herwig} are known 
to give much too small a cross section for $Z + $ hard jets~\cite{Wjjjj_MC}. 
Thus, input from parton-level studies with full tree-level matrix elements is 
required, which this dissertation provides. 

While the calculations are more reliable when using tree level matrix elements 
for a hard scattering process, one is then forced to make an approximation in 
place of full parton shower simulation. For the simulations here I equate 
final-state colored particles, which are required to be well-separated, with 
final-state jets of a given radius in the lego plot~\footnote{Rectilinear plot 
of the convenient detector variables pseudorapidity ($\eta$) v. azimuthal angle 
($\phi$).} centered around the single particle. This parton-level-only 
approximation is sufficient for proof-of-concept of search strategies and for 
developing tools to enhance a signal relative to its backgrounds. More 
sophisticated parton showering calculations with complete matrix elements for 
the hard scattering processes will eventually be required for comparison of 
theory with experimental data.


\section{Monte Carlo Techniques}
\label{sec:MC}

Calculation of a cross section in field theory is an integration of a squared 
matrix element for a given process, summed over final-state quantum numbers such 
as helicity or color, averaged over the same initial-state quantum numbers if 
the initial-state is unpolarized, and integrated over the phase space of the 
initial- and final-state particles~\cite{crossec}. A total cross section is the 
sum of cross sections for all distinct initial/final-state subprocceses. This 
may be written as
\bq
\label{eq:sigma}
\sigma \, = \, \int dx_1 dx_2 \sum_{subproc} f_{a_1}(x_1) f_{a_2}(x_2) 
            {1\over 2\hat{s}} \int d\Phi_n \Theta(cuts)
            \sum^{\Huge\_} |{\cal M}|^2(subproc)
\, ,
\eq
where the summed/averaged/squared matrix element is
\bq
\label{eq:M2}
\sum^{\Huge\_}|{\cal M}|^2 \, = \, {1\over 4} {1\over \#{\rm colors}(a_1 a_2)}
                                \sum_{\rm colors}\sum_{\rm pol'zn}|{\cal M}|^2
\eq
(the ${1\over 4}$ comes from both initial-state particles $a_1$ and $a_2$ 
being partons with two polarization degrees of freedom), the Lorentz-invariant 
phase space (LIPS) is
\bq
\label{eq:PS}
d\Phi_n(P;p_1...p_n) \, = \, \prod^n_{i=1}
                          \left( {d^3 p_i \over (2\pi)^3 2E_i} \right)
                          (2\pi)^4 \delta^4(P - \sum_i p_i)
\, ,
\eq
the $f_{a_i}(x_i)$ are the parton distribution functions, which give the 
probability to find parton $a_i$ with momentum fraction $x_i$ inside the 
incoming hadron, and $\hat{s} = s \cdot x_1 \cdot x_2$.

For hadron colliders, such as the LHC, it is impossible to calculate any cross 
section analytically, even for total rates, because the squared amplitude must 
be folded with structure functions for the incoming (anti-)protons. In addition, 
the calculational goal is typically not a total rate, but instead an integration 
over only the part of the phase space where the final-state particles are 
visible in the detector. Further cuts on phase space may also be enforced to 
extract a signal from the total background. As such, hadron collider 
calculations require the use of a computer to perform the integration over a 
very complicated phase space configuration. We may approximate the integration 
over the LIPS by transforming to a more convenient set of variables, preferably 
across which the differential cross section is fairly flat, and sampling this 
phase space with quasi-random numbers. This transformation involves a Jacobian 
that must be reevaluated at every phase space point:
\bq
\label{eq:jacobian}
{1\over 2\hat{s}} dx_1 dx_2 d\Phi_n \, = \, J \prod^{3n-2}_{i=1} dr_i \, .
\eq
We then instead evaluate the expression
\bq
\label{eq:compint}
\sigma \, \approx \, {1\over N} \sum_{\{r_i\}} J \Theta(cuts) 
                     \sum_{subproc} f_{a_1}(x_1) f_{a_2}(x_2)
                     \sum^{\Huge\_} |{\cal M}|^2(subproc)
\, ,
\eq
where the matrix element squared is evaluated numerically at each phase space 
point. Typically, the phase space variables $dp_{1x}dp_{1y}dp_{1z}...$ are very 
inefficient, and an expression for the Jacobian transform to detector variables 
such as $p_T,\eta,\phi$ is performed by hand before coding. $p_T$ is simply a 
convenient variable for detector capability, and $\eta$ is useful because it 
expands the angular region where small angle scattering results in a divergent 
cross section, thus making it more efficient for the integration routine to 
avoid. The goal is to choose a set of phase space variables such that the cross 
section is nearly flat across $0 < r_i < 1$ for each variable. In practice, this 
is difficult to achieve, but a transformation to detector variables is vastly 
more efficient than using rectilinear variables. I outline the general structure 
of our Monte Carlo programs and our technique for generating matrix elements in 
Appendix~\ref{app:programs}.


%% file: body/WBF.tex

\section{Introduction}

\begin{figure}[t]
\begin{picture}(0,0)(0,0)
\includegraphics{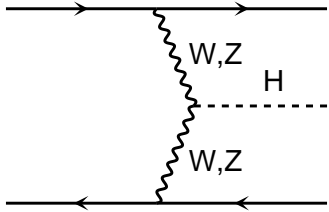}
\end{picture}
\vspace{3cm}
\caption{Weak boson fusion Higgs production.}
\label{fig:Hjj}
\end{figure}

For the intermediate mass range, $M_H \approx 100-200$~GeV, most of the 
literature has focussed on Higgs production via gluon fusion~\cite{reviews} and 
$t\bar{t}H$~\cite{ttH} or $WH(ZH)$~\cite{WH} associated production. Cross 
sections for Higgs production at the LHC are well-known~\cite{reviews}, and 
while production via gluon fusion has the largest cross section by almost one 
order of magnitude, there are substantial QCD backgrounds but few handles to 
distinguish them from the signal.  Essentially, only the decay products' 
transverse momentum and the resonance in their invariant mass distribution (if 
it can be constructed) can be used. The second largest production cross section 
for the standard model (SM) Higgs boson is predicted for weak-boson fusion 
(WBF), $qq \to qqVV \to qqH$, shown in Fig.~\ref{fig:Hjj}. WBF events contain 
additional information in the observable quark jets. Techniques like forward jet 
tagging~\cite{Cahn,BCHP,DGOV} can then be exploited to reduce the backgrounds. 
I discuss the basic signal process and characteristics, including the results 
of the forward jet tagging technique, in Sec.~\ref{sec:WBF}.

It is necessary to study other production channels for several reasons. For 
instance, electroweak symmetry breaking (ESB) and fermion mass generation may be 
less intimately connected than in the Standard Model (SM) and the coupling of 
the lightest Higgs resonance to fermions might be severely suppressed. In this 
case, neither $gg \to H$ fusion nor $t\bar{t}H$ associated production would be 
observed. Once the Higgs is observed in both $gg \to H$ and the WBF process 
$qq \to qqH$, where the Higgs is radiated off virtual $W$'s 
or $Z$'s, the cross section ratio of these modes measures the ratio of the Higgs 
coupling to the top quark and to W,Z. This value is fixed in the SM, but 
deviations are expected in more general models, like supersymmetry with its two 
Higgs doublets~\cite{susycoupl}. As I shall demonstrate, the WBF channel may 
yield a quicker discovery, requiring only 2-30~fb$^{-1}$, depending on the Higgs 
mass, which is less than the integrated luminosity needed for discovery in the 
$gg \to H$ channel in some decay modes. Since the Higgs can be discovered in the 
WBF channel at relatively low integrated luminosity, a fairly precise 
measurement of various Higgs couplings can be obtained with 100~fb$^{-1}$ of 
data or more.

Another feature of the WBF signal is the lack of color exchange between the 
initial-state quarks. Color coherence between initial- and final-state gluon 
bremsstrahlung leads to suppressed hadron production in the central region, 
between the two tagging-jet candidates of the signal~\cite{bjgap}. This is in 
contrast to most background processes, which typically involve color exchange 
in the $t$-channel and thus lead to enhanced hadronic activity in the central 
region. We may exploit these features, via a veto on additional soft jet 
activity in the central region~\cite{bpz_minijet}. I describe the idea of a 
minijet veto in Sec.~\ref{sec:minijet}, including different techniques for 
estimating veto probabilities.


\section{The WBF Signature}
\label{sec:WBF}

The signal can be described, at lowest order, by two single-Feynman-diagram
processes, $qq \to qq(WW,ZZ) \to qqH$, i.e. $WW$ and $ZZ$ fusion where the
weak bosons are emitted from the incoming quarks~\cite{qqHorig}, as shown in 
Fig.~\ref{fig:Hjj}. As the Higgs is a spin-0 particle, its decay may be treated 
separately from production, either as a branching ratio or via a decay matrix 
element which is squared and summed over helicities separately, then simply 
multiplied to the production cross section. Thus, we can discuss general 
features of WBF Higgs production independent of the decay channel. 
\footnote{Physical parameters, parton distribution functions and the 
factorization scale are chosen as in Appendix~\ref{app:parms}. Detector 
simulation is included to the extent of energy/momentum resolution of the 
final-state particles via Gaussian smearing prescriptions based on the LHC 
detector expectations; see Appendix~\ref{app:smear} for details.}

The first task is to identify the search region for these hard $Hjj$ events. 
Features of the signal are a centrally produced Higgs which tends to yield 
central decay products, and two jets which enter the detector at large rapidity 
compared to the decay products. I start out by discussing three levels of 
general cuts on the $qq\to qqH$ signal, before considering decay products and 
their identification. This procedure makes explicit the source of the major 
signal reduction factors encountered.

\begin{figure}[t]
\vspace*{0.5in}            
\begin{picture}(0,0)(0,0)
\includegraphics{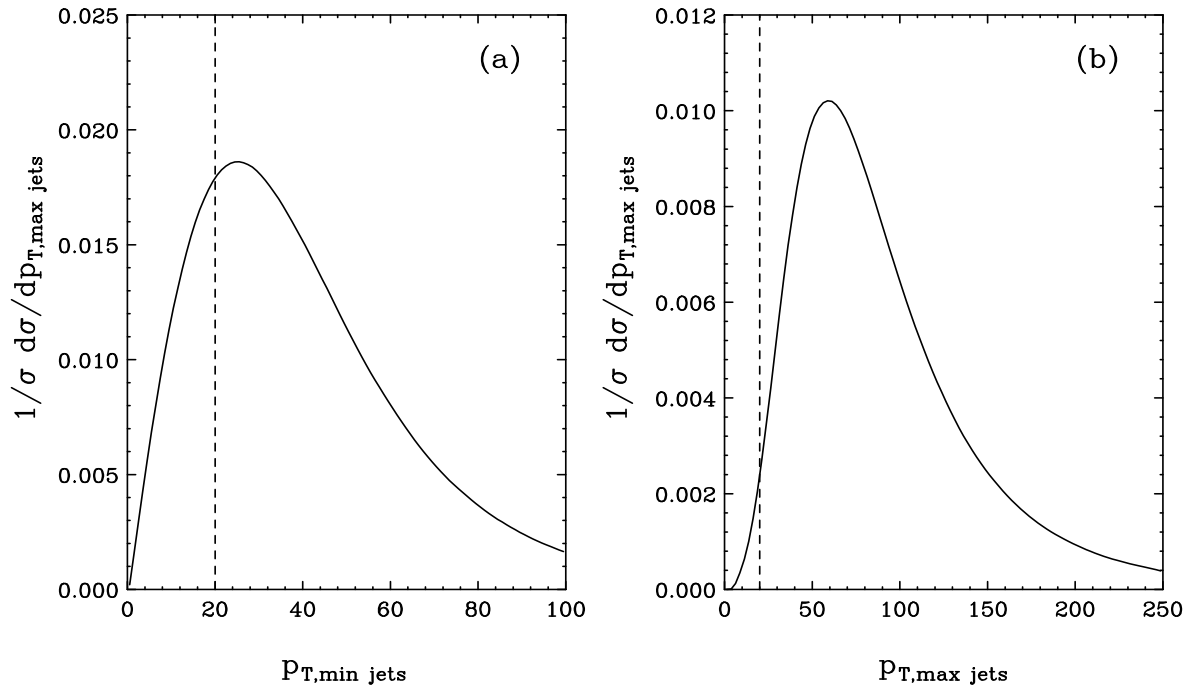}
\end{picture}
\vspace{8cm}
\caption{Normalized minimum and maximum $p_T$ distributions of the two jets in 
signal $qq\to qqH,H\to\gamma\gamma$ events at the LHC, with $M_H = 120$~GeV. 
No cuts are imposed; the vertical dashed lines represent the minimum detector 
acceptance $p_{T_j}$ cut of Eq.~\protect\ref{eq:basic}.}
\label{fig:pT_jets_incl}
\end{figure}
\begin{figure}[t]
\vspace*{0.5in}            
\begin{picture}(0,0)(0,0)
\includegraphics{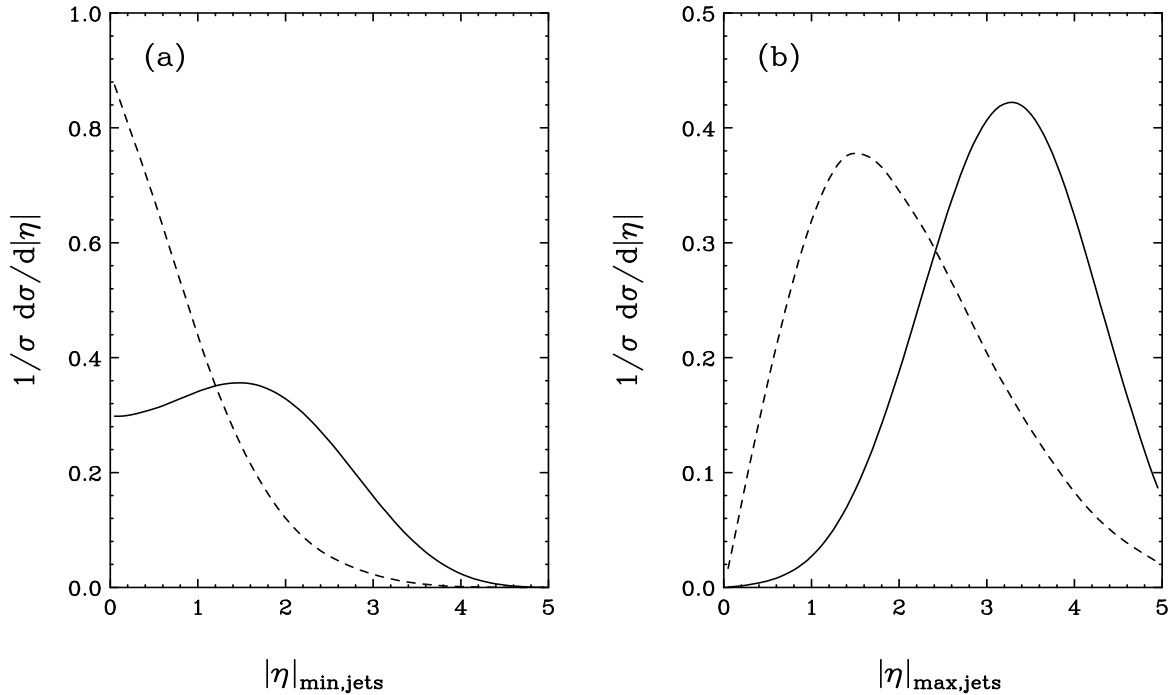}
\end{picture}
\vspace{8cm}
\caption{Normalized pseudo-rapidity distributions of (a) the most central and 
(b) the most forward of the two tagging jets in $\gamma\gamma jj$ events at the 
LHC. The generic acceptance cuts of Eq.~(\protect\ref{eq:basic}) are imposed. 
Results are shown for the $qq\to qqH$ signal at $m_H=120$~GeV (solid line) and 
for the irreducible QCD $\gamma\gamma jj$ background (dashed line).}
\label{fig:y_jets_basic}
\end{figure}
\begin{figure}[t]
\vspace*{0.5in}            
\begin{picture}(0,0)(0,0)
\includegraphics{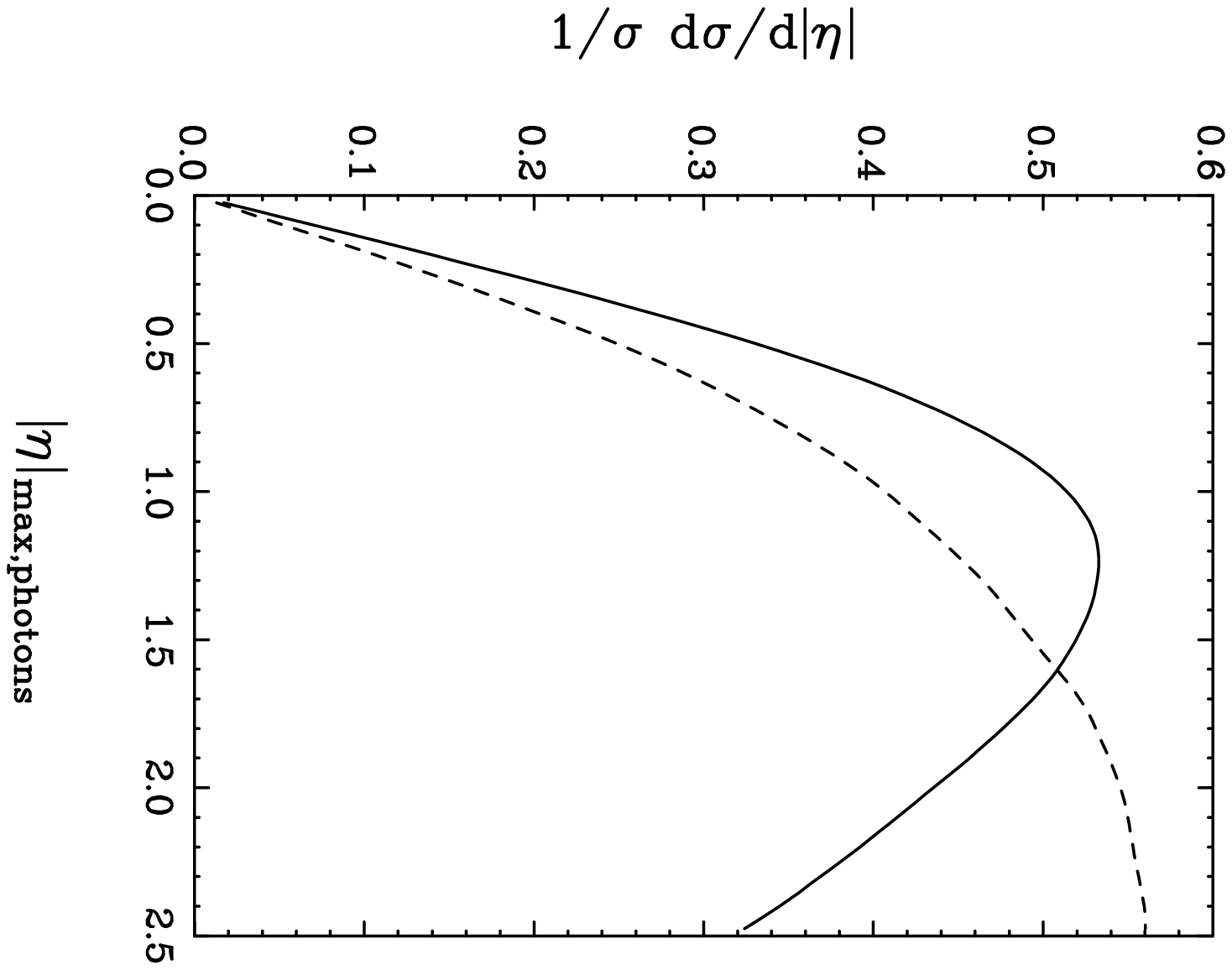}
\end{picture}
\vspace{8cm}
\caption{Normalized pseudo-rapidity distributions of the most forward of the two 
photons in $\gamma\gamma jj$ events at the LHC. The generic acceptance cuts of 
Eq.~(\protect\ref{eq:basic}) are imposed. Results are shown for the $qq\to qqH$ 
signal at $m_H=120$~GeV (solid line) and for the irreducible QCD 
$\gamma\gamma jj$ background (dashed line).}
\label{fig:y_pho_basic}
\end{figure}

The basic acceptance requirements must ensure that the two jets and the Higgs 
decay products are observed inside the detector. The outgoing quarks/gluons must 
lie within the reach of the hadronic calorimeter, with sufficient $p_T$ to be 
identified as jets. As the decay products are typically leptons or photons, they 
are required to lie within the electromagnetic calorimeter. In addition, the 
jets and observable decay products (generically denoted by $X$) must be 
well-separated from each other:
\ba
\label{eq:basic}
& p_{T_j} \geq 20~{\rm GeV} \, ,\qquad |\eta_j| \leq 5.0 \, ,\qquad 
\triangle R_{jj} \geq 0.7 \, , & \nonumber\\
& |\eta_X| \leq 2.5 \, , \qquad \triangle R_{jX} \geq 0.7 \, . &
\ea
The maximum and minimum jet $p_T$ and $\eta$ distributions of the $Hjj$ signal 
in Figs.~\ref{fig:pT_jets_incl} and \ref{fig:y_jets_basic} demonstrate 
that these basic acceptance requirements remove little of the signal. The 
figures are shown for $M_H = 120$~GeV, but this is true regardless of mass, 
which is not surprising as the $p_T$'s of the jets are governed by the scale of 
the weak boson masses and are thus rarely small. Similarly, the maximum rapidity 
distribution of the Higgs decay products, shown for the case of photons for 
simplicity in Fig.~\ref{fig:y_pho_basic}, tend to be very central, well within 
the reach of the electromagnetic calorimeters, which will extend to about 2.5 
units in $\eta$ for both CMS and ATLAS. The QCD, backgrounds, on the other hand, 
seen in the same figures, typically have more central jets and more forward 
photons. Slightly more than half of all signal events pass these basic cuts, as 
seen in the first two columns of Table~\ref{table_base}.

\begin{figure}[t]
\vspace*{0.5in}            
\begin{picture}(0,0)(0,0)
\includegraphics{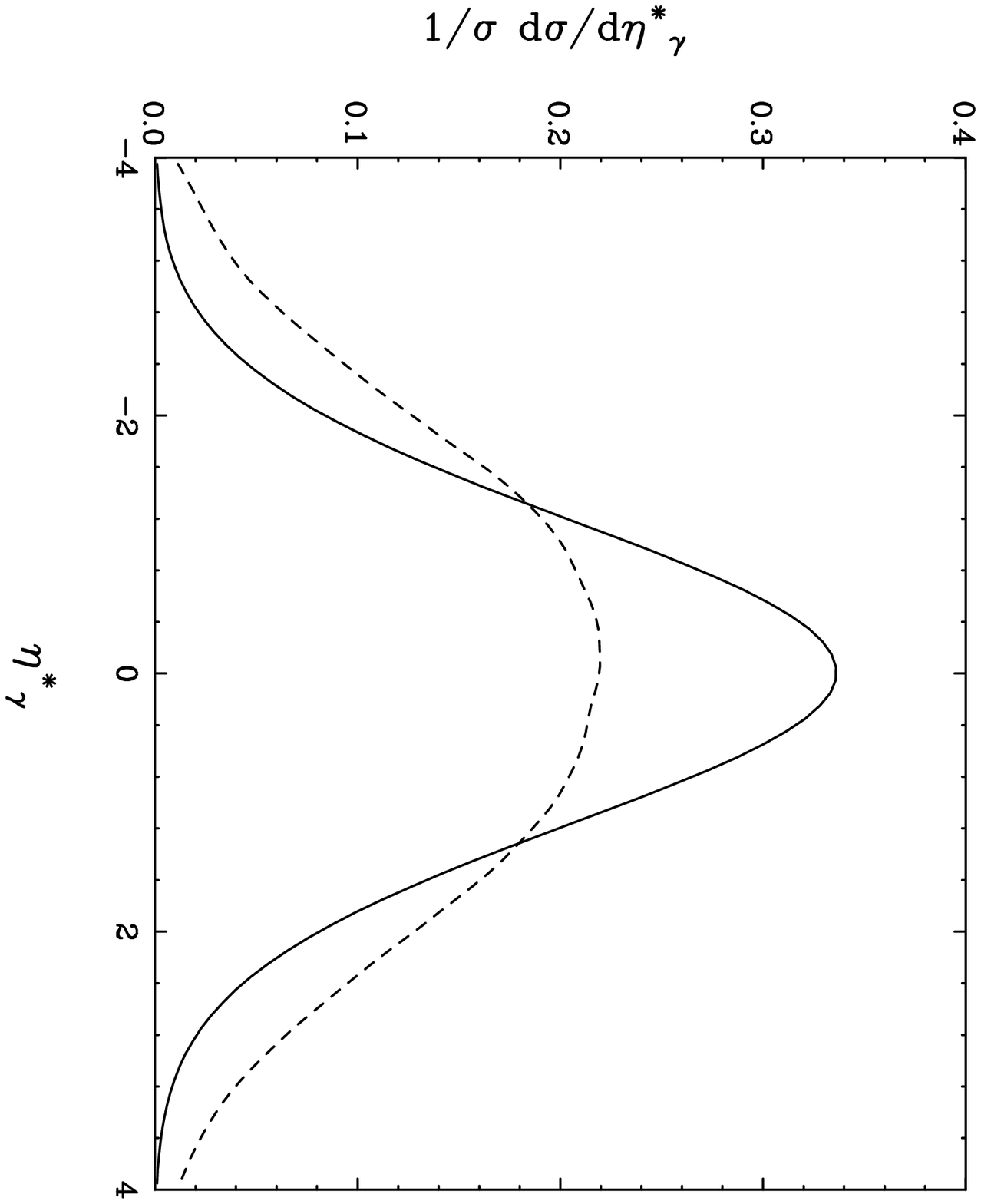}
\end{picture}
\vspace{8.5cm}
\caption{Normalized pseudo-rapidity distributions of both photon rapidities 
with respect to the center of the tagging jets (as discussed in the text) in 
$\gamma\gamma jj$ events at the LHC. The acceptance cuts of 
Eq.~(\protect\ref{eq:basic}) are imposed. Results are shown for the $qq\to qqH$ 
signal at $m_H=120$~GeV (solid line) and for the irreducible QCD 
$\gamma\gamma jj$ background (dashed line).}
\label{fig:ystar_pho}
\end{figure}

The fact that irreducible QCD backgrounds typically give higher-rapidity $X$'s 
as compared to the Higgs signal is due to the bremsstrahlung nature of $W$, $Z$ 
and $\gamma$ emission, which occur at small angles with respect to the parent 
quarks, producing $\gamma$'s or leptons forward of the jets. To illustrate this 
more clearly, we may define a new variable, a shifted photon rapidity, 
$\eta^*_{\gamma}$, which is the rapidity of the photons with respect to the 
center of the tagging jets, $\eta^*_{\gamma_i} \, = \, 
\eta_{\gamma_i} - {1\over 2}(\eta_{tag 1}+\eta_{tag 2})$, which is shown for 
$Hjj,H\to\gamma\gamma$ signal and QCD $\gamma\gamma jj$ background events with 
the basic cuts of Eq.~(\ref{eq:basic}) in Fig.~\ref{fig:ystar_pho}. An obvious 
cut is to reject events that typically have Higgs decay products outside the 
central region defined by the tagging jets. Thus, at the second level of cuts 
both $X$'s are required to lie between the jets with a minimum separation from 
each jet in pseudorapidity $\triangle \eta_{jX} > 0.7$; and the jets are 
required to occupy opposite hemispheres:
\bq
\label{eq:Xcen}
\eta_{j,min} + 0.7 < \eta_{X_{1,2}} < \eta_{j,max} - 0.7 \, , \qquad
\eta_{j_1} \cdot \eta_{j_2} < 0 \: .
\eq
\begin{figure}[t]
\vspace*{0.5in}            
\begin{picture}(0,0)(0,0)
\includegraphics{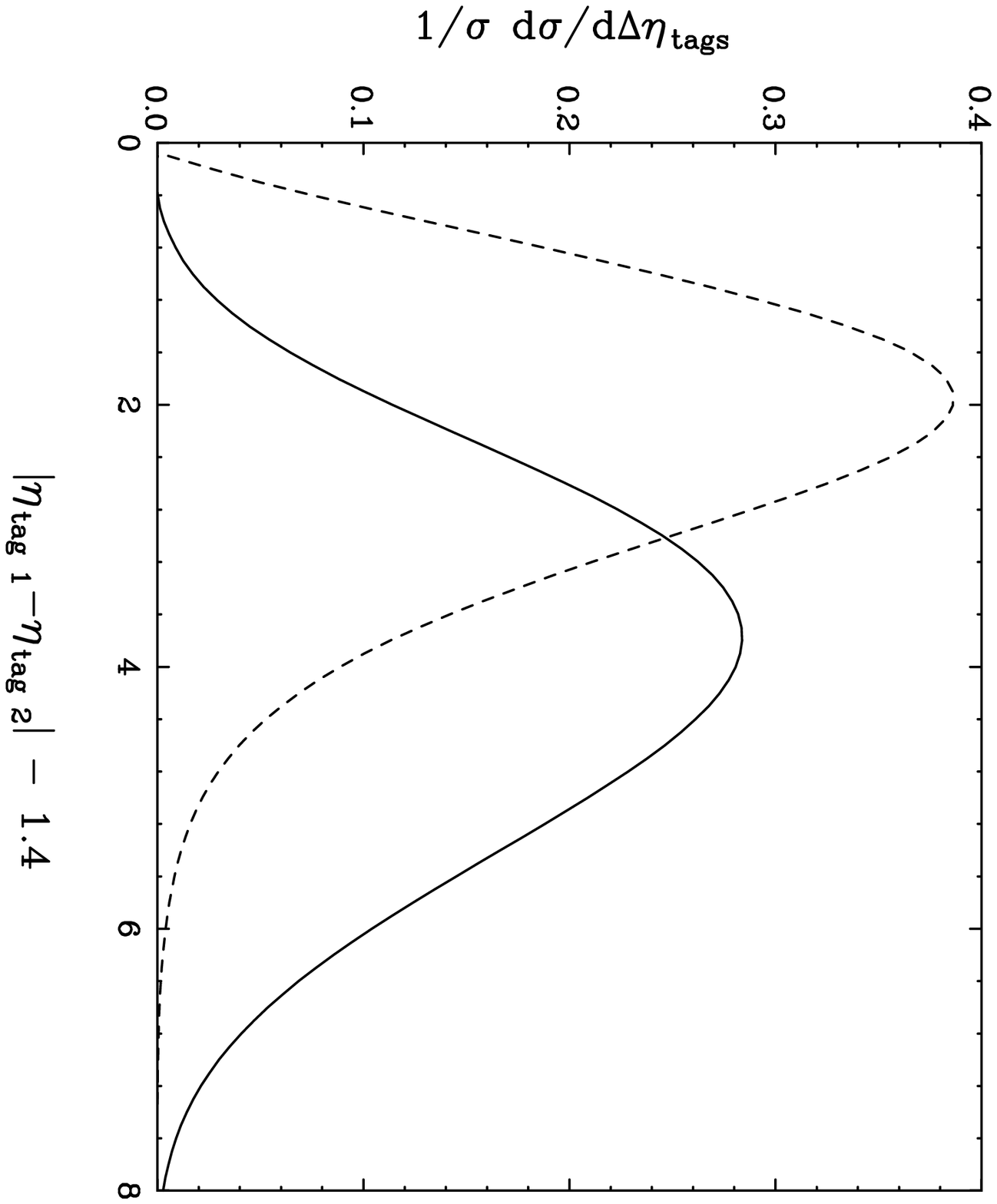}
\end{picture}
\vspace{8.5cm}
\caption{Normalized pseudo-rapidity distributions of the width of the allowable 
Higgs decay product observable gap region $\eta$, 
$|\eta_{tag 1}-\eta_{tag 2}|-1.4$, in $\gamma\gamma jj$ events at the LHC. The 
acceptance cuts of Eqs.~(\protect\ref{eq:basic},\protect\ref{eq:Xcen}) are 
imposed. Results are shown for the $qq\to qqH$ signal at $m_H=120$~GeV (solid 
line) and for the irreducible QCD $\gamma\gamma jj$ background (dashed line).}
\label{fig:dely_tags}
\end{figure}

Finally, Fig.~\ref{fig:dely_tags}, which shows the tagging jet separation in 
$\eta$ for $Hjj,H\to\gamma\gamma$ and QCD $\gamma\gamma jj$ events, reveals that 
the separation is drastically wider for the signal. The narrow separation of the 
tagging jets is a general feature of QCD processes at the LHC, which are largely 
gluon-initiated; large separations require higher invariant masses, and thus 
restrict the processes to the region of Feynman $x$ where quarks dominate. Thus, 
at the third level of cuts, which is also the starting point for our 
consideration of the various backgrounds, a wide separation in pseudorapidity is 
required between the two forward tagging jets, 
\bq
\label{eq:gap}
\triangle \eta_{tags} = |\eta_{j_1}-\eta_{j_2}| \geq 4.4 \, ,
\eq
leaving a gap of at least 3 units of pseudorapidity in which the $X$'s can be 
observed. This technique to separate weak boson scattering from various 
backgrounds is 
well-established~\cite{Cahn,BCHP,DGOV,bpz_minijet,RZ_gamgam,RZ_tautau,RZ_WW}, 
in particular for heavy Higgs boson searches. Table~\ref{table_base} shows the 
effect of these cuts on the signal for a SM Higgs boson over the mass range 
$m_H = 100-200$~GeV. 
Overall, about $29\%$ of all $Hjj$ events generated in weak boson fusion are 
accepted by the cuts of Eqs.~(\ref{eq:basic}-\ref{eq:gap}). These cuts form the 
core identification technique for a WBF Higgs signal. Additional, unique cuts 
used to distinguish the very different cases $X = \gamma,W,\tau$ will be 
discussed in the appropriate Chapters.

\begin{table}
\caption{Signal inclusive cross sections (pb) for $Hjj$ events of various Higgs 
masses in $pp$ collisions at $\protect\sqrt{s}=14$~TeV. Results are given for no 
cuts and successive cuts of Eqs.~(\protect\ref{eq:basic}-\protect\ref{eq:gap}). 
For the angular cuts of Eq.~(\protect\ref{eq:basic}) on the Higgs decay 
products, $H$ is assumed to decay into two massless particles ($b\bar{b}$, 
$\tau^+\tau^-$, $\gamma\gamma$, etc.).}
\label{table_base}
\begin{center}
\begin{tabular}{|p{0.8in}|p{0.8in}|p{0.8in}|p{0.8in}|p{0.8in}|}
\hline\hline
Higgs mass (GeV) & inclusive &
   + Eq.~(\ref{eq:basic}) & + Eq.~(\ref{eq:Xcen}) & + Eq.~(\ref{eq:gap}) \\
\hline
100 & 4.8 & 2.7 & 1.62 & 1.21 \\
120 & 4.1 & 2.4 & 1.45 & 1.10 \\
140 & 3.5 & 2.1 & 1.30 & 1.00 \\
160 & 3.1 & 1.8 & 1.18 & 0.92 \\
180 & 2.7 & 1.6 & 1.07 & 0.84 \\
200 & 2.4 & 1.4 & 0.97 & 0.78 \\
\hline\hline
\end{tabular}
\end{center}
\end{table}

I should note that this calculation of the $Hjj$ signal process has very low 
uncertainties, as reflected by its stability against factorization scale 
uncertainty: variation of the factorization scale by a factor of two changes the 
2-jet cross section at the level of forward tagging cuts 
(Eqs.~(\ref{eq:basic}-\ref{eq:gap})), shown in the last column of 
Table~\ref{table_base}, by $\leq \pm 10\%$. This implies that next-to-leading 
order (NLO) corrections to this process are expected to be small, which is borne 
out by calculations found in Ref.~\cite{Hjj_NLO}.


\section{Distribution Patterns of Additional Radiation}
\label{sec:minijet}

A further characteristic of EW vs. QCD scattering can be exploited, namely the 
absence of color exchange between the two scattering quarks in the $qq\to qqH$ 
signal process. $t$-channel color singlet exchange in the EW case leads to soft 
gluon emission mainly in the very forward and very backward directions, whereas 
QCD processes are dominated by $t$-channel color octet exchange which results 
in soft gluon radiation mainly in the central detector. It was hoped that 
resulting rapidity gaps in signal events (large regions in pseudorapidity 
without observed hadrons) could be used for background suppression~\cite{bjgap}. 
Unfortunately, in $pp$ collisions of $\sqrt{s}=14$~TeV at the LHC, overlapping 
events in a single bunch crossing will likely fill a rapidity gap even if it is 
present at the level of a single $pp$ collision. Very low luminosity running is 
not useful because of the small signal cross section.

\begin{figure}[t]
\vspace*{0.5in}            
\begin{picture}(0,0)(0,0)
\includegraphics{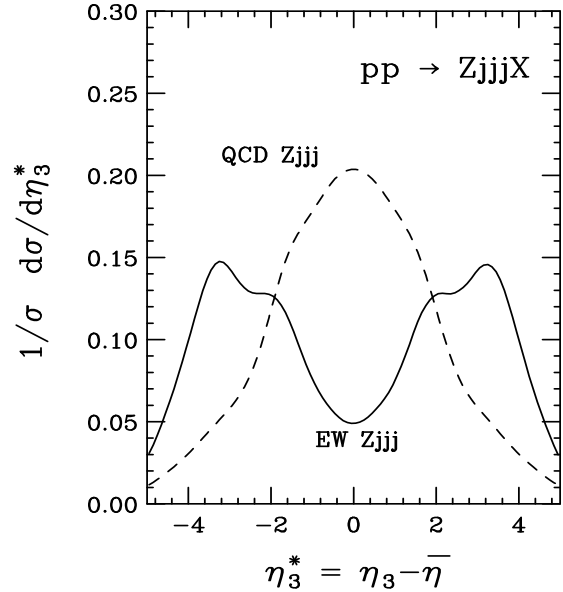}
\end{picture}
\vspace{6.5cm}
\caption{Characteristics of the third (soft) jet in EW $Zjjj$ ``signal'' (solid 
lines) and QCD $Zjjj$ ``background'' (dashed lines) events at the LHC. 
Acceptance cuts are as given in the text. The pseudorapidity $\eta^*_3$ is 
measured with respect to the center of the two tagging jets, 
$\bar\eta = {1\over 2}( \eta^{tag 1}_j + \eta^{tag 2}_j )$ and the distributions 
are normalized to the total respective cross sections.}
\label{fig:Zjjj}
\end{figure}

The different color structures of signal and background processes can be 
exploited even at high luminosity, however, if one defines rapidity gaps in 
terms of minijets ($p_{Tj} \approx$ 15-40~GeV) instead of soft 
hadrons~\cite{bpz_minijet}. As has been shown for the analogous EW $Zjj$ 
process~\cite{RSZ_vnj}, with its very similar kinematics, minijet emission in 
EW exchange occurs mainly in the very forward and very backward regions, and 
even here is substantially softer than in the QCD $Zjj$ analogue. This is 
shown in Fig.~\ref{fig:Zjjj}, where we impose the cuts of 
Eqs.~(\ref{eq:basic}-\ref{eq:gap}) and additionally require
\bq
\label{eq:addcuts}
p_{Tj} > 40 \: {\rm GeV} \, ; \; 
p_{T\ell} > 20 \: {\rm GeV} \, , \; |\eta_\ell| < 2 \, ,
\eq
and
\bq
M_Z - 10 \: {\rm GeV} < m_{\ell\ell} < M_Z + 10 \: {\rm GeV} \, . 
\eq
Fig.~\ref{fig:Zjjj} shows the rapidity distribution of the additional radiation 
(jet of $p_T > 20$~GeV) with respect to the center of the two tagging jets. It 
is clear that the additional radiation in QCD processes tends to be much more 
central than in EW processes; the knee in the EW $Zjj$ curve at 
$|\eta|\approx 2.5$ is a phase space effect: that is the average rapidity of the 
tagging jets, and loss of rate is experienced due to the finite cone sizes of 
the tagging and third jets. This suggests that a veto on these central minijets 
will substantially improve the signal-to-background ratio. Following the 
analysis of Ref.~\cite{RSZ_vnj}, a veto is imposed on additional central jets, 
{\em i.e.}, in the region 
\ba
\label{eq:mjveto}
p_{Tj}^{\rm veto} & > & p_{T,{\rm veto}} \;, \label{eq:ptveto} \\
\eta_{j,min}^{\rm tag} & < & \eta_j^{\rm veto}
< \eta_{j,max}^{\rm tag} \: , \label{eq:etaveto}
\ea
where $p_{T,\rm veto}$ may be chosen based on the capability of the detector; 
it is expected to be about 20~GeV for the LHC.

Sizable background reduction via a minijet veto requires lowering of the 
$p_{T,\rm veto}$ threshold to a range where the probability for additional 
parton emission becomes order unity. In a perturbative calculation the 
resulting condition, $\sigma(n+1\;{\rm jets})\approx \sigma(n\;{\rm jets})$, 
indicates that one is leaving the validity range of fixed-order perturbation 
theory, and it becomes difficult to provide reliable theoretical estimates of 
minijet emission rates. This happens at much higher values of $p_{T,\rm veto}$ 
for QCD processes than for their EW analogues. Gluon emission is governed by 
very different scales in EW as compared to QCD processes, due to their different 
color structures. Thus, a parton shower approach does not immediately give 
reliable answers unless both color coherence and the choice of scale are 
implemented correctly, corresponding to the answer given by a complete QCD 
calculation.

For the study of a central jet veto, the emission of at least one extra parton 
must be simulated. This is achieved by calculating the cross sections for the 
relevant ``2-jet'' process under consideration with additional gluon radiation 
and all crossing related processes. For the $Hjj$ signal this includes
\bq
\label{procsig}
q\bar{q} \to q\bar{q}Hg \, ,\qquad \;
\bar q\bar q \to \bar q\bar qHg \, , \qquad
qg \to qq\bar{q}H \, , \qquad
\bar q g \to \bar qq\bar{q}H \, ,
\eq
which can be found in Ref.~\cite{qqHjorig}. For this case with three 
final-state partons, the factorization scale is chosen as $\mu_f =$ min($p_T$) 
of the tagging jets and the renormalization scale $\mu_r$ is set to the 
transverse momentum of the non-tagging parton (minijet). Different scale 
choices or different input parameters will, of course, affect our numerical 
results. 

While the necessary information on angular distributions and hardness of 
additional radiation is available in the ``3-jet'' and $t\bar t + jets$ 
processes discussed in Chapters~\ref{ch:gammagamma}-\ref{ch:tautau}, one must 
either regulate or reinterpret these divergent cross sections. I use the 
truncated shower approximation (TSA)~\cite{TSA} for the former, treating the 
``2-jet'' cross sections as the inclusive rate. For the latter I assume that 
additional soft parton emission, which will be dominated by soft gluons, 
exponentiates like soft photon emission. A Poisson distribution can then be used 
to estimate the veto probability. Details of these procedures can be found in 
Appendix~\ref{app:minijet} and Refs.~\cite{RZ_tautau,RSZ_vnj}. Once the veto 
probability $P_{veto}$ for a given process is known, the total rate is estimated 
by multiplying this by the ``2-jet'' cross section, 
\bq
\label{eq:surv}
\sigma_{surv} = \sigma_{2-jet} \cdot (1-P_{veto}) 
              = \sigma_{2-jet} \cdot P_{surv} \, .
\eq
The values for $P_{surv}$ I use in the following Chapters for the various 
processes considered are shown in Table~\ref{vetosum.}.

\begin{table}
\caption{Summary of veto survival probabilities for $p_T^{veto} = 20$~GeV used 
in Chapters~\protect\ref{ch:gammagamma}-\protect\ref{ch:tautau}.}
\label{vetosum.}
\begin{center}
\begin{tabular}
{|p{0.85in}|p{0.35in}|p{0.35in}|p{0.35in}|p{0.55in}|p{0.55in}|p{0.45in}|p{0.4in}
 |p{0.4in}|p{0.4in}|p{0.35in}|}
\hline\hline
search & $Hjj$ & $t\bar{t}$ & $t\bar{t}j,$ & QCD      & EW       & QCD    
& QCD          & DPS               \\
       &       &            & $t\bar{t}jj$ & $V(V)jj$ & $V(V)jj$ & $Wjjj$ 
& $b\bar{b}jj$ & $\gamma\gamma jj$ \\
\hline
$\gamma\gamma jj$  & 0.89 &  -   &  -   & 0.30 & 0.75 &  -   &  -   & 0.30 \\
$W^{(*)}W^{(*)}jj$ & 0.89 & 0.46 & 0.29 & 0.29 & 0.75 &  -   &  -   &  -   \\
$\tau\tau jj$      & 0.87 &  -   &  -   & 0.28 & 0.80 & 0.28 & 0.28 &  -   \\
\hline\hline
\end{tabular}
\end{center}
\end{table}
%


\section{Summary}

I have identified and discussed the important characteristics of the core 
signal process, $H$ production via weak boson fusion, in association with two 
quark jets, and established base kinematic cuts useful in enhancing the signal 
relative to a large class of backgrounds. These are the far forward and 
backward tagging jets of moderate $p_T$, and a centrally produced Higgs with 
considerable $p_T$, which will tend to yield decay products with relatively 
high $p_T$. Additional cuts will be necessary to yield an observation of the 
Higgs, but will uniquely depend on the decay final state of the Higgs, and are 
discussed in the following Chapters.

I have also identified an additional important distinction between the signal 
WBF process and general QCD backgrounds, the collective hardness and rapidity 
distributions of additional partonic activity in the central region between the 
tagging jets, attributable to higher-order gluon emission from the hard 
scattering process. I have discussed the method for calculating these 
distributions, which are only marginally perturbative, mentioned two techniques 
for extracting a perturbative result, and provided a summary of the results. The 
techniques are discussed in detail in Appendix~\ref{app:minijet}.


%% file: body/gammagamma.tex

\section{Introduction}

Fits to precision electroweak (EW) data have for some time suggested a 
relatively low Higgs boson mass, in the 100~GeV range~\cite{EWfits} and this is 
one of the reasons why the search for an intermediate mass Higgs boson is 
particularly important~\cite{reviews}. Beyond the reach of LEP, for masses in 
the $100-150$~GeV range, the $H\to \gamma\gamma$ decay channel at the CERN LHC 
is very promising. Consequently, LHC detectors are designed with excellent 
photon detection capabilities, resulting in the case of the CMS detector in a 
di-photon mass resolution of order 1~GeV for a Higgs boson mass around 
120~GeV~\cite{CMS-ATLAS}.  Another advantage of the $H\to \gamma\gamma$ channel, 
in particular compared to the dominant $H\to b\bar b$ mode, is the lower 
background from QCD processes. 

In this Chapter I demonstrate the observation potential of $H\to\gamma\gamma$ in 
WBF, outlining the additional cuts needed to enhance the signal relative to the 
principal backgrounds, and including estimates of minijet veto survival 
probability. This discussion closely follows that in Ref.~\cite{RZ_gamgam}, but 
here I reanalyze this mode to include detector resolution effects, 
identification efficiencies and improved mass resolution estimates, corrections 
to the $H\to\gamma\gamma$ branching ratio, and estimates of the minijet veto 
survival probabilities, found in Appendix~\ref{app:minijet}. A programming bug 
was also fixed that affected the QCD and EW $\gamma\gamma jj$ backgrounds, 
increasing them by $\approx 20\%$.


\section{Calculational Tools}
\label{sec:gam_tools}

Higgs production and its characteristics has already been described in 
Section~\ref{sec:WBF}. We must now consider Higgs decay to the final state 
$\gamma\gamma$. As the Higgs is a spin-0 particle, it is sufficient to treat its 
decay simply by multiplying the total cross section by the appropriate branching 
ratio and using phase space distributions for the decay products. For the 
$H\to\gamma\gamma$ partial decay width I use input from the program 
{\sc hdecay}~\cite{Hdecay}. In the mass region of interest, 100-150~GeV, 
$B(H\to\gamma\gamma)$ varies from $0.13\%$ to $0.22\%$, with the maximum value 
at about 125~GeV. As the photons are electromagnetic in nature, their momenta 
are smeared according to the resolution prescribed by Eq.~\ref{eq:smear_em}.

\begin{figure}[t]
\vspace*{0.5in}            
\begin{picture}(0,0)(0,0)
\includegraphics{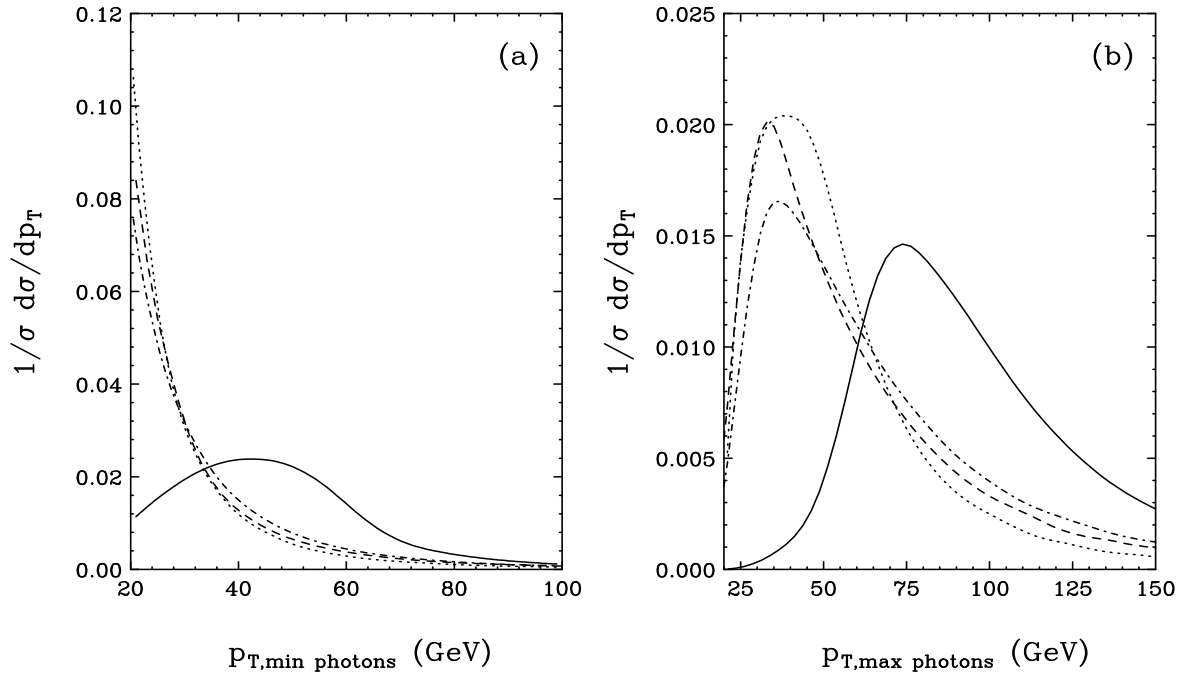}
\end{picture}
\vspace{8cm}
\caption{Normalized transverse momentum distributions of (a) the minimum-$p_T$ 
photon and (b) the maximum-$p_T$ photon in $jj\gamma\gamma$ events at the LHC. 
The core acceptance cuts of Eqs.~(\protect\ref{eq:basic}-\protect\ref{eq:gap}) 
are imposed. Results are shown for the $qq\to qqH$ signal at $m_H = 120$~GeV 
(solid line), the irreducible QCD background (dashed line), the irreducible EW 
background (dot-dashed line), and for the double parton scattering (DPS) 
background (dotted line).}
\label{fig:pT_pho}
\end{figure}

In addition to the basic features of the $Hjj$ signal, a centrally produced 
Higgs and two far forward, semi-hard tagging jets, the $H\to\gamma\gamma$ 
process will produce hard, central, isolated photons. These additional 
characteristics are shown in Fig.~\ref{fig:pT_pho}. The two photons tend to be 
emitted with considerable $p_T$, in the 20-50~GeV range, due to the massive 
nature of the Higgs, relative to massless photons.

\begin{figure}[t]
\vspace*{0.5in}            
\begin{picture}(0,0)(0,0)
\includegraphics{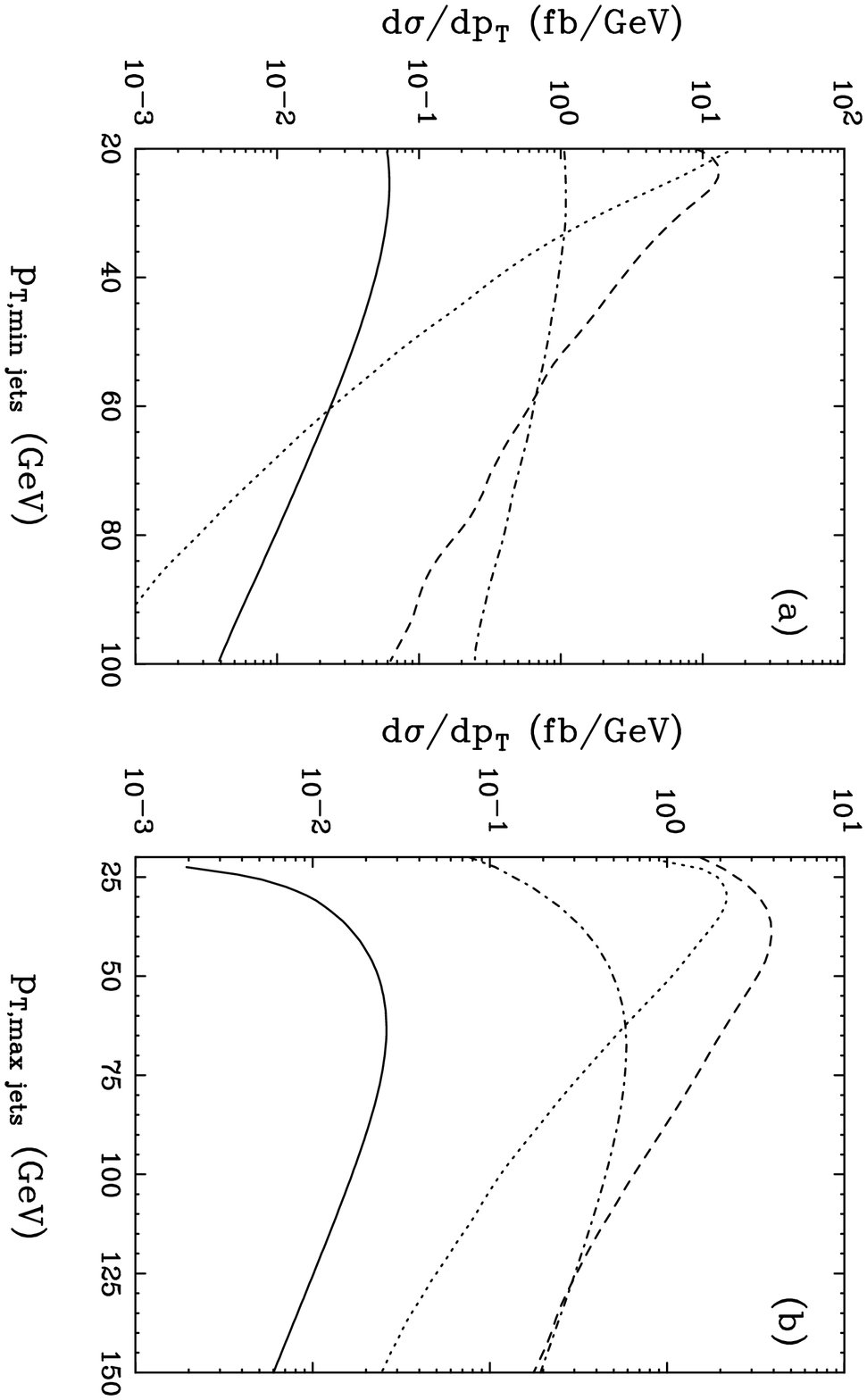}
\end{picture}
\vspace{8cm}
\caption{Transverse momentum distributions of (a) the softer and (b) the 
harder of the two tagging jets in $jj\gamma\gamma$ events. Generic acceptance
cuts (Eq.~(\protect\ref{eq:basic})) and forward jet tagging cuts 
(Eq.~(\protect\ref{eq:Xcen})) are imposed.}
\label{fig:pT_jets}
\end{figure}

Given the features of the signal, one must consider background processes that 
can lead to events with two hard, isolated photons and two forward jets. The 
largest background consists of all QCD $2\to 2$ processes which contain one or 
two quark lines, from which the two photons are radiated. Examples are 
$q \bar Q\to q\bar Q\gamma\gamma$ or $qg\to qg\gamma\gamma$. Matrix elements 
for the irreducible QCD processes are available in the literature~\cite{VVjj}. 
For this irreducible QCD background, the renormalization scale and factorization 
scales are chosen as described in Appendix~\ref{app:parms}.

The EW analog to the above QCD processes is an additional background. The 
irreducible EW background consists of $qQ\to qQ$ processes mediated by
$t$-channel $Z$, $\gamma$, or $W$ exchange, with additional radiation of two 
photons. $\gamma$ and $Z$ exchange processes have amplitudes which are 
proportional to those of analogous gluon exchange processes, but with smaller 
couplings. These are ignored because, in all regions of phase space, they 
constitute only a tiny correction to the irreducible QCD background.  All 
charged current $qQ\to qQ\gamma\gamma$ (and crossing related) processes are 
included, however; while formally of order $\alpha^4$ and thus suppressed 
compared to the order $\alpha^3$ Higgs signal, the small $H\to\gamma\gamma$ 
branching ratio leads to comparable event rates. Furthermore, $W$ exchange 
processes can produce central photons by emission from the exchanged $W$ and, 
therefore, are kinematically similar to the signal. This signal-like component 
remains after forward jet tagging cuts, as can readily be seen in the $p_T$ 
distribution of the jets in Fig.~\ref{fig:pT_jets}. I am not aware of previous 
calculations of the irreducible EW background, and construct the matrix elements 
for it using {\sc madgraph}~\cite{Madgraph}. The irreducible EW background is 
determined with the same choice of factorization scale as the signal.

Double parton scattering (DPS), with pairs of jets and/or photons arising from 
two independent partonic collisions in one $pp$ interaction, must also be 
considered. I do not, however, consider reducible backgrounds, where e.g. a 
jet fragmenting into a leading $\pi^0$ is misidentified as a photon. Reducible 
backgrounds were shown to be small compared to irreducible ones in the analysis 
of the $gg\to H\to\gamma\gamma$ signal~\cite{CMS-ATLAS} and I assume the same 
to hold for the cleaner signal considered here. 

With jet transverse momenta as low as 20~GeV, double parton scattering (DPS) is 
a potential source of backgrounds that must also be considered. DPS is the
occurrence of two distinct hard scatterings in the collision of a single pair
of protons.  Following Ref.~\cite{DPS_theory}, I calculate the cross section 
for two distinguishable processes happening in one $pp$ collision, as 
\bq
\label{eq:DPS}
\sigma_{DPS} = \frac{\sigma_{1}\sigma_{2}}{\sigma_{eff}} \;,
\eq
with the additional constraint that the sum of initial parton energies from one 
proton be bounded by the beam energy. $\sigma_{eff}$ parameterizes the 
transverse size of the proton.  It has been measured by the Fermilab CDF 
collaboration to be $\sigma_{eff}=14.5$~mb~\cite{CDF_DPS}. I assume the same 
value to hold for LHC energies. 

One DPS background arises from simultaneous $\gamma\gamma j$ and $jj$ events, 
where the jet in the $\gamma\gamma j$ hard scattering is observed as a tagging 
jet, together with one of the two jets in the dijet process. In order to avoid 
a three-jet signature, one might want to require the second jet of the dijet 
process to fall outside the acceptance region of Eq.~(\ref{eq:basic}). However, 
this would severely underestimate this DPS contribution, since soft gluon 
radiation must be taken into account in a more realistic simulation. Soft 
radiation destroys the $p_T$ balance of the two jets in the dijet process, 
leading to the possibility of only one of the two final state partons to be 
identified as a jet, even though both satisfy the pseudo-rapidity requirements 
of Eq.~(\ref{eq:basic}). Since our tree-level calculation cannot properly take 
into account such effects, I conservatively select the higher-energy jet of the 
dijet process in the hemisphere opposite that of the jet from the 
$\gamma\gamma j$ event, and allow the third jet to be anywhere, completely 
ignoring it for the purposes of imposing further cuts. I choose the 
factorization scale for DPS to be $\mu_f = {1\over n_{part}}\sum p_{T_{all}}$, 
and the renormalization scale to be $\mu_r = {1\over n_{jet}}\sum p_{T_j}$.

A second DPS mode consists of two overlapping $\gamma j$ events. All final-state 
particles must be observed above threshold in the detector. With full acceptance 
cuts this background is found to be insignificant compared to the others, and 
will not be considered further. I do not consider DPS backgrounds from 
overlapping $\gamma\gamma$ and $jj$ events since the double forward jet tagging 
requirements of Eq.~(\ref{eq:gap}) forces the dijet invariant mass to be very 
large, effectively eliminating this background.


\section{Separation of Signal and Background}

As a reference starting point, our signal cross section times branching ratio 
with the cuts of Eqs.~(\ref{eq:basic}-\ref{eq:gap}) for $m_H = 120$~GeV is 
$\sigma_{Hjj}\cdot B(H\to\gamma\gamma) = 2.4$~fb. To ensure a clean signal the 
search must first establish observation of the Higgs decay products 
well-separated from the tagging jets. Thus our first additional cut, beyond the 
basic requirements of Eqs.~(\ref{eq:basic}-\ref{eq:gap}), is a minimum $p_T$ 
requirement for the photons:
\bq
\label{eq:pho_pT_base}
p_{T_\gamma} > 20~{\rm GeV} \, .
\eq
This still leaves a 2.2~fb signal for $m_H = 120$~GeV, better than $90\%$ 
acceptance. The backgrounds, however, are still overwhelming, as shown in the 
first line of Table~\ref{tab:phocuts}. This is not surprising, as the presence 
of $p_T=20$~GeV jets is a common occurrence in hard scattering events at the 
LHC.

Prominent features of the irreducible QCD background are the steeply falling 
transverse momentum distributions of both the jets and photons, as shown in 
Figs.~\ref{fig:pT_pho} and~\ref{fig:pT_jets}. These distributions are typical 
for bremsstrahlung processes and the low $p_{T_\gamma}$ nature of the 
backgrounds is enhanced by the soft photons being emitted typically at high 
rapidity, forward of their parents quark jet. This allows one to suppress the 
backgrounds further by harder $p_T$ cuts; I find that the following asymmetric 
$p_T$ cuts bring the backgrounds down another factor of three, while accepting 
$\approx 85\%$ of the signal:
\ba
\label{eq:phofinal}
p_{T_{j1}} \geq 40~{\rm GeV}\; , && \qquad p_{T_{j2}} \geq 20~{\rm GeV}\; , 
\nonumber \\ 
p_{T_{\gamma 1}} \geq 50~{\rm GeV}\; , && \qquad 
p_{T_{\gamma 2}} \geq 25~{\rm GeV}\; .
\ea
For $m_H = 120$~GeV, the resulting signal cross section is 1.9~fb. One could 
require even higher $p_{T_{\gamma}}$ cuts, making the backgrounds negligible, 
but this would come at the expense of a sizeable reduction in signal rate.

The SM Higgs resonance for this mass range is extremely narrow, 18~MeV at the 
most, which is much smaller than either of the expected photon resolutions of 
the CMS or ATLAS detectors, $\approx \pm 0.6 \cdots 1$~GeV and 
$\approx \pm 1.5 \cdots 1.8$~GeV, respectively. I use the expected resolution 
of the CMS detector, and examine 2~GeV mass bins:
\bq
\label{eq:massbin}
m_H - 1 \, {\rm GeV} \; \leq \; 
m_{\gamma\gamma} \; \leq \; m_H + 1 \, {\rm GeV} \, .
\eq
For this optimization, $70\%$ of the Higgs signal's Gaussian profile would be 
retained~\cite{profile}. This effect is included on line three of 
Table~\ref{tab:phocuts}. Line four of the same table further introduces common 
CMS and ATLAS efficiencies for identification of tagging jets and photons: each 
tagging jet is a factor 0.86 and each photon is a factor 0.80, for both the 
Higgs signal and all backgrounds.

\begin{table}
\caption{Signal $m_H = 120$~GeV and background $\gamma\gamma jj$ cross sections 
(fb) for successive levels of cuts given by the Equations in parenthesis, 
application of CMS expected tagging jet and photon ID efficiencies, mass 
resolution effects for 2 GeV bins, and application of a minijet veto with 
$p_T^{veto} = 20$~GeV.}
\label{tab:phocuts}
\begin{center}
\begin{tabular}{|p{2.4in}|p{0.7in}|p{0.75in}|p{0.75in}|p{0.7in}|}
\hline\hline
cuts & $\sigma_{Hjj}$ & QCD $jj\gamma\gamma$ & EW $jj\gamma\gamma$ & DPS \\
\hline
forward tagging + ID (\ref{eq:basic}-\ref{eq:gap},\ref{eq:pho_pT_base}) 
                                                  & 2.2  & 215  & 62   & 83   \\
+ staggered $p_{T(j,\gamma)}$ (\ref{eq:phofinal}) & 1.9  & 66   & 29   & 17   \\
+ 2 GeV mass bin (\ref{eq:massbin})               & 1.3  & 0.87 & 0.34 & 0.24 \\
+ efficiencies ($\epsilon = 0.473$)               & 0.63 & 0.41 & 0.16 & 0.12 \\
$P_{surv,20}$ & ${\it\times 0.89}$ & ${\it\times 0.30}$ & ${\it\times 0.80}$ 
  & ${\it\times 0.30}$ \\
+ minijet veto (\ref{eq:mjveto}) & 0.56 & 0.12 & 0.12 & 0.04 \\
\hline\hline
\end{tabular}
\end{center}
\end{table}

Including these detector efficiencies and resolutions is an improvement on the 
$VV\to H\to\gamma\gamma$ analysis first presented in Ref.~\cite{RZ_gamgam}. I 
also correct here a programming error that affected the QCD and EW 
$\gamma\gamma jj$ cross sections; the result is an approximate $20\%$ increase 
in rate for those processes. The efficiencies and mass resolution demonstrate an 
overall signal rate loss of about two-thirds, which can fortunately be partly 
compensated for by exploiting another feature of the $qq\to qqH$ signal, namely 
the absence of color exchange between the two scattering quarks. From the 
discussion of the minijet veto in Section~\ref{sec:minijet} and summarized 
results for the Higgs signal, and general QCD and EW backgrounds using the TSA 
and detailed in Appendix~\ref{app:minijet}, I apply expected minijet veto 
survival probabilities to the signal and background rates, summarized in line 
five of Table~\ref{tab:phocuts}. The veto is on additional radiation in the 
region between the two tagging jets, with $p_T > 20$~GeV.

\begin{table}
\caption{Signal and total background $\gamma\gamma jj$ cross sections (fb) for 
various Higgs masses, after application of all cuts, including a 2~GeV mass bin 
centered around the expected Higgs mass, application of CMS expected tagging 
jet and photon ID efficiencies ($\epsilon = (0.86)^2\cdot (0.8)^2 = 0.473$), 
mass resolution effects for 2 GeV bins, and application of a minijet veto with 
$p_T^{veto} = 20$~GeV. Gaussian equivalent Poisson statistical signal 
significances are given for 50~fb~$^{-1}$ of data at low luminosity.}
\label{tab:phosum}
\begin{center}
\begin{tabular}{|p{1.7in}|p{0.5in}|p{0.5in}|p{0.5in}|p{0.5in}|p{0.5in}|p{0.5in}|}
\hline\hline
Higgs mass (GeV)                     & 100  & 110  & 120  & 130  & 140  & 150  \\
\hline
$\epsilon\cdot\sigma_{Hjj}\cdot B(H\to\gamma\gamma)$ (fb) 
                                     & 0.37 & 0.48 & 0.56 & 0.56 & 0.48 & 0.33 \\
$\epsilon\cdot\sigma_{QCD}$ (fb)     & 0.14 & 0.13 & 0.12 & 0.11 & 0.10 & 0.08 \\
$\epsilon\cdot\sigma_{EW}$  (fb)     & 0.14 & 0.13 & 0.12 & 0.11 & 0.10 & 0.09 \\
$\epsilon\cdot\sigma_{DPS}$ (fb)     & 0.05 & 0.04 & 0.04 & 0.03 & 0.03 & 0.02 \\
$\epsilon\cdot\sigma_{bkg,tot}$ (fb) & 0.33 & 0.31 & 0.28 & 0.25 & 0.22 & 0.19 \\
\hline
S/B                     & 1.1 & 1.6 & 2.0 & 2.3 & 2.2 & 1.8 \\
$\sigma_{\rm Gaus}$     & 3.8 & 5.0 & 6.0 & 6.2 & 5.7 & 4.3 \\
\hline\hline
\end{tabular}
\end{center}
\end{table}

The minijet veto improves signal-to-background (S/B) rates to better than 1/1 
in all cases, and better than 2/1 over much of the mass range examined. 
Gaussian equivalent statistical significances for Poisson statistical treatment 
of background fluctuations is at the $5\sigma$ level for almost the entire mass 
range with about 40-50~fb$^{-1}$ of data at low luminosity. The slightly lower 
significances at the lower end of the mass range are of no concern, as this 
region will already have been explored by both the CERN LEP and Fermilab 
Tevatron Higgs searches.


\section{Discussion}

\begin{figure}[htb]
\vspace*{0.6in}            
\begin{picture}(0,0)(0,0)
\includegraphics{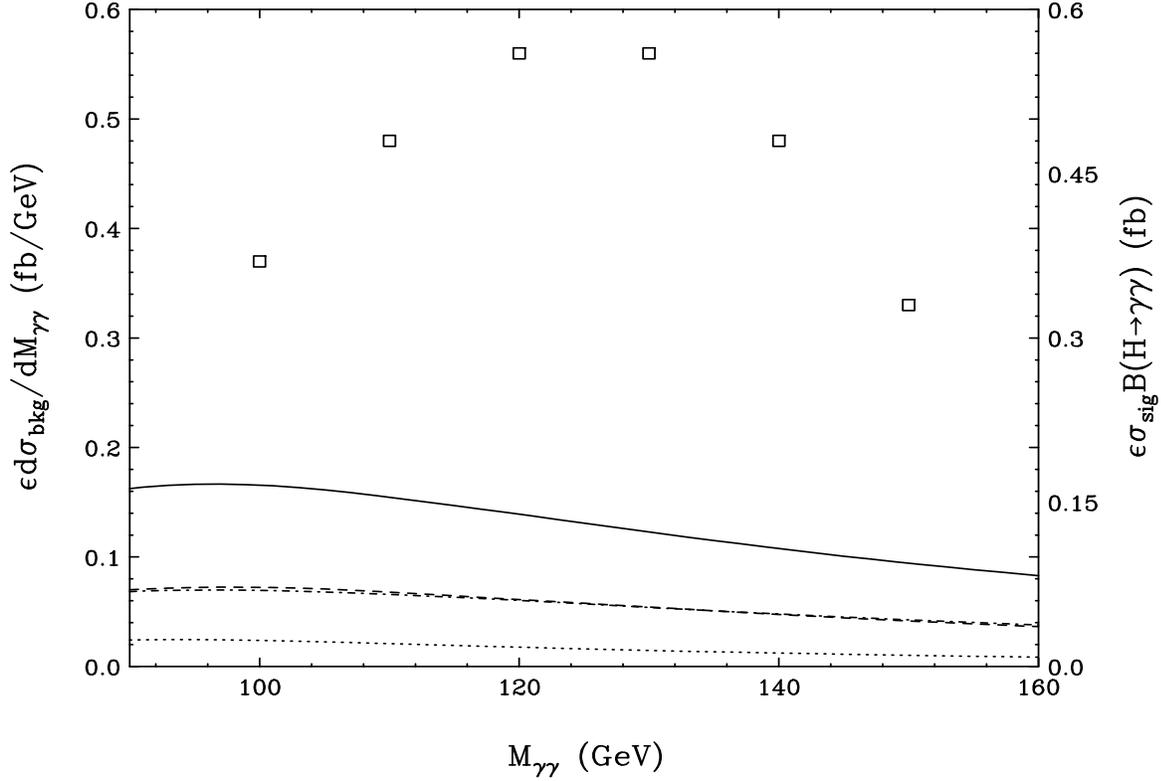}
\end{picture}
\vspace{8.5 cm}
\caption{Higgs signal cross section (fb) and diphoton invariant mass 
distribution (in fb/GeV) for the backgrounds after application of all cuts, 
Eqs.~(\protect\ref{eq:basic}-\protect\ref{eq:gap},\protect\ref{eq:pho_pT_base},
\protect\ref{eq:phofinal}), 
including CMS expected tagging jet and photon ID efficiencies 
($\epsilon = (0.86)^2\cdot (0.8)^2 = 0.473$), mass resolution effects, and 
application of a minijet veto with $p_T^{veto} = 20$~GeV. The squares are the 
Higgs signal for $m_H = 100,110,120,130,140,150$~GeV. The solid line represents 
the sum of all backgrounds, with individual components from the irreducible QCD 
background (dashed line), the irreducible EW background (dot-dashed line), and 
the DPS background (dotted line) shown below.}
\label{fig:Mgamgam_veto}
\end{figure}

Fig.~\ref{fig:Mgamgam_veto} shows the results after the application of all cuts, 
Eqs.~(\ref{eq:basic}-\ref{eq:gap},\ref{eq:pho_pT_base}-\ref{eq:massbin}), 
inclusion of tagging jet and photon ID efficiencies, mass resolution effects for 
2 GeV bins, and a minijet veto with $p_{T,{\rm veto}} = 20$~GeV. This plot 
compares the total signal cross section, in fb, to the di-photon invariant mass 
distribution, $\epsilon\cdot d\sigma/dm_{\gamma\gamma}$ in fb/GeV and thus 
indicates the relative size of signal and background. For our cuts, with 
50~fb$^{-1}$ of data at low luminosity, we thus expect anywhere from 16 to 28 
$H\to\gamma\gamma$ events on a background of 9 to 14 events, corresponding to a 
4.3 to 6.2 standard deviation signal, depending on the Higgs mass. Thus, Higgs 
observation with a modest 40-50~fb$^{-1}$ of data appears quite feasible in the 
$qq\to qqH\to \gamma\gamma jj$ channel. We note that this channel is thus 
competitive with the inclusive $H\to\gamma\gamma$ search, which is predicted to 
require about 20~fb$^{-1}$ ($> 100$~fb$^{-1}$) of data to reach $5\sigma$ 
significance coverage over the mass range 100-150~GeV by the CMS (ATLAS) 
collaboration~\cite{CMS-ATLAS}.

A more detailed analysis is warranted because more than $50\%$ of the signal 
events have at least one jet with $|\eta|\leq 2.4$ (see 
Fig.~\ref{fig:y_jets_basic}), leading to charged particle tracks in the central 
detector. As a result, the position of the interaction vertex can be more 
accurately obtained, leading to improved photon invariant mass resolution. We 
leave detailed studies of detector performance to the detector collaborations.

%


%% file: body/WW.tex

\section{Introduction}

While I have established an additional technique to search for the Higgs in the 
mass range $100 < m_H < 150$~GeV, just above this range the branching fraction 
$B(H\to\gamma\gamma)$ falls off sharply due to the increasing availability of 
the $H\to W^{(*)}W^{(*)}$ mode which dominates - the photon decay mode is no 
longer viable, even with very large luminosity. A search for the very clean 
four-lepton signature from $H\to ZZ$ decay can find a Higgs boson in the mass 
region $130 \lesssim m_H \lesssim 190$~GeV, but due to the small branching 
fraction of this mode, very large integrated luminosities, up to 100~fb$^{-1}$ 
or more, are required. Since $H\to W^{(*)}W^{(*)}$ decays dominate in this 
region, I turn our attention to the cleanest $WW$ decay signature, 
$e^\pm\mu^\mp\sla{p}_T$. One can search for this signature in inclusive 
$gg\to H$ Higgs production, and while some attention has been given to this for 
the LHC~\cite{glo_will_hww,DittDrein}, production via WBF for the same decay 
mode has only recently been discussed by us in the literature~\cite{RZ_WW}. 
This discussion closely follows that work, but includes additional detector 
efficiencies for tagging jet identification. I will show that WBF is comparable 
to gluon fusion for discovery in this mode, and that it overlaps considerably 
with the photon decay region, allowing for additional coupling ratios to be 
determined with low luminosity, on the order of 10~fb$^{-1}$. 

In Section~\ref{sec:WW_tools} I describe the calculational tools, the methods 
employed in the simulation of the various processes, and important parameters. 
Extra minijet activity is simulated by adding the emission of one extra parton 
to the basic signal and background processes. Generically I call the basic 
signal process (with its two forward tagging jets) and the corresponding 
background calculations ``2-jet'' processes, and refer to the simulations with 
one extra parton as ``3-jet'' processes. In Section~\ref{sec:WW_cuts}, using 
the 2-jet programs for the backgrounds, I demonstrate forward jet tagging, a 
$b$ veto and other important cuts which combine to yield an $\approx \;$2/1 to 
1/2 signal-to-background (S/B) ratio, depending on the Higgs mass. 

In Section~\ref{sec:WW_mj} I analyze the different minijet patterns in signal 
and background, using both the truncated shower approximation (TSA)~\cite{TSA} 
to regulate the cross sections, and the gluon exponentiation model to estimate 
the minijet multiplicity~\cite{CDFjets}. By exploiting the two most important 
characteristics of the extra radiation, its angular distribution and its 
hardness, the QCD backgrounds can be suppressed substantially by a veto on extra 
central jet emission. Within the TSA and exponentiation models, probabilities 
are estimated for vetoing signal and background events, and are combined with 
the production cross sections of the previous section to predict signal and 
background rates in Table~\ref{WW_sum}. These rates demonstrate the feasibility 
of extracting a very low background $H\to W^{(*)}W^{(*)}$ signal at the LHC.

The signal selection is not necessarily optimized yet. The variables I identify 
for cuts are the most distinctive, but deserve a multivariate analysis with 
detector simulation. I do demonstrate an additional variable in 
Section~\ref{sec:WW_disc} which is not used for cuts, but rather can be used to 
extract the Higgs boson mass from the final event sample.


\section{Calculational Tools}
\label{sec:WW_tools}

All signal and background cross sections are determined in terms of full 
tree level matrix elements for the contributing subprocesses and are 
discussed in more detail below.

For all our numerical results I have chosen the values listed in 
Appendix~\ref{app:parms}, including the {\sc hdecay}-corrected effective 
branching ratios presented in Table~\ref{table_BR}. Unless otherwise noted the 
factorization scale is chosen as $\mu_f =$ min($p_T$) of the defined jets. 


\subsection{The $qq\to qqH(g)$,$H\to W^{(*)}W^{(*)} \to e^\pm\mu^\mp\sla{p}_T$ 
signal process}

An important additional tool for distinguishing the $H\to e^\pm\mu^\mp\sla{p_T}$ 
signal from various backgrounds is the anti-correlation of the $W$ spins, as 
pointed out in Ref.~\cite{DittDrein}. This is due to the preservation of 
angular momentum in the decay of the spin-0 Higgs boson. Of course, one can 
observe only the angular distributions of the charged decay leptons, but this 
is sufficient. The decay rate is proportional to 
$(p_{\ell^-}\cdot p_{\nu})(p_{\ell^+}\cdot p_{\bar\nu})$. In the rest frame of 
the Higgs boson, in which the $e^-\bar\nu$ or $e^+\nu$ pairs are emitted 
back-to-back for $W^+W^-$ production at threshold, this product is a maximum 
for the charged leptons being emitted parallel. This characteristic is 
preserved and even enhanced when boosted to the lab frame, as the Higgs 
boson in weak boson fusion is typically emitted with $p_T \approx 60-120$~GeV. 

To model this one cannot simply apply the $H\to WW$ branching ratio to our 
production cross section and generate the Higgs decay products by phase space 
distributions as in the $H\to\gamma\gamma$ case. Instead, one must calculate the 
value of the matrix element squared for $H\to WW \to e^\pm \mu^\mp \sla{p_T}$, 
separately for each phase space point. This result can then be multiplied 
directly into the production cross section, thus maintaining angular 
correlations of the $W$ decays. 


\subsection{The QCD $t \bar{t} + jets$ backgrounds}

Given the H decay signature, the main physics background to our 
$e^{\pm}\mu^{\mp}\sla p_T$ signal arises from $t\bar t + jets$ production, due 
to the large top production cross section at the LHC and because the branching 
ratio $B(t\to Wb)$ is essentially $100\%$. 

The basic process I consider is $pp\to t\bar{t}$, which can be either $gg$- or 
$q\bar q$-initiated, with the former strongly dominating at the LHC. 
QCD corrections to this lead to additional real parton emission, i.e., to 
$t\bar{t} + j$ events. Relevant subprocesses are 
\bq
\label{QCD_tt}
g q  \to t \bar{t} q \, , \qquad g \bar{q} \to t \bar{t} \bar{q} \, , \qquad 
q \bar{q} \to t \bar{t} g \, , \qquad g g \to t \bar{t} g \, ,
\eq
and the subprocesses for $t\bar{t} + jj$ events can be obtained similarly. For 
the case of no additional partons, the $b$'s from the decaying top quarks may be 
identified as the tagging jets. In this case, calculating the cross section for 
$t\bar{t} + j$ where the $b$'s are explicitly identified as the tagging jets 
serves to estimate the effect of additional soft parton emission, {\em i.e.}, 
minijet activity in the central detector; this is described in detail in 
Sec.~\ref{sec:WW_mj}. At the same time, one can identify a distinctly different, 
perturbative region of phase space, where the final-state light quark or gluon 
gives rise to one tagging jet, and one of the two decay $b$'s is identified as 
the other tagging jet. In this case, $t\bar{t} + jj$ may be used to estimate 
minijet activity for the hard process $pp\to t\bar{t} + j$. Finally, there is a 
third distinct region of phase space, for the perturbative hard process 
$pp\to t\bar{t} + jj$, where the final state light quarks or gluons are the two 
tagging jets.

Thus, the ``$t\bar{t} j$'' and ``$t\bar{t}jj$'' calculations serve a dual 
purpose: to obtain the cross sections for the contribution of the perturbative 
processes where light quark or gluon jets lie in the region of phase space 
where they are experimentally identified as far-forward/backward tagging jets; 
and to estimate the additional QCD radiation patterns for the next-lower-order 
perturbative $t\bar{t} + jets$ process. The $t\bar{t}$ and $t\bar{t}j$ matrix 
elements were constructed using {\sc madgraph}~\cite{Madgraph}, while the 
$t\bar{t}jj$ matrix elements are from Ref.~\cite{Stange}.

Decays of the top quarks and $W$'s are included in the matrix elements; however, 
while the $W$'s are allowed to be off-shell, the top quarks are required to be 
on-shell. Energy loss from $b\to\ell\nu X$ is included to generate more accurate 
$\sla{p}_T$ distributions. In all cases, the factorization scale is chosen as 
$\mu_f =$ min($E_T$) of the massless partons/top quarks. The overall strong 
coupling constant factors are taken as 
$(\alpha_s)^n = \prod_{i=1}^n \alpha_s(E_{T_i})$, where the product runs over 
all light quarks, gluons and top quarks; {\em i.e.} the transverse momentum of 
each additional parton is taken as the relevant scale for its production, 
irrespective of the hardness of the underlying scattering event. This procedure 
guarantees that the same $\alpha_s^2$ factors are used for the hard part of a 
$t\bar{t} + jets$ event, independent of the number of additional minijets, and 
at the same time the small scales relevant for soft-parton emission are 
implemented.


\subsection{The QCD $WW+jj$ background}

The next obvious background arises from real-emission QCD corrections to 
$W^+W^-$ production. For $W^+ W^- jj$ events these background processes 
include~\cite{VVjj}
\bq
\label{QCD_WW}
q g \to q g W^+ W^- \, , \qquad  q q' \to q q' W^+ W^- \, ,
\eq
which are dominated by $t$-channel gluon exchange, and all crossing
related processes, such as
\bq
q \bar{q} \to g g W^+ W^- \, , \qquad g g \to q \bar{q} W^+ W^- \;.
\eq
I call these processes collectively the ``QCD $WWjj$'' background. I do not 
calculate cross sections for the corresponding $WW+3$-jet processes, but instead 
apply the minijet veto probability estimate for QCD $Zjj(j),Z\to\tau\tau$ 
processes, for the acceptance cuts of this search, found in 
Appendix~\ref{app:minijet}. The QCD $\tau\tau jj$ and QCD $WWjj$ backgrounds are 
quite similar kinematically, which justifies the use of the same veto 
probabilities for central jets.

The factorization scale is chosen as for the Higgs boson signal. The strong 
coupling constant factor is taken as 
$(\alpha_s)^2 = \alpha_s(p_{T_1})\alpha_s(p_{T_2})$, {\em i.e.}, the transverse 
momentum of each additional parton is taken as the relevant scale for its 
production. Variation of the scales by a factor of 2 or $1\over 2$ reveals scale 
uncertainties of $\approx 35\%$, however, which emphasizes the need for 
experimental input or NLO calculations.

The $WW$ background lacks the marked anti-correlation of $W$ spins seen in the 
signal. As a result the momenta of the charged decay leptons will be more 
widely separated than in $H\to W^{(*)}W^{(*)}$ events.


\subsection{The EW $WW+jj$ background}

These backgrounds arise from $W^+W^-$ bremsstrahlung in quark--(anti)quark 
scattering via $t$-channel electroweak boson exchange, with subsequent decay 
$W^+W^-\to\ell^+\ell^-\sla p_T$:
\bq
qq' \to qq' W^+W^-
\label{eq:EW_WW}
\eq
Na\"{\i}vely, this EW background may be thought of as suppressed compared to 
the analogous QCD process in Eq.~(\ref{QCD_WW}). However, as in the analogous 
case for $\gamma\gamma jj$ events discussed in Chapter~\ref{ch:gammagamma}, it 
includes electroweak boson fusion, $VV \to W^+W^-$ via $s$- or $t$-channel 
$\gamma/Z$-exchange or via $VVVV$ 4-point vertices, which has a momentum and 
color structure identical to the signal. Thus, it cannot easily be suppressed 
via cuts.

The matrix elements for these processes were constructed using 
{\sc madgraph}~\cite{Madgraph}. I include charged-current (CC) and 
neutral-current (NC) processes, but discard s-channel EW boson and t-channel 
quark exchange processes as their contribution was found to be $\approx 1\%$ 
only, while adding significantly to the CPU time needed for the calculation. 
In general, for the regions of phase space containing far-forward and -backward 
tagging jets, s-channel processes are severely suppressed. I refer collectively 
to these processes as the ``EW $WWjj$'' background. Both $W$'s are allowed to 
be off-shell, and all off-resonance graphs are included. In addition, the 
Higgs boson graphs must be included to make the calculation well-behaved at 
large $W$-pair invariant masses. However, these graphs include our signal 
processes and might lead to double counting. Thus, I set $m_H$ to 60~GeV in 
the EW $WWjj$ background to remove their contribution. A clean separation of 
the Higgs boson signal and the EW $WWjj$ background is possible because 
interference effects between the two are negligible for the Higgs boson mass 
range of interest.

Again an estimate of additional gluon radiation patterns is needed. This was 
first done for EW processes in Ref.~\cite{DZ_IZ_minijet}, but for different 
cuts on the hard process. Here I use the results for EW $Zjj(j),Z\to\tau\tau$ 
processes found in Appendix~\ref{app:minijet}. The EW $\tau\tau jj$ and EW 
$WWjj$ backgrounds are quite similar kinematically, which justifies the use of 
the same veto probabilities for central jets.


\subsection{The QCD $\tau^+\tau^-$ background}
\label{sec:QCD_tau}

The leptonic decay of $\tau$'s provides a source of electrons, muons and 
neutrinos which can be misidentified as $W$ decays. Thus, I need to study 
real-emission QCD corrections to the Drell-Yan process 
$q\bar{q} \to (Z,\gamma) \to \tau^+\tau^-$. 
For $\tau^+ \tau^- jj$ events these background processes include~\cite{Kst}
\bq
\label{QCD_tautau}
q g \to q g \tau^+ \tau^- \, , \qquad  q q' \to q q' \tau^+ \tau^- \, ,
\eq
which are dominated by $t$-channel gluon exchange, and all crossing-related 
processes, such as
\bq
q \bar{q} \to g g \tau^+ \tau^- \, , \qquad g g \to q \bar{q} \tau^+ \tau^- \;.
\eq
All interference effects between virtual photon and $Z$-exchange are included.
I call these processes collectively the ``QCD $\tau\tau jj$'' background. The 
cross sections for the corresponding $\tau\tau+3$-jet processes, which I need 
for our modeling of minijet activity in the QCD $\tau\tau jj$ background, have 
been calculated in Refs.~\cite{HZ,BHOZ,BG}. Similar to the treatment of the 
signal processes, I use a parton-level Monte-Carlo program based on the work 
of Ref.~\cite{BHOZ} to model the QCD $\tau\tau jj$ and $\tau\tau jjj$ 
backgrounds.

The factorization scale is chosen as for the Higgs boson signal. With $n=2$ 
and $n=3$ colored partons in the final state, the overall strong-coupling 
constant factors are taken as $(\alpha_s)^n = \prod_{i=1}^n \alpha_s(p_{T_i})$, 
{\em i.e.} the transverse momentum of each additional parton is taken as the 
relevant scale for its production, irrespective of the hardness of the 
underlying scattering event. This procedure guarantees that the same 
$\alpha_s^2$ factors are used for the hard part of a $Zjj$ event, independent 
of the number of additional minijets, and at the same time the small scales 
relevant for soft-gluon emission are implemented.

The momentum distributions for the $\tau$ decay products are generated as for 
the Higgs boson signal. Because of the (axial-)vector coupling of the virtual 
$Z,\gamma$ to $\tau$'s, the produced $\tau^+$ and $\tau^-$ have the same 
chirality. This correlation of the $\tau$ polarizations is taken into account 
by calculating individual helicity amplitudes and folding the corresponding 
cross sections with the appropriate $\tau^+$ and $\tau^-$ decay distributions, 
i.e. the full $\tau$ polarization information is retained in the energy 
distribution of the $\tau$ decay products.

The dual leptonic decays of the $\tau$'s are simulated by multiplying the 
$\tau^+\tau^-jj$ cross section by a branching ratio factor of $(0.3518)^2/2$ 
and by implementing collinear tau decays with helicity correlations as 
discussed in Appendix~\ref{app:taudecay}. I also use the results from this 
study for our minijet emission estimates, summarized in 
Appendix~\ref{app:minijet}.


\subsection{The EW $\tau^+\tau^-$ background}
\label{sec:EW_tau}

These backgrounds arise from $Z$ and $\gamma$ bremsstrahlung in 
quark--(anti)quark scattering via $t$-channel electroweak boson exchange, 
with subsequent decay $Z,\gamma\to \tau^+\tau^-$: 
\bq
qq' \to qq' \tau^+\tau^-
\label{eq:qQqQZ}
\eq
Naively, this EW background may be thought of as suppressed compared to the 
analogous QCD process in Eq.~(\ref{QCD_tautau}). However, the EW background 
includes electroweak boson fusion, $VV \to \tau^+\tau^-$, either via $t$-channel 
$\tau/\nu$-exchange or via $s$-channel $\gamma/Z$-exchange, and the latter has a 
momentum and color structure which is identical to the signal and cannot easily 
be suppressed via cuts. 

I use the results of Ref.~\cite{CZ_gap} for our calculation which ignore 
$s$-channel EW boson exchange contributing to $q\bar{q}$ production, and Pauli 
interference of identical quarks. When requiring a large rapidity separation 
between the two quark jets (tagging jets) the resulting large dijet invariant 
mass severely suppresses any $s$-channel processes which might give rise to the 
dijet pair, and the very different phase space regions of the two scattered 
quarks make Pauli interference effects small. All charged-current (CC) and 
neutral-current (NC) subprocesses are included. The CC process dominates over NC 
exchange, however, mainly because of the larger coupling of the quarks to the 
$W$ as compared to photon and $Z$ interactions. I will refer to these EW 
processes as the ``EW $\tau\tau jj$'' background.

The $\tau$ decay distributions are generated according to the prescription in 
Appendix~\ref{app:taudecay}. Since the programs of Ref.~\cite{CZ_gap} generate 
polarization averaged $\tau^+\tau^-$ cross sections, I must assume 
unpolarized $\tau$'s. However, as for the QCD $\tau\tau jj$ background, the 
$\tau^+\tau^-$ pair arises from virtual vector boson decay, resulting in a 
$\tau^+$ and $\tau^-$ of the same chirality.  This correlation of the $\tau$ 
polarizations is taken into account. 

In order to determine the minijet activity in the EW $Zjj$ background I need 
to evaluate the ${\cal O}(\alpha_s)$ real parton emission corrections. The 
corresponding ${\cal O}(\alpha^4\alpha_s)$ diagrams for
\bq
qq'\to qq'g\;\tau^+\tau^- \; ,
\label{eq:qQqQgtautau}
\eq
and all crossing related subprocesses, have first been calculated in 
Ref.~\cite{RSZ_vnj}. Production of the $\tau$-pair via $Z$ and $\gamma$ exchange 
is considered. The factorization and renormalization scales are chosen to be the 
same as for the $Hjj$ signal, as this is also a hard EW process.

I have previously examined other scale choices for the $\tau\tau$ 
backgrounds~\cite{RSZ_vnj}, and found small uncertainties ($\approx \pm 15 \%$) 
for the EW component, while variations for the QCD component reach a factor 1.5. 
I thus expect the signal and EW $\tau\tau jj$ background cross sections to be 
fairly well determined at leading order, while the much larger theoretical 
uncertainty for the QCD $\tau\tau jj$ background emphasizes the need for 
experimental input.


\section{Separation of Signal and Background}
\label{sec:WW_cuts}

The $qq\to qqH, \; H\to W^{(*)}W^{(*)} \to e^\pm \mu^\mp \nu \bar{\nu}$ dual 
leptonic decay signal is characterized by two forward jets and the $W$ decay 
leptons ($e,\mu$). To begin, I impose the core acceptance requirements of 
forward tagging found in Eqs.~(\ref{eq:basic}-\ref{eq:gap}) and additionally a 
minimum observability cut on the leptons, 
\bq
\label{eq:Wlepmin}
p_{T_l} \, > \, 20 \: {\rm GeV} \, .
\eq
This ensures observation and isolation of the two tagging jets and Higgs 
final-state decay products, as well as imposes severe restrictions on the 
bremsstrahlung nature of the backgrounds: the generally higher rapidity of the 
$W$'s in the QCD $WWjj$ background as compared to the Higgs signal, for example, 
is due to weak boson bremsstrahlung occurring at small angles with respect to 
the parent quarks, producing $W$'s forward of the jets. Line 1 of 
Table~\ref{WW_data} shows the effect of these cuts on the signal and backgrounds 
for a SM Higgs boson of mass $m_H = 160$~GeV. 

Somewhat surprisingly, the EW $WWjj$ background reaches $68\%$ of the QCD $WWjj$ 
background already at this level. This can be explained by the contribution from 
$W,Z,\gamma$ exchange and fusion processes which can produce central $W$ pairs 
and are therefore kinematically similar to the signal. This signal-like 
component remains after the forward jet tagging cuts.

As is readily seen from the first line of Table~\ref{WW_data}, the most 
worrisome background is $W$ pairs from $t\bar{t} + jets$ production. Of the 
1080~fb at the basic cuts level, 12~fb are from $t\bar{t}$, 310~fb are from 
$t\bar{t}j$, and the remaining 760~fb arise from $t\bar{t}jj$ production. 
The additional jets (corresponding to massless partons) are required to be 
identified as far forward tagging jets. The $t\bar{t}jj$ cross section is 
largest because the $t\bar{t}$ pair is not required to have as large an 
invariant mass as in the first two cases, where one or both $b$'s from the 
decay of the top quarks are required to be the tagging jets. 

For the events where one or both of the $b$'s are not identified as the tagging 
jets, they will most frequently lie between the two tagging jets, in the region 
where I search for the $W$ decay leptons. Vetoing events with these additional 
$b$ jets provides a powerful suppression tool to control the top background. I 
discard all events where a $b$ or $\bar b$ jet with $p_T > 20$~GeV is observed 
in the gap region between the tagging jets,
\bq
\label{eq:bveto}
p_{T_b} > 20 {\rm GeV} \, , \qquad
\eta_{j,min} < \eta_{b} < \eta_{j,max} \, .
\eq
This leads to a reduction of $t\bar{t}j$ events by a factor 7 while $t\bar{t}jj$ 
events are suppressed by a factor 100. This results in cross sections of 43 and 
7.6 fb, respectively, at the level of the forward tagging $+$ lepton ID cuts of 
Eqs.~(\ref{eq:basic}-\ref{eq:gap},\ref{eq:Wlepmin}), which are now comparable to 
the other individual backgrounds. Note that the much higher $b$ veto probability 
for $t\bar{t}jj$ events results in a lower cross section than that for 
$t\bar{t}j$ events, an ordering which will remain even after final cuts have 
been imposed (see below).

\begin{figure}[t]
\vspace*{0.5in}
\begin{picture}(0,0)(0,0)
\includegraphics{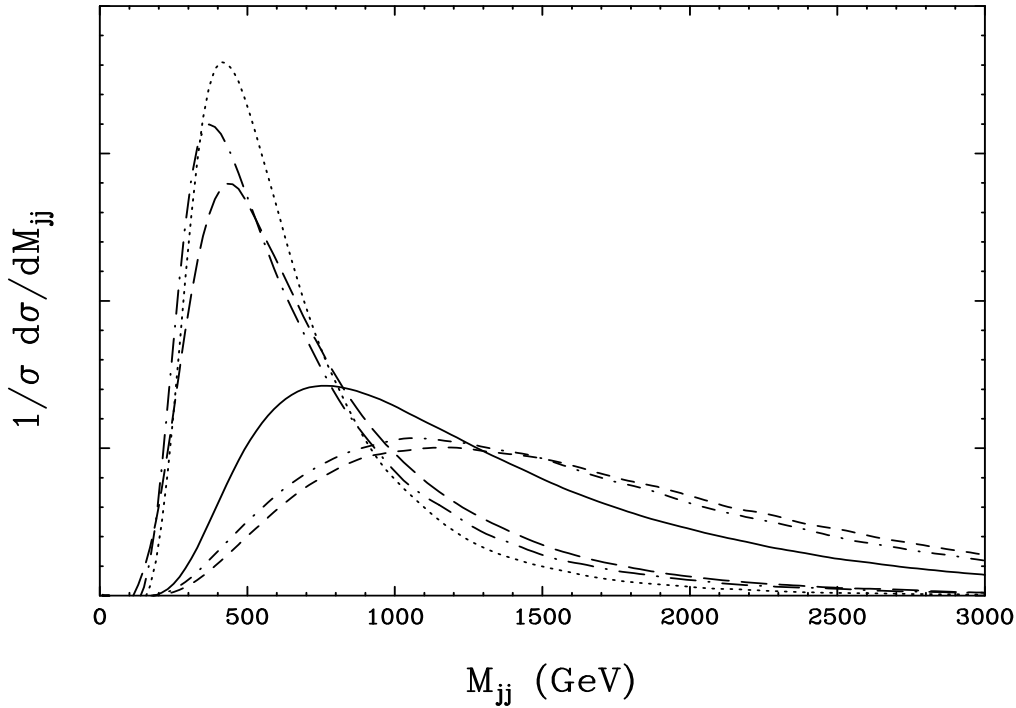}
\end{picture}
\vspace{7.5cm}
\caption{Normalized invariant mass distribution of the two tagging jets for 
the signal (solid) and various backgrounds: $t\bar{t} + jets$ (dotted), 
QCD $WWjj$ (long dashed), EW $WWjj$ (short dashed), 
QCD $\tau\tau jj$ (long dash-dotted) and EW $\tau\tau jj$ (short dash-dotted). 
The cuts of Eqs.~(\protect\ref{eq:basic}-\protect\ref{eq:gap},
\protect\ref{eq:Wlepmin}-\protect\ref{eq:bveto}) are imposed.}
\label{fig:Mjj_W}
\end{figure}

QCD processes at hadron colliders typically occur at smaller invariant masses 
than EW processes, due to the dominance of gluons at small Feynman $x$ in the 
incoming protons. I observe this behavior here, as shown in 
Fig.~\ref{fig:Mjj_W}. The three $t\bar{t} + jets$ backgrounds have been combined 
for clarity, even though their individual distributions are slightly different. 
I can thus significantly reduce much of the QCD background by imposing a lower 
bound on the invariant mass of the tagging jets: 
\bq
\label{eq:mjj_W}
m_{jj} > 650~{\rm GeV} \; .
\eq

\begin{figure}[t]
\vspace*{0.5in}
\begin{picture}(0,0)(0,0)
\includegraphics{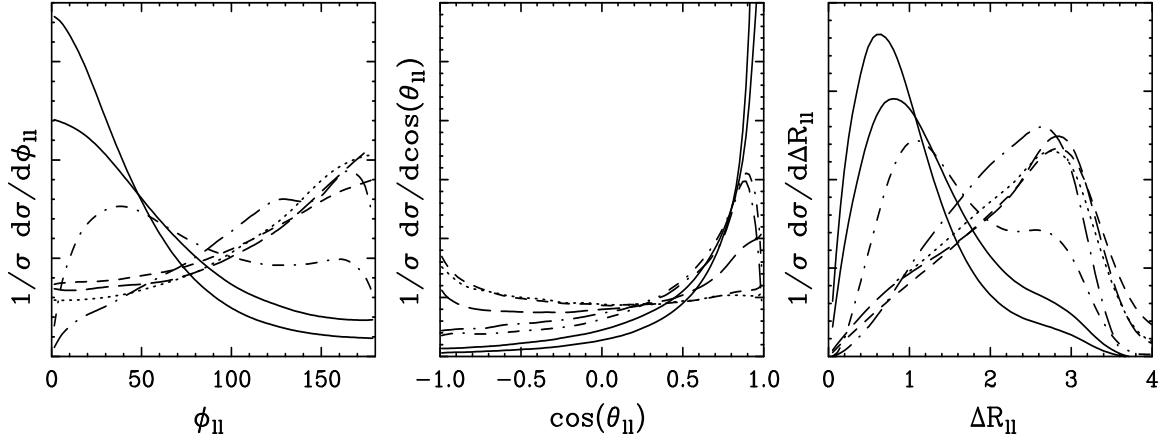}
\end{picture}
\vspace{4.5cm}
\caption{Normalized angular distributions of the charged leptons: azimuthal 
opening angle, lab opening angle, and separation in the lego plot. Results 
are shown for a Higgs boson mass of 160~GeV and 190~GeV (solid lines) and for 
the various backgrounds as in Fig.~\protect\ref{fig:Mjj_W}. Lepton angular 
separation is clearly smaller for the $m_H = 160$~GeV scenario. The cuts of 
Eqs.~(\protect\ref{eq:basic}-\protect\ref{eq:gap},
\protect\ref{eq:Wlepmin}-\protect\ref{eq:bveto}) are imposed.}
\label{fig:angdist}
\end{figure}

Another significant difference is the angular distribution of the charged decay 
leptons, $e^\pm$ and $\mu^\mp$, relative to each other. In the case of the Higgs 
signal, the $W$ spins are anti-correlated, so the leptons are preferentially 
emitted in the same direction, close to each other. A significant fraction of 
the various backgrounds does not have anti-correlated $W$ spins. 
These differences are demonstrated in Fig.~\ref{fig:angdist}, which shows the 
azimuthal (transverse plane) opening angle, polar (lab) opening angle, and 
separation in the lego plot. I exploit these features by establishing the 
following lepton-pair angular cuts:
\bq
\label{eq:ang}
\phi_{e\mu} < 105^{\circ} \, , \, \, \, 
{\rm cos} \; \theta_{e\mu} > 0.2 \, , \, \, \, 
\triangle R_{e\mu} < 2.2 \, .
\eq
It should be noted that while these cuts appear to be very conservative, for 
higher Higgs boson masses the $\phi_{e\mu}$ and $\triangle{R}_{e\mu}$ 
distribution broadens out to higher values, overlapping the backgrounds more. 
For $m_H \sim 180-200$~GeV these cuts are roughly optimized and further 
tightening would require greater integrated luminosity for discovery at this 
upper end of the mass range. Because of the excellent signal-to-background 
ratio achieved below, I prefer to work with uniform acceptance cuts, instead 
of optimizing the cuts for specific Higgs boson mass regions. 

\begin{figure}[t]
\vspace*{0.5in}
\begin{picture}(0,0)(0,0)
\includegraphics{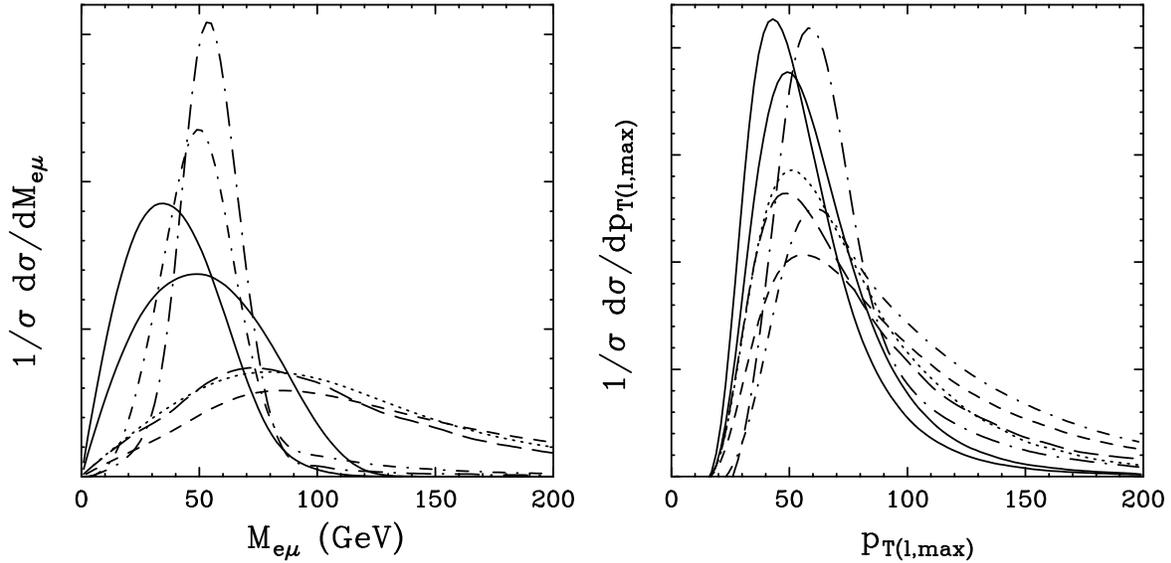}
\end{picture}
\vspace{6cm}
\caption{Normalized distributions of the dilepton invariant mass and maximum 
charged lepton momentum after the cuts of 
Eqs.~(\protect\ref{eq:basic}-\protect\ref{eq:gap},
\protect\ref{eq:Wlepmin}-\protect\ref{eq:bveto}). Results are shown for a Higgs 
boson mass of 160~GeV and 190~GeV (solid lines) and for the various backgrounds 
as in Fig.~\protect\ref{fig:Mjj_W}. The $m_H = 160$~GeV curve peaks at lower 
values of $m_{e\mu}$ and $p_{T_{\ell,max}}$. }
\label{fig:mllptl}
\end{figure}

I also examine the distributions for lepton-pair invariant mass, $m_{e\mu}$, 
and maximum lepton $p_T$, as shown in Fig.~\ref{fig:mllptl} for the case 
$m_H = 160,\; 190$~GeV. As is readily seen, the QCD backgrounds and EW $WWjj$ 
background prefer significantly higher values for both observables. Thus, in 
addition to the angular variables, I find it useful to restrict the 
individual $p_T$ of the leptons, as well as the invariant mass of the pair: 
\bq
\label{eq:adv}
m_{e\mu} < 110 \; {\rm GeV} \, , \, \, \, 
p_{T_{e,\mu}} < 120 \; {\rm GeV}  \, .
\eq
These are particularly effective against the top backgrounds, where the large 
top mass allows for very high-$p_T$ leptons far from the tagging jets, and 
against the EW $WWjj$ background, where the leptons tend to be well-separated 
in the lego plot. Again, the cuts are set quite conservatively so as not to 
bias a lower Higgs boson mass. Results after cuts~(\ref{eq:mjj_W}-\ref{eq:adv}) 
are shown on the third line of Table~\ref{WW_data}, for the case of a 160~GeV 
Higgs boson. 

\footnotesize
\begin{table}
\caption{Signal rates $\sigma\cdot B(H\to e^\pm\mu^\mp\sla{p_T})$ for 
$m_H = 160$~GeV and corresponding background cross sections, in fb. Results 
are given for various levels of cuts and are labeled by equation numbers 
discussed in the text. The expected tagging jet identification efficiency is 
shown on line 5. In the last line the minijet veto in included. Line six gives 
the survival probabilities for each process, with $p_T^{veto} = 20$~GeV.}
\label{WW_data}
\begin{center}
\begin{tabular}
{|p{1.8in}|p{0.4in}|p{0.5in}|p{0.45in}|p{0.45in}|p{0.4in}|p{0.4in}|p{0.4in}|}
\hline\hline
cuts & $Hjj$ & $t\bar{t} + jets$ & QCD & EW & QCD & EW & S/B \\
     &  &  & $WWjj$ & $WWjj$ & $\tau\tau jj$ & $\tau\tau jj$ & \\
\hline
forward tagging (\ref{eq:basic}-\ref{eq:gap},\ref{eq:Wlepmin})
& 17.1 & 1080 & 4.4  &  3.0 & 15.8 & 0.8  & $\approx$1/65 \\
+ $b$ veto (\ref{eq:bveto})
&      &  63  &      &      &      &      & 1/5.1 \\
+ $M_{jj}$, ang. cuts (\ref{eq:mjj_W}-\ref{eq:adv})
& 11.8 &  2.8 & 0.54 & 0.50 & 3.6  & 0.4  & 1.5/1 \\
+ real $\tau$ rejection (\ref{eq:tau})
& 11.4 &  2.6 & 0.50 & 0.45 & 0.6  & 0.08 & 2.7/1 \\
+ tag ID efficiency (${\it\times 0.74}$)
&  8.4 &  1.9 & 0.37 & 0.33 & 0.45 & 0.06 & 2.7/1 \\
$P_{surv,20}$
& ${\it\times 0.89}$ & ${\it\times 0.29}$ & ${\it\times 0.29}$ 
& ${\it\times 0.75}$ & ${\it\times 0.29}$ & ${\it\times 0.75}$ & - \\
+ minijet veto (\ref{eq:mjveto})
&  7.5 & 0.56 & 0.11 & 0.25 & 0.13 & 0.05 & 6.9/1 \\
\hline\hline
\end{tabular}
\end{center}
\end{table}
\normalsize

At this level of cuts the combined QCD and EW $\tau\tau jj$ backgrounds exceed 
all other individual backgrounds, contributing $50\%$ of the total. One can take 
advantage of the fact that in these backgrounds, the $Z$ or $\gamma$ is emitted 
with quite high $p_T$, on the order of 100~GeV, which contributes to large 
$\tau$ boosts and causes the $\tau$ decay products to be nearly collinear in the 
lab frame. Within the collinear approximation, the $\tau$ momenta can be 
reconstructed knowing the charged lepton momenta and the missing transverse 
momentum vector~\cite{RZ_tautau,ATLAS_tau}. Labeling by $x_{\tau_1},x_{\tau_2}$ 
the fraction of $\tau$ energy each charged lepton takes with it in the $\tau$ 
decay, $\sla{p}_{T,x},\sla{p}_{T,y}$ can be used to solve the two equations 
(x,y transverse directions) for the two unknowns $x_{\tau_{1,2}}$ 
(see Appendix~\ref{app:taudecay}). For real $\tau$ decays, the $\sla{p_T}$ 
vector must lie between the two leptons, and apart from finite detector 
resolution the reconstruction must yield $0 < x_{\tau_{1,2}} < 1$. For the $Hjj$ 
signal and other backgrounds, the collinear approximation is not valid because 
the $W$'s receive modest boosts in the lab only. In this case, the $\sla{p_T}$ 
vector will rarely lie between the two leptons, and an attempt to reconstruct a 
$\tau$ pair will result in $x_{\tau_1} < 0$ or $x_{\tau_2} < 0$ for $95\%$ of 
the events~\footnote{Conversely, requiring $x_{\tau_1} > 0$, $x_{\tau_2} > 0$ 
largely eliminates $WW$ backgrounds and promises clean isolation of 
$H\to\tau\tau\to e^\pm \mu^\mp \sla{p}_T$~\cite{RPZ_prep}.}. Additionally, the 
``$\tau$ pair'' invariant mass that is reconstructed does not peak at $m_Z$, 
even when it is positive. One can therefore apply a highly efficient cut against 
the QCD and EW $\tau\tau jj$ backgrounds by vetoing events where an attempt to 
reconstruct a $\tau$ pair in the collinear decay approximation results in two 
``real'' $\tau$'s near the Z pole:
\bq
\label{eq:tau}
x_{\tau_1} , \; x_{\tau_2} > 0 \; , \qquad
m_Z - 25\ {\rm GeV}\; < m_{\tau\tau} < \; m_Z + 25\ {\rm GeV} \, .
\eq
The results of this final cut are shown in line four of Table~\ref{WW_data}. 
The $\tau$ backgrounds are virtually eliminated, while the signal and the 
other backgrounds each lose $\approx 5\%$.


\section{Minijet Veto}
\label{sec:WW_mj}

If we are to veto central $b$ jets to reduce the $t\bar{t} + jets$ background 
to a manageable level, I must take care to correctly estimate higher-order 
additional central partonic emission in the signal and backgrounds. 
Fortunately, due to the absence of color exchange between the two scattering 
quarks in EW processes, which includes our $Hjj$ signal, one expects soft gluon 
emission mainly in the very forward and very backward directions. However, for 
QCD processes, which are dominated by $t$-channel color octet exchange, soft 
gluon radiation occurs mainly in the central detector. Thus, when I estimate 
additional central radiation with $p_T \geq 20$~GeV to match our $b$ veto 
condition, I will reject QCD background events with much higher probability 
than the EW processes. Our $b$ veto is then automatically also a minijet veto, 
a tool for QCD background suppression which has been previously studied in 
great detail for $Hjj$ production at hadron 
colliders~\cite{bpz_minijet,RZ_tautau,DZ_IZ_minijet}. 

Following the analysis of Ref.~\cite{RSZ_vnj} for the analogous EW $Zjj$ 
process which would be used to ``calibrate'' the tool at the LHC, I veto 
additional central jets in the central region between the two tagging jets 
according to Eq.~(\ref{eq:mjveto}), again using the cutoff 
$p_{T,\rm veto} = 20$~GeV as for the $H\to\gamma\gamma$ case in 
Chapter~\ref{ch:gammagamma}.

While the necessary information on angular distributions and hardness of 
additional radiation is available in the ``3-jet'' and $t\bar{t} + jets$ 
processes discussed in Section~\ref{sec:WW_tools}, one must either regulate 
or reinterpret these divergent cross sections. As discussed in 
Section~\ref{sec:minijet}, I use the TSA~\cite{TSA} for the former, and 
directly apply the results, summarized in Appendix~\ref{app:minijet}, to the 
QCD and EW cases for the $WWjj$ and $\tau\tau jj$ backgrounds. For the 
$t\bar{t} + jets$ backgrounds, it is simpler instead to reinterpret the 
divergent higher-order cross sections using the exponentiation model, also 
discussed in Appendix~\ref{app:minijet}. However, the estimated value 
$P_{surv} = 0.12$ for the $t\bar{t} + jets$ background is much lower than the 
values experienced via either the TSA method or the exponentiation method for 
other QCD processes, and this difference is not yet fully understood. Hence, I 
apply conservatively the same value of $P_{surv} = 0.29$ for this background as 
for the other QCD backgrounds. The veto probabilities are summarized in line six 
of Table~\ref{WW_data}. I emphasize that while these probabilities are estimates 
only, they can be independently determined at the LHC in processes like $Zjj$ 
and $Wjj$ production~\cite{RSZ_vnj,CZ_gap}. For a Higgs mass of 160~GeV we are 
left with a signal cross section of 7.5~fb compared to a total background of 
1.09~fb.


\section{Discussion}
\label{sec:WW_disc}

So far I have considered a single Higgs boson mass of 160~GeV only. Since I 
have largely avoided mass-specific cuts, I can immediately extend our results 
to a larger range of $m_H$. The expected number of signal events for 
115~GeV~$\leq m_H \leq 200$~GeV and an integrated luminosity of 5~fb$^{-1}$ are 
shown in Table~\ref{WW_sum}. For the same luminosity, 5.5 background events are 
expected. Thus, the signal-to-background rate, S/B, is better than 1/1 for 
$\approx 125 < M_H < 200$~GeV, almost the entire mass range. In the second row 
of Table~\ref{WW_sum} the Poisson probabilities for this background to fluctuate 
up to the signal level are given, in terms of the equivalent Gaussian 
significances which can be expected in the experiment, on average. For the mass 
range $M_H = 155-180$~GeV, our technique is slightly superior to that of gluon 
fusion~\cite{DittDrein}. The gluon fusion and WBF modes together should then be 
able to measure coupling ratios quite accurately with moderate additional 
luminosity.

\begin{table}
\caption{Number of expected events for the $Hjj$ signal, for 5~${\rm fb}^{-1}$ 
integrated luminosity and application of all efficiency factors and cuts 
including a minijet veto, but for a range of Higgs boson masses. The total 
background is 5.5 events. As a measure of the Poisson probability of the 
background to fluctuate up to the signal level, the second line gives 
$\sigma_{Gauss}$, the number of Gaussian equivalent standard deviations.}
\vspace{0.15in}
\label{WW_sum}
\begin{center}
\begin{tabular}
{|p{0.8in}|p{0.35in}|p{0.35in}|p{0.35in}|p{0.35in}|p{0.35in}|p{0.35in}
|p{0.35in}|p{0.35in}|p{0.35in}|p{0.35in}|}
\hline\hline
$m_H$
& 115 & 120 & 130 & 140  & 150  & 160  & 170  & 180  & 190  & 200  \\
\hline
no. events       
& 1.6 & 3.6 & 8.8 & 15.8 & 24.0 & 37.5 & 36.2 & 29.9 & 20.8 & 16.3 \\
$\sigma_{Gauss}$ 
& 0.6 & 1.2 & 3.0 &  5.0 &  7.1 & 10.0 &  9.8 &  8.4 &  6.3 &  5.1 \\
\hline\hline
\end{tabular}
\end{center}
\end{table}

These results show that it is possible to isolate a virtually background free 
$qq\to qqH,\;H\to WW$ signal at the LHC, with sufficiently large counting rate 
to obtain a $5\sigma$ signal (or much better) with a mere 5~fb$^{-1}$ of data 
for the mass range 140-200~GeV. Extending the observability region down to 
130~GeV requires at most 15~fb$^{-1}$. This nicely overlaps the regions of 
observability for $H\to\gamma\gamma$ (100-150~GeV)~\cite{RZ_gamgam} discussed 
in Chapter~\ref{ch:gammagamma}. To reach 120~GeV would require 
$\approx 65$~fb$^{-1}$ at low luminosity, and to reach 115~GeV would require 
$\approx 165$~fb$^{-1}$.

\begin{figure}[htb]
\vspace*{0.5in}
\begin{picture}(0,0)(0,0)
\includegraphics{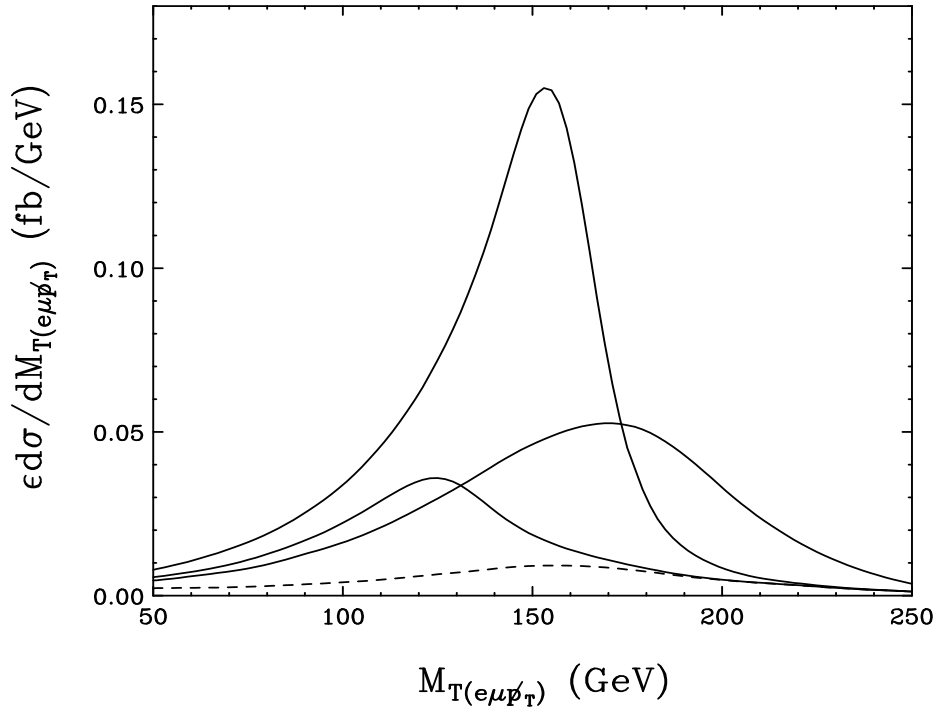}
\end{picture}
\vspace{8cm}
\caption{Dilepton-$\sla{p_T}$ transverse mass distributions expected for a 
Higgs of mass $m_H$ = 130, 160, and 190~GeV (solid) after the cuts of 
Eqs.~(\protect\ref{eq:basic}-\protect\ref{eq:gap},
\protect\ref{eq:Wlepmin}-\protect\ref{eq:adv}) and 
application of all detector efficiencies ($\epsilon = (0.86)^2 = 0.74$) and a 
minijet veto with $p_{T,{\rm veto}} = 20$~GeV. Also shown is the background 
only (dashed).}
\label{fig:M_T}
\end{figure}

As the $H\to WW$ mode is likely to be the discovery channel for the mass range 
130~GeV~$< m_H <$~200~GeV, we wish to be able to reconstruct the Higgs boson 
mass. At threshold, the two (virtual) $W$'s are at rest in the Higgs boson 
center-of-mass frame, resulting in $m_{e\mu} = m_{\nu\bar\nu}$, so we can 
calculate the transverse energy of both the charged lepton and invisible 
neutrino systems,
\bq
\label{eq:E_T}
E_{T_{e\mu}} = \sqrt{\vec{p}_{T_{e\mu}}^2 + m_{e\mu}^2} \; , \qquad
\sla{E}_T = \sqrt{\vec\sla{p}_T^2 + m_{e\mu}^2} \; .
\eq
Using these results for the transverse energies, we may compute a transverse 
mass of the dilepton-$\vec\sla{p}_T$ system, 
\bq
\label{eq:M_T}
M_{T_{WW}} \; = \; 
\sqrt{(\sla{E_T}+E_{T_{e\mu}})^2 - ({\vec{p}}_{T_{e\mu}}+\vec\sla{p_T})^2}
\; ,
\eq
At threshold this is exactly the Higgs boson transverse mass. Below threshold, 
the relation $m_{e\mu} = m_{\nu\bar\nu}$ is still an excellent approximation, 
while above threshold it begins to lose validity as the $W$ bosons acquire a 
non-zero velocity in the Higgs boson rest frame. But even at $m_H=200$~GeV 
this ``pseudo'' transverse mass remains extremely useful for mass 
reconstruction. I show the dramatic results in Fig.~\ref{fig:M_T}, for Higgs 
boson masses of 130, 160 and 190~GeV. Clearly visible is the Jacobian peak at 
$M_{T_{WW}} = m_H$, in particular for $m_H = 160$~GeV. The combined backgrounds 
are added to the Higgs signal, and are shown after application of all cuts and 
detector efficiencies, as well as both the $b$ and minijet vetoes discussed in 
the previous Sections. The very low background, in the absence of a Higgs 
signal, is also shown. The high purity of the signal is made possible because 
the weak boson fusion process, together with the 
$H\to W^+ W^-\to e^\pm \mu^\mp \sla{p_T}$ decay, provides a complex signal with 
a multitude of characteristics which distinguish it from the various 
backgrounds. 

An important point to note is that once a Higgs signal has been identified and 
its mass approximately determined, even further enhancement of the signal can 
be had by binning events in $M_{T_{WW}}$. This would be especially useful for 
a low-mass Higgs, since Fig.~\ref{fig:M_T} shows clearly that the peak of the 
distribution for background events occurs at $M_{T_{WW}} \approx 160$~GeV.

For $H\to WW$ decays, lepton angular distributions are extremely useful for 
reducing the backgrounds even further. The anti-correlation of $W$ spins in 
$H$ decay forces the charged leptons to be preferentially emitted in the same 
direction, close together in the lego plot. This happens for a small fraction 
of the background only. I have identified the most important distributions 
for enhancing the signal relative to the background, and set the various cuts 
conservatively to avoid bias for a certain Higgs boson mass range. There is 
ample room for improvement of our results via a multivariate analysis of a 
complete set of signal and background distributions, which I encourage the 
LHC collaborations to pursue. Additional suppression of the $t\bar{t} + jets$ 
background may be possible with $b$ identification and veto in the 
$p_T < 20$~GeV region. 

In addition to various invariant mass and angular cuts, I can differentiate 
between the $W$'s of the signal and $W,t$ backgrounds and the real $\tau$'s 
in the QCD and EW $\tau\tau jj$ backgrounds. This is possible because the 
high energy of the produced $\tau$'s makes their decay products almost 
collinear. Combined with the substantial $p_T$ of the $\tau^+\tau^-$ system 
this allows for $\tau$-pair mass reconstruction. The $W$ decays do not exhibit 
this collinearity due to their large mass, thus the angular correlation 
between the $\sla{p_T}$ vector and the charged lepton momenta is markedly 
different. Our real-$\tau$ rejection makes use of these differences and 
promises to virtually eliminate the $\tau\tau jj$ backgrounds.

A final step is to veto additional soft jet activity in the central region 
between the two tagging jets, as additional gluon radiation in QCD processes is 
characteristically central and hard, whereas for WBF/EW processes it tends to be 
more soft and forward/backward. I expect a typical $70\%$ reduction in QCD 
backgrounds for a central jet veto implemented for $p_T > 20$~GeV, and about a 
$25\%$ reduction for EW backgrounds but only about a $10\%$ suppression for the 
WBF Higgs production.


%% file: body/tautau.tex

\section{Introduction}

For Higgs searches in the portion of the intermediate-mass range where decay to 
fermion pairs has a non-negligible branching fraction, $\approx 110-150$~GeV, I 
show that observation of the $H\to\tau\tau$ decay channel is quite promising. 
An advantage of the $H\to\tau\tau$ channel, in particular compared to the 
dominant $H\to b\bar{b}$ mode, is the lower background from QCD processes. The 
$H\to\tau\tau$ channel thus also offers the best prospects for a direct 
measurement of the Higgs boson's couplings to fermions.

While some attention has been given to $A/H \to \tau\tau$ searches at the 
LHC~\cite{CMS-ATLAS,Cavalli,RW} in the framework of the MSSM, where the 
increased couplings of A/H to $\tau$ predicted for $\tan\beta \gg 1$ lead to 
higher production rates, conventional wisdom says that the chance of seeing the 
SM Higgs via this decay mode is nil, and it had been ignored in the literature 
until our recent analysis~\cite{RZ_tautau}. I discuss those results here, and 
update the analysis with corrected $H\to\tau\tau$ branching ratios, inclusion of 
a tagging jet identification efficiency, and an improved estimate of minijet 
veto probabilities. The results apply to an intermediate-mass SM $H\to\tau\tau$ 
at the LHC, covering the main physics and reducible backgrounds, and demonstrate 
the feasibility of Higgs boson detection in this channel with only modest 
luminosity. $H \to \tau\tau$ event characteristics are analyzed for one $\tau$ 
decaying leptonically and the other decaying hadronically, because of the high 
trigger efficiency and good branching ratio of this mode; Ref.~\cite{Cavalli} 
found the dual leptonic decay mode to be considerably more difficult due to 
higher backgrounds.

In Section~\ref{sec:tau_tools} I describe our calculational tools, the methods 
employed in the simulation of the various processes, and important parameters. 
Extra minijet activity is again simulated by adding the emission of one extra 
parton to the basic signal and background processes. In 
Sections~\ref{sec:tau_phybkg}~\&~\ref{sec:tau_redbkg}, using the 2-jet programs 
for physics and reducible backgrounds, respectively, I demonstrate forward jet 
tagging and $\tau$ identification and reconstruction criteria which yield an 
$\approx$2/1 signal-to-background (S/B) ratio. Both the $Wj+jj$ and $b\bar{b}jj$ 
reducible backgrounds intrinsically are much larger than the $Z\to\tau\tau$ and 
Drell-Yan $\tau$-pair production backgrounds. I explain and emphasize the 
cuts crucial to reducing these backgrounds to a manageable level.

In Section~\ref{sec:tau_mj} I analyze the different minijet patterns in signal 
and background, using the TSA~\cite{TSA} to regulate the cross sections. Within 
the TSA, probabilities are estimated for vetoing signal and background events, 
and are combined with the production cross sections of the previous section to 
predict signal and background rates in Table~\ref{tau_sum}. These rates 
demonstrate the possibility to extract a very low background $H\to\tau\tau$ 
signal at the LHC.

The signal selection is not necessarily optimized yet. Additional observables 
are available to distinguish the signal from background. The final discussion 
in Section~\ref{sec:tau_disc} includes a survey of distributions which can be 
used, e.g. in neural-net algorithms, to further improve the signal significance.


\section{Calculational Tools}
\label{sec:tau_tools}

Physical constants are chosen as in Appendix~\ref{app:parms}, including 
{\sc hdecay}-corrected $H\to\tau\tau$ branching ratios. I employ CTEQ4L parton 
distribution functions~\cite{CTEQ4_pdf} throughout. Unless otherwise noted the 
factorization scale is chosen as $\mu_f =$ min($p_T$) of the defined jets.


\subsection{The $qq\to qqH(g); H\to\tau^+\tau^-$ signal process}

In the following I consider only $\tau$-pair decays with one $\tau$ decaying 
leptonically, $\tau \to e\nu_e\nu_\tau,\; \mu\nu_\mu\nu_\tau$, and the other 
decaying hadronically, $\tau^\pm\to h^\pm X$, with a combined branching 
fraction of $45\%$. 

Positive identification of the hadronic $\tau^\pm\to h^\pm X$ decay requires 
severe cuts on the charged hadron isolation. Possible strategies have been 
analyzed by Cavalli {\it et al.}~\cite{Cavalli} and I base my simulations 
on their result. Considering hadronic jets of $E_T>40$~GeV in the ATLAS 
detector, they find non-tau rejection factors of 400 or more (see below) while 
retaining true hadronic $\tau$ decays with an identification efficiency
\bq
\label{eq:epstauh}
\epsilon_\tau(\tau\to\,\nu+{\rm hadrons}) = 0.26 \; .
\eq
This estimate includes the requirement of seeing a single charged hadron track, 
of $p_T>2$~GeV, pointing in the $\tau$ direction, and thus effectively singles 
out 1-prong $\tau$ decays. Accordingly, only the 1-prong hadronic branching 
ratios are considered in our mixture of $\pi$, $\rho$ and $a_1$ modes. 
Since the overall efficiency includes 3-prong events, which have negligible 
acceptance, the effective efficiency for 1-prong events is larger and taken as 
0.337 in the following, which reproduces the overall efficiency of 
Eq.~(\ref{eq:epstauh}).


\subsection{The QCD and EW $\tau^{+}\tau^{-}+jj(j)$ physics backgrounds}

These backgrounds are identical to the QCD and EW $\tau\tau jj(j)$ backgrounds 
discussed in Chapter~\ref{ch:WW}, except that here I have one tau decaying 
leptonically and the other hadronically; the decays are discussed in 
Appendix~\ref{app:taudecay}.


\subsection{The QCD $Wj+jj(j)$ reducible background}
\label{sec:tau_Wj}

Reducible backgrounds to $\tau^+\tau^-\to \ell^\pm h^\mp\sla p_T$ events
can arise from any process with a hard, isolated lepton, missing $p_T$,
and an additional narrow jet in the final state which can be mistaken as a
hadronically decaying $\tau$. A primary reducible background thus arises from
leptonic $W$ decays in $Wj$ events, where additional QCD radiation supplies
the two tagging jet candidates. At lowest order I need to consider
$Wj+jj$ production as the hard process, which is very similar to the
simulation of the QCD $Zjjj$ background discussed before,
with the bremsstrahlung $Z$ replaced by a $W$. $W\to e\nu_e,\;\mu\nu_\mu$
decays only are considered and are treated as a fake $\tau$ decaying
leptonically. Real leptonic $\tau$ decays from 
$W\to\tau\nu_\tau\to \ell\nu_\ell\nu_\tau\nu_\tau$ are relatively suppressed 
by the $\tau$ leptonic branching ratio of $35\%$ and the severity of the
transverse momentum cuts on the softer charged lepton spectrum. They will
be ignored in the following.

Two of the jets in $Wj+jj$ events are identified as tagging jets, and
fluctuations of the third
into a narrow jet are considered, resembling a hadronically-decaying $\tau$.
In Ref.~\cite{Cavalli} the probability for misidentifying a gluon or 
light-quark jet as a hadronic $\tau$ decay was estimated as 
\bq
\label{eq:epsjtau}
\epsilon_\tau({\rm jet}\to\,''\nu+{\rm hadrons}'') = 0.0025 \; ,
\eq
and I assign this probability to each of the final state jets. 
In each event one of the hard partons is randomly assigned to be the $\tau$. 
To mimic the signal, this jet and the identified charged lepton must be of
opposite charge. Thus, I reduce the $Wj+jj$ background by an additional 
factor of two to simulate the opposite charge requirement for 
the single track allowed in the $\tau$-like jet. 
As the $Wj+jj$ events are a QCD background, I use the same 
factorization and renormalization scales as for the QCD $Zjj$ case.

To simulate additional minijet emission, I need to add one more parton to
the final state. The code for $W+4j$ matrix elements has been available since 
the work of Berends et al.~\cite{tausk}. Here I use the program developed
in Ref.~\cite{W4j}, which was generated via {\sc madgraph}~\cite{Madgraph}. 
Since $W+4j$ production produces a six-particle final state, with up to 516 
graphs for the most complicated processes, it takes considerable CPU time to 
obtain good statistics. I modified the {\sc madgraph} code to
do random helicity summation, speeding up the calculation by approximately a
factor of 3 for a given statistical error in the final cross section.
As before, $\alpha_s$ is taken as the geometric mean of
$\alpha_s(p_T)$ factors for each of the partons, including the parton
which fakes the hadronic $\tau$ decay.


\subsection{The QCD $b\bar{b}jj$ reducible background}
\label{sec:tau_bb}

The semileptonic decay of $b$-quarks provides another source of leptons
and neutrinos which can be misidentified as tau decays. Even though $b$-quark
decays are unlikely to lead to isolated charged leptons and very narrow
tau-like jets in a single event, the sheer number of $b\bar b$ pairs produced
at the LHC makes them potentially dangerous. Indeed, the analysis of
Ref.~\cite{Cavalli} found that $b\bar b$ pairs lead to a reducible
$\tau^+\tau^-$ background which is similar in size to $Wj$ production.
I therefore study $b\bar{b}jj$ production as our second reducible
background and neglect any other sources like $t\bar t$ events which
were shown to give substantially smaller backgrounds to
$\tau^+\tau^-$-pairs in Ref.~\cite{Cavalli}.

I only consider $b$-production events where both $b$-quarks have large
transverse momentum. In addition, two forward tagging jets will be
required as part of the signal event selection. The relevant leading
order process therefore is the
production of $b\bar{b}$ pairs in association with two jets, which
includes the subprocesses
\ba
    gg          & \rightarrow & b\bar{b} gg  \nonumber \\
    qg          & \rightarrow & b\bar{b} qg   \\
    q_{1} q_{2} & \rightarrow & b\bar{b} q_{1} q_{2} \,. \nonumber
\ea
The exact matrix elements for the ${\cal O} (\alpha_{s}^{4})$ processes
are evaluated, including all the crossing related subprocesses, and retaining
a finite $b$-quark mass~\cite{Stange}.
The Pauli interference terms between identical quark flavors in the process
$q_{1}q_{2}\rightarrow b\bar{b}q_{1}q_{2}$ are neglected, with little effect
in the overall cross section, due to the large differences in the rapidity
of the final state partons.
The factorization scale is chosen as the smallest transverse energy
of the final state partons before the $b$-quark decay.
The strong coupling constant $\alpha_{s}$
is evaluated at the corresponding transverse energy of the final
state partons, i.e.,
$\alpha_{s}^{4} = \alpha_{s}(E_{T}(b)) \alpha_{s}(E_{T}(\bar{b}))
                  \alpha_{s}(p_{T,{\rm jet}_{1}})
                  \alpha_{s}(p_{T,{\rm jet}_{2}})$.

The semileptonic decay $b\to\ell\nu c$ of one of the $b$-quarks is simulated by 
multiplying the $b\bar{b}jj$ cross section by a branching ratio factor of 0.395 
(corresponding to at least one semileptonic $b$-decay to occur) and by 
implementing a three-body phase space distribution for the decay momenta. This 
part of the simulation is performed in order to estimate the effects of the 
lepton isolation cuts on the transverse momentum distributions of the $b$-decay 
leptons. Since these are kinematic effects I use the lightest meson masses in 
the simulation and set $m_b = 5.28$~GeV and $m_c=1.87$~GeV. In 
Ref.~\cite{Cavalli} a factor 100 reduction of the $b\bar{b}$ background was 
found as a result of lepton isolation, requiring $E_T<5$~GeV in a cone of radius 
0.6 around the charged lepton. In the simulation, after energy smearing of the 
charm quark jet (see below), I find a reduction factor of 52 due to lepton 
isolation with a cone of radius 0.7. However, the simulation does not include 
parton showers or hadronization of the $b$-quark, effectively replacing the 
$b$-quark fragmentation function by a delta-function at one, and thus 
underestimates the effect of lepton isolation cuts on the $b$-quark background. 
Since I cannot model the lepton isolation efficiency, I multiply the 
$b\bar{b}jj$ rates by another factor 0.52, thus effectively implementing the 
factor 100 suppression found by Cavalli {\it et al.}~\cite{Cavalli}. This value 
should probably be reanalyzed in light of the different signal signature here.

In addition to an isolated lepton, the $b\bar{b}jj$ events must produce a
narrow jet which is consistent with a hadronic $\tau$ decay, and has charge
opposite the identified charged lepton. This may either be one of the light 
quark or gluon jets, for which the misidentification probability of $0.25\%$ 
of Eq.~(\ref{eq:epsjtau}) will be used, or it may be the $b$-quark jet. 
In Ref.~\cite{Cavalli} the probability for misidentifying
a $b$-quark jet as a hadronic $\tau$ decay was estimated as
\bq
\label{eq:epsbtau-cavalli}
\epsilon_\tau(b\to\,''\nu+{\rm hadrons}'') \approx 0.0005 \; .
\eq
However, due to limited Monte Carlo statistics, this number was based on a
single surviving event only. Since we are really interested in an upper
bound on the $b\bar{b}jj$ background I follow the ATLAS
proposal~\cite{CMS-ATLAS} instead, and use the upper bound,
\bq
\label{eq:epsbtau-ATLAS}
\epsilon_\tau(b\to\,''\nu+{\rm hadrons}'') < 0.0015 \; ,
\eq
for our analysis. Thus, all our $b\bar{b}jj$ cross sections, after $\tau$
identification, should be considered conservative estimates. A more precise
analysis of $b\to \tau$ misidentification probabilities in the LHC detectors 
is clearly needed, which is beyond the scope of the present work. Finally,
an additional overall factor of two reduction is applied, as in the $Wj+jj$
case, for the lepton-jet opposite charge requirement.

The purpose of our $b$-analysis is to verify that $b$ semileptonic decays do not 
overwhelm the signal. The above procedures are adequate for this purpose, since 
I obtain final $b\bar{b}jj$ backgrounds (in Table~\ref{tau_sum}) which are 20 to 
40 times smaller than the signal. I do not calculate additional $b$ quark 
backgrounds arising from intrinsic $b$ contributions (processes like 
$gb\to bggg$). The matrix elements for these processes are of the same order 
$\left(\alpha_s^4\right)$ as for the $b\bar{b}jj$ subprocesses discussed above, 
but they are suppressed in addition by the small $b$-quark density in the 
proton. Also, I do not simulate additional soft gluon emission for the 
$b\bar{b}jj$ background. This would require $b\bar{b}+3$~jet matrix elements 
which are not yet available. Rather, I assume the probability for extra minijet 
emission to be the same as for the other reducible QCD background, $Wj+jj$ 
production.


\section{Higgs Signal and Real $\tau^{+}\tau^{-}$ Backgrounds}
\label{sec:tau_phybkg}

I begin as in Chapters~\ref{ch:gammagamma}~\&~\ref{ch:WW}, adopting the core 
forward tagging requirements of Eqs.~(\ref{eq:basic}-\ref{eq:gap}). As in the 
$H\to\gamma\gamma$ analysis I adopt asymmetric $p_T$ cuts for the jets, a useful 
tool to discriminate against the steeply falling $p_{T_j}$ distributions typical 
of QCD backgrounds:
\bq
\label{eq:pT_j}
p_{T_{j(1,2)}} \geq 40, 20~{\rm GeV} \, .
\eq

The resulting $Hjj,\;H\to\tau\tau$ cross section is compared with the 
irreducible $Zjj,\;Z\to\tau\tau$ backgrounds in the first row of 
Table~\ref{tau_data}. Somewhat surprisingly, the EW $Zjj$ background reaches 
$5\%$ of the QCD $Zjj$ background already at this level, while na{\"\i}vely one 
might expect suppression by a factor 
$(\alpha_{QED}/\alpha_s)^2 \approx 4\times 10^{-3}$. In the EW $Zjj$ background, 
$W$ exchange processes can produce central $\tau$ pairs by $Z$ emission from the 
exchanged $W$ and are therefore kinematically similar to the signal. This 
signal-like component remains after the forward jet tagging cuts, and, as we 
will see, will grow in relative importance as the overall signal/background 
ratio is improved.

\footnotesize
\begin{table}
\caption{Signal and background cross sections $B\sigma$ (fb) for $m_H = 120$~GeV 
$Hjj$ events. Results are given after increasingly stringent cuts given by the 
Equation numbers in parenthesis, and all values include the efficiency for 
tagging jet identification $\epsilon = 0.74$. The last column gives the ratio of 
the signal to the background cross sections listed in the previous columns.}
\label{tau_data}
\begin{center}
\begin{tabular}
{|p{2.3in}|p{0.3in}|p{0.6in}|p{0.52in}|p{0.5in}|p{0.4in}|p{0.35in}|}
\hline\hline
cuts & $Hjj$ & QCD $Zjj$ & EW $Zjj$ & $Wj+jj$ & $b\bar{b}+jj$ & S/B \\
\hline
~forward tagging ~(\ref{eq:basic}-\ref{eq:gap},\ref{eq:pT_j}) &
                      50.6 & 1240 & 67   &      &      &       \\
$+ \; \tau$ identification (\ref{eq:tauID}) &
                      1.47 & 14.8 & 1.07 & 19.5 & 5.6  & 1/28  \\
$+ \; 110 < m_{\tau\tau} < 130 {\rm GeV}$ (\ref{eq:tauphys}) &
                      0.97 & 0.70 & 0.05 & 1.31 & 0.44 & 1/2.6 \\
$+ \; m_{jj}>1$~TeV, $m_T(\ell,\sla p_T) < 30$~GeV & & & & & & \\
(\ref{eq:mjj_tau},\ref{eq:mTlnu}) &
                      0.51 & 0.12 & 0.03 & 0.08 & 0.11 & 1.5/1 \\
$+ \; x_{\tau_l} < 0.75$, $x_{\tau_h} < 1.0$ (\ref{eq:x1x2}) &
                      0.40 & 0.11 & 0.02 & 0.02 & 0.04 & 2.1/1 \\
\hline\hline
\end{tabular}
\end{center}
\end{table}
\normalsize

So far I have not considered $\tau$ decays. In order to get more realistic rate 
estimates and to include the reducible backgrounds ($Wj+jj$ and $b\bar{b}jj$, 
see Section~\ref{sec:tau_redbkg}) I need to study definite $\tau$ decay 
channels. I consider $\tau^+\tau^-$ decays with one $\tau$ decaying 
leptonically ($e$ or $\mu$) and the other decaying hadronically in the 
following, since previous studies have shown that dual leptonic decay is more 
difficult to observe~\cite{Cavalli}. With a hadronic branching ratio 
$B(\tau\to\nu+{\rm hadrons})=0.65$ and the overall hadronic $\tau$-decay 
identification efficiency of Eq.~(\ref{eq:epstauh}), the selection of this 
$\tau$-pair decay channel immediately reduces all $\tau^+\tau^-$ rates by a 
factor 
\ba
\epsilon B & = & 2 \epsilon_\tau(\tau\to\,\nu+{\rm hadrons}) \;
B(\tau\to\nu+{\rm hadrons})\; B(\tau\to\ell\nu_\ell\nu_\tau) \nonumber \\
  & = & 2\cdot 0.26\cdot 0.65\cdot 0.35 = 1/8.5 \, .
\ea
In addition, triggering the event via the isolated $\tau$-decay lepton and 
identifying the hadronic $\tau$ decay as discussed in Ref.~\cite{Cavalli} 
requires sizable transverse momenta for the observable $\tau$ decay products. 
In the following I require 
\bq
\label{eq:tauID}
p_{T_{\tau,lep}} > 20~{\rm GeV} \, , \qquad
p_{T_{\tau,had}} > 40~{\rm GeV} \, ,
\eq
where the second requirement is needed to use the results of Ref.~\cite{Cavalli} 
on hadronic $\tau$ identification. These transverse momentum requirements are 
quite severe and reduce the Higgs signal by another factor of $3.8$. Resulting 
signal and background cross sections are given in the second row of 
Table~\ref{tau_data}.

Crucial for further background reduction is the observation that the $\tau$-pair 
invariant mass can be reconstructed from the observable $\tau$ decay products 
and the missing transverse momentum vector of the event~\cite{tautaumass}. 
Details of the reconstruction are found in Appendix~\ref{app:taudecay}. 
The assumption of collinear tau decays is satisfied to an excellent degree 
because of the high $\tau$ transverse momenta needed to satisfy 
Eq.~(\ref{eq:tauID}). $\tau$ pair mass reconstruction is possible only as long 
as the the decay products are not back-to-back. This last condition is met in 
our case because the $H$ and $Z$ bosons are typically produced with high $p_T$, 
on the order of 150~GeV for all processes except the $b\bar{b}jj$ background 
(in which case the average $p_T \approx 85$~GeV is still sufficient). 

Mismeasured transverse momenta (smearing effects) can still lead to unphysical 
solutions for the reconstructed $\tau$ momenta. In order to avoid these, I 
impose a cut on the angle between the $\tau$ decay products and require 
positivity of the calculated $x_{\tau_i}$: 
\bq
\label{eq:tauphys}
\cos\theta_{\tau\tau} > -0.9 \, , \qquad  x_{\tau_{l,h}} > 0 \, .
\eq

The resulting $\tau$-pair invariant mass resolution is somewhat narrower than 
the one found in Ref.~\cite{Cavalli}, the 1-$\sigma$ half-width for the $H$ peak 
ranging from about 7 GeV for $m_H = 110$~GeV to about 10 GeV for $m_H = 150$~GeV 
(see Fig.~\ref{fig:Mtautau} below). This improved resolution is an effect of the 
higher average $p_T$ of the underlying process: in our case, the two forward 
tagging jets from weak boson scattering impart a higher $p_T$ on the $H$ or $Z$ 
than is the case from QCD radiation in gluon fusion. The smaller $\tau^+\tau^-$ 
opening angle then leads to a better $\tau$ momentum reconstruction via 
Eq.~(\ref{eq:taurecon}). Given this $\tau$-pair mass resolution, I choose 
$\pm 10$~GeV mass bins for analyzing the cross sections. Signal and background 
cross sections in a 20~GeV mass bin centered at 120~GeV, after the 
reconstruction conditions of Eq.~(\ref{eq:tauphys}), are listed in the third row 
of Table~\ref{tau_data}. QCD and EW $Zjj$ backgrounds are reduced by a factor of 
20, while about 2/3 of the signal survives the mass reconstruction cuts.

\begin{figure}[t]
\vspace*{0.5in}
\begin{picture}(0,0)(0,0)
\includegraphics{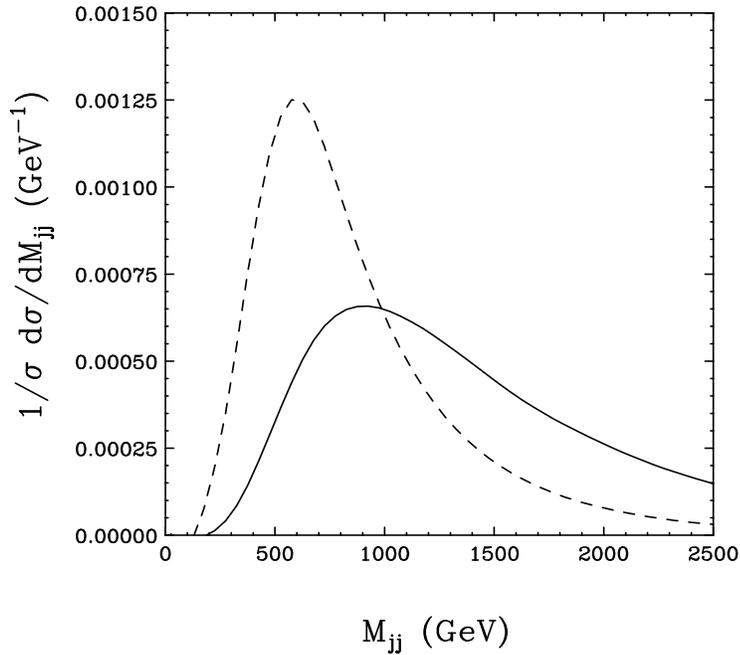}
\end{picture}
\vspace{7.0cm}
\caption{Invariant mass distribution of the two tagging jets for the 
$M_H = 120$~GeV $Hjj$ signal (solid line) and the QCD $Zjj$ background (dashed 
line), at the level of forward tagging cuts and $\tau$ reconstruction, 
Eqs.~(\protect\ref{eq:basic}-\protect\ref{eq:gap},
\protect\ref{eq:pT_j}-\protect\ref{eq:tauphys},\protect\ref{eq:mTlnu}).}
\label{fig:Mjj_tau}
\end{figure}

Because the QCD backgrounds typically occur at small invariant masses, I 
can further reduce them by imposing a cut on the invariant mass of the 
tagging jets, 
\bq
\label{eq:mjj_tau}
m_{jj} > 1~{\rm TeV}.
\eq
Fig.~\ref{fig:Mjj_tau} shows the tagging jets' invariant mass distribution for 
the signal and QCD $Zjj$ background to illustrate the effect of the cut.


\section{Fake $\tau^{+}\tau^{-}$ Events: Reducible Backgrounds}
\label{sec:tau_redbkg}

Reducible backgrounds to the $H\to\tau\tau$ signal, with subsequent leptonic 
decay of one of the $\tau$'s, arise from any source of isolated, single hard 
leptons. As discussed in Section~\ref{sec:tau_tools}, I consider $Wj+jj$ events 
and heavy quark production, in the form of $b\bar{b}jj$ events. Intrinsically, 
these reducible backgrounds are enormous and overwhelm even the physics 
backgrounds before $\tau$ identification and tight lepton isolation cuts are 
made. Crucial for the reduction of these backgrounds to a manageable level is 
the requirement of a narrow $\tau$-like jet, which leads to a factor 400 
suppression for the $Wj+jj$ background (see Section~\ref{sec:tau_Wj}). The 
probability for a $b$-quark to fluctuate into a narrow $\tau$-like jet is even 
smaller, below 0.0015, and another large reduction, by a factor 100 
(see Section~\ref{sec:tau_bb}), is expected from requiring the $b$-decay lepton 
to be well isolated. An additional factor of two reduction is achieved by 
requiring opposite charges for the isolated lepton and the tau-like jet. The 
resulting background rates, for charged leptons and $\tau$-like jets satisfying 
the transverse momentum requirements of Eq.~(\ref{eq:tauID}), are listed in the 
second row of Table~\ref{tau_data}. 

Unlike the Higgs signal or the $Zjj$ backgrounds, the reducible backgrounds show 
no resonance peaks in the $m_{\tau\tau}$ distribution. As a result, another 
reduction by an order of magnitude is achieved when comparing rates in a Higgs 
search bin of width 20~GeV (third row of Table~\ref{tau_data}). Additional 
reductions are possible by making use of specific properties of the reducible 
backgrounds. Analogous to the QCD $Zjj$ background, the $Wj+jj$ and $b\bar{b}jj$ 
backgrounds are created at smaller parton center of mass energies than the 
signal. As a result, the $m_{jj}>1$~TeV cut of Eq.~(\ref{eq:mjj_tau}) reduces 
both of them by roughly a factor of 4. 

\begin{figure}[t]
\vspace*{0.5in}
\begin{picture}(0,0)(0,0)
\includegraphics{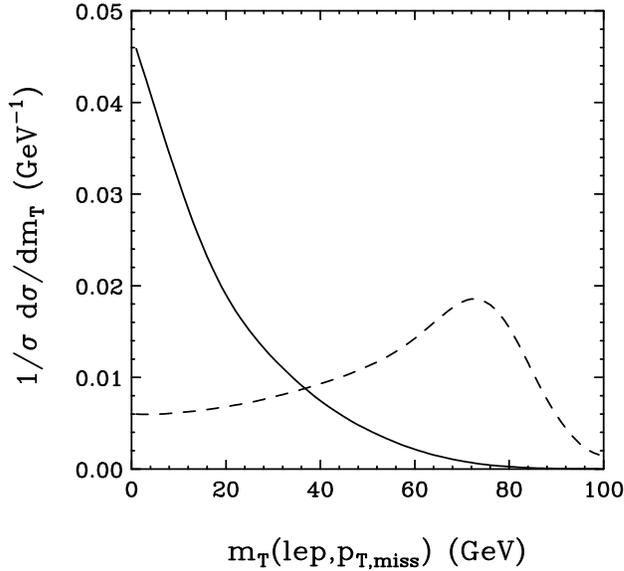}
\end{picture}
\vspace{6.0cm}
\caption{Transverse mass distribution of the $\ell$-$\sla p_T$ system for the 
$M_H = 120$~GeV $Hjj$ signal (solid line) and the $Wj+jj$ reducible background 
(dashed line), at the level of far forward tagging cuts, $\tau$-reconstruction, 
and $m_{jj} > 1$~TeV (Eqs.~\protect\ref{eq:basic}-\protect\ref{eq:gap},
\protect\ref{eq:pT_j}-\protect\ref{eq:mjj_tau}).}
\label{fig:mT}
\end{figure}

Further suppression of the $Wj+jj$ background can be achieved by taking 
advantage of the Jacobian peak in the lepton-$\sla p_T$ transverse mass 
distribution~\cite{Cavalli}, a feature which is otherwise used to measure the 
mass of the $W$. I compare the $m_T$ distribution for the signal and the 
$Wj+jj$ background in Fig.~\ref{fig:mT}. A cut 
\bq
\label{eq:mTlnu}
m_T(\ell,\sla p_T) < 30~{\rm GeV} \,
\eq
reduces the $Wj+jj$ background by a factor of 5 while reducing the signal 
acceptance by only $15\%$. Similar to the signal, the other backgrounds are 
affected very little by the transverse mass cut. 

\begin{figure}[t]
\vspace*{0.5in}
\begin{picture}(0,0)(0,0)
\includegraphics{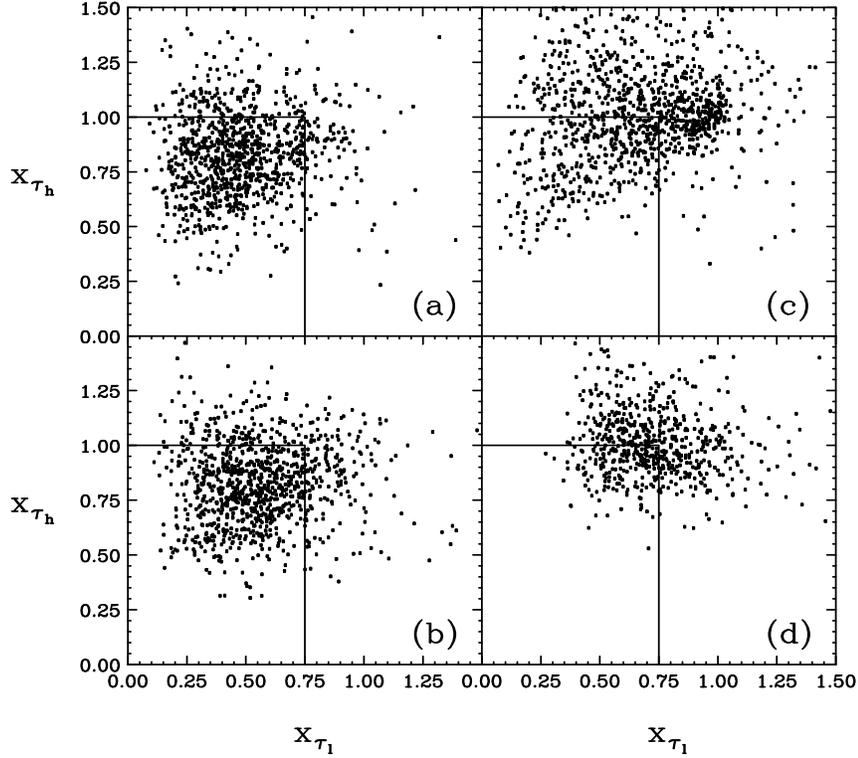}
\end{picture}
\vspace{8.5cm}
\caption{Scatter plots of $x_{\tau_l}$ v. $x_{\tau_h}$ with the cuts of
Eqs.(\protect\ref{eq:basic}-\protect\ref{eq:gap},\protect\ref{eq:tauID},
\protect\ref{eq:tauphys}-\protect\ref{eq:mTlnu}), 
for: (a) the 120 GeV $Hjj$ signal; (b) the combined QCD and EW $Zjj$ irreducible 
backgrounds; (c) the $Wj+jj$ and (d) the $b\bar{b}jj$ reducible backgrounds. 
The number of points in each plot is arbitrary and corresponds to 
significantly higher integrated luminosities than expected for the LHC. 
The solid lines indicate the cuts of Eq.~(\protect\ref{eq:x1x2}).}
\label{fig:x1x2}
\end{figure}

At this level the S/B ratio is nearly 1/1, and I study additional event 
characteristics, such as the missing momentum. As discussed previously in 
Section~\ref{sec:WW_cuts}, in real $\tau$-pair events, the missing momentum is 
a vector combination of neutrino momenta, which carry away a significant 
fraction of the $\tau^+$ and $\tau^-$ energies. In the reducible backgrounds it 
is purely from the leptonically decaying parent particle, either the $W$ or one 
of the $b$'s. As such, one should reconstruct $x_{\tau_h} = 1$ for the narrow 
$\tau$-like jet, except for smearing effects. The effect is clearly observable 
in the distribution of events in the $x_{\tau_l}$--$x_{\tau_h}$ plane, which is 
shown in Fig.~\ref{fig:x1x2}. The $x_{\tau_l}$ distribution of the leptonically 
decaying $\tau$-candidate also is softer for real $\tau$'s than for the 
reducible backgrounds, because the charged lepton shares the parent $\tau$ 
energy with two neutrinos. A cut 
\bq
\label{eq:x1x2}
x_{\tau_l} < 0.75 \, , \qquad   x_{\tau_h} < 1 \, ,
\eq
proves very effective in suppressing the reducible backgrounds. For the $Wj+jj$ 
background I find suppression by another factor of 4.5 and the $b\bar{b}jj$ 
background is reduced by a factor of 3, while retaining $80\%$ of the signal 
rate. One should note that these cuts are not optimized, they are merely chosen 
to demonstrate the usefulness of the $x_{\tau_l}$--$x_{\tau_h}$ distributions 
in restricting the otherwise troublesome reducible backgrounds to a manageable 
level. Cross sections including these cuts are given in the last row of 
Table~\ref{tau_data}. 

In principle, the $x_\tau$ distributions contain information on $\tau$ 
polarization and $x_{\tau_l}$--$x_{\tau_h}$ correlations allow one to 
distinguish between the decay of a spin-0 object, like the Higgs which results 
in opposite $\tau^+$ and $\tau^-$ chiralities, and the decay of the spin-1 $Z$ 
boson, with equal $\tau^\pm$ chiralities~\cite{taupolzn}. Comparison of the two 
scatter plots in Figs.~\ref{fig:x1x2}(a) and \ref{fig:x1x2}(b) shows, however, 
that the remaining correlations are very weak. This may partially be due to the 
stringent transverse momentum cuts (\ref{eq:tauID}) on the $\tau$ decay products 
which needed to be imposed for background reduction. In addition, the visible 
$\tau$ energy fractions in $\tau\to\ell\bar{\nu_\ell}\nu_\tau$ and 
$\tau\to\rho\nu_\tau$ decays are mediocre polarization analyzers only (measuring 
the splitting of the $\rho$'s energy between its two decay pions would improve 
the situation for the latter~\cite{HMZ}). A dedicated study is needed to decide 
whether a $\tau$ polarization analysis is feasible at the LHC, but because of the 
small rates implied by Table~\ref{tau_data} I do not pursue this issue here.

%


\section{Minijet Veto}
\label{sec:tau_mj}

Finally, I apply the results of a veto on additional central radiation, as for 
the $H\to\gamma\gamma$ and $H\to WW$ cases. Details of our calculations for the 
$H\to\tau\tau$ case may be found in Appendix~\ref{app:minijet}. In general, the 
QCD backgrounds here are rejected about three times as often as the Higgs 
signal, while the EW background is rejected only marginally more often, 
reflecting the component that shares the kinematic and color structure of the 
signal.

Table~\ref{tau_sum} applies the survival probabilities found for the 
$\eta$-method of selecting tagging jets to the cross sections after final cuts, 
for Higgs boson masses ranging from 110 to 150~GeV. A constant size of the mass 
bins of 20~GeV is kept for simplicity. In the actual experiment, the mass window 
will need to be optimized depending on the predicted width of the signal and 
background distributions, and may have to be asymmetric for low values of $m_H$. 
Our table merely shows how observing a light Higgs boson is quite feasible, even 
in the mass window close to the smeared $Z$ peak. As $m_H$ approaches 150~GeV, 
however, the $H\to\tau\tau$ branching ratio drops rapidly in the SM and the 
signal gets low for integrated luminosities of order 60~fb$^{-1}$ at low 
machine luminosity. It should be noted that with higher integrated luminosity, 
this channel is still very effective to make a direct measurement of the 
$H\tau\tau$ coupling; it would take order 200~fb$^{-1}$ to make a $5\sigma$ 
observation of a $M_H = 150$~GeV Higgs.

\begin{table}
\caption{Number of expected events for the signal and backgrounds, for 
60~fb$^{-1}$ at low luminosity and cuts as in the last line of 
Table~\protect\ref{tau_data} and a minijet veto with $p_T^{veto} = 20$~GeV, 
including an efficiency factor for tagging jet identification 
($\epsilon = 0.74$), for a range of Higgs boson masses. Mass bins of 
$\pm 10$~GeV around a given central value are assumed.}
\label{tau_sum}
\begin{center}
\begin{tabular}
{|p{0.7in}|p{0.4in}|p{0.7in}|p{0.6in}|p{0.6in}|p{0.4in}|p{0.45in}|}
\hline\hline
$m_H$(GeV) & $Hjj$ & QCD $Zjj$
       & EW $Zjj$ & $Wj+jj$ & $b\bar{b}jj$ & $\sigma_{Gauss}$ \\
\hline
%
%
110 & 24.2 & 6.3 & 3.4 & 0.3 & 0.8 & 5.7 \\
120 & 20.6 & 1.8 & 1.2 & 0.3 & 0.7 & 7.4 \\
130 & 16.0 & 0.9 & 0.7 & 0.3 & 0.6 & 6.3 \\
140 & 10.0 & 0.6 & 0.5 & 0.4 & 0.5 & 4.7 \\
150 &  4.8 & 0.4 & 0.4 & 0.3 & 0.4 & 2.6 \\
\hline\hline
\end{tabular}
\end{center}
\end{table}


\section{Discussion}
\label{sec:tau_disc}

\begin{figure}[htb]
\vspace*{0.5in}
\begin{picture}(0,0)(0,0)
\includegraphics{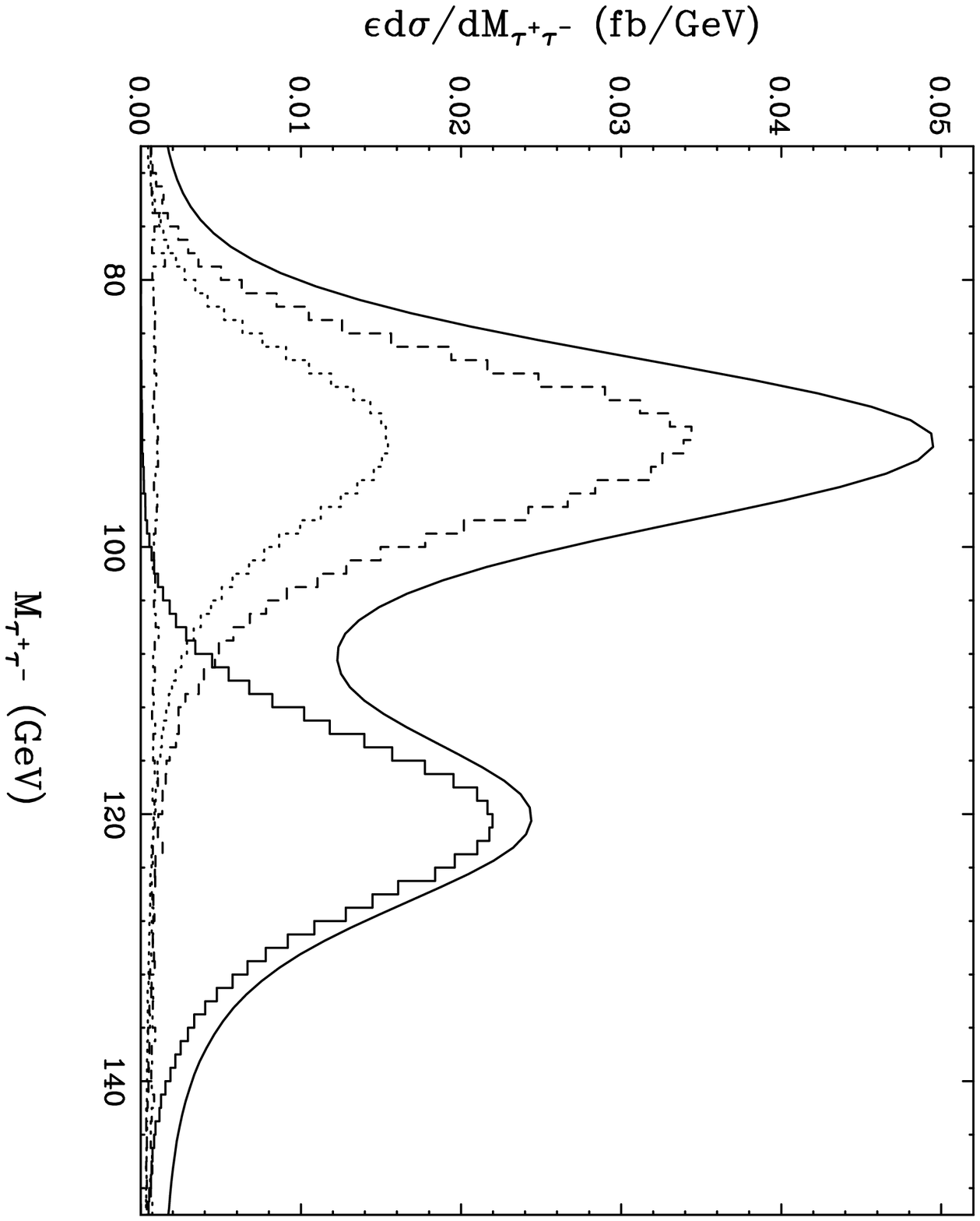}
\end{picture}
\vspace{8.7cm}
\caption{Reconstructed $\tau$ pair invariant mass distribution for the signal 
and backgrounds after the cuts of 
Eqs.~(\protect\ref{eq:basic}-\protect\ref{eq:gap},
\protect\ref{eq:pT_j}-\protect\ref{eq:x1x2}) and multiplication of the Monte 
Carlo results by the overall detector efficiencies (tagging jet identification 
efficiency $\epsilon_{tag} = (0.86)^2 = 0.74$, times $\tau$ ID or jet/$b$ 
rejection efficiencies $\epsilon_\tau$ (see 
Eqs.~(\protect\ref{eq:epstauh}, \protect\ref{eq:epsjtau},
\protect\ref{eq:epsbtau-ATLAS})) and expected 
survival probabilities. The solid line represents the sum of the signal and all 
backgrounds. Individual components are shown as histograms: the $Hjj$ signal 
(solid), the irreducible QCD $Zjj$ background (dashed), the irreducible EW $Zjj$ 
background (dotted), and the combined $Wj+jj$ and $b\bar{b}jj$ reducible 
backgrounds (dash-dotted).}
\label{fig:Mtautau}
\end{figure}

The results summarized in Table~\ref{tau_sum} show that it is possible to 
isolate a virtually background free $qq\to qqH,\;H\to\tau\tau$ signal at the 
LHC, with sufficiently large counting rate to obtain a $4-5\sigma$ signal with a 
modest 60~fb$^{-1}$ of data at low luminosity over most of the mass range. To 
reach $5\sigma$ for $M_H = 150$~GeV would require about 200~fb$^{-1}$. The 
expected purity of the signal is demonstrated in Fig.~\ref{fig:Mtautau}, where 
the reconstructed $\tau\tau$ invariant mass distribution for a SM Higgs boson 
of mass 120~GeV is shown, together with the various backgrounds, after 
application of all cuts discussed in the previous Sections. This purity is made 
possible because the weak boson fusion process, together with the 
$H\to\tau^+\tau^-\to\ell^\pm {\rm hadrons}^\mp \sla{p_T}$ decay, provides a 
complex signal, with a multitude of characteristics which distinguish it from 
the various backgrounds.

\begin{figure}[t]
\vspace*{0.5in}
\begin{picture}(0,0)(0,0)
\includegraphics{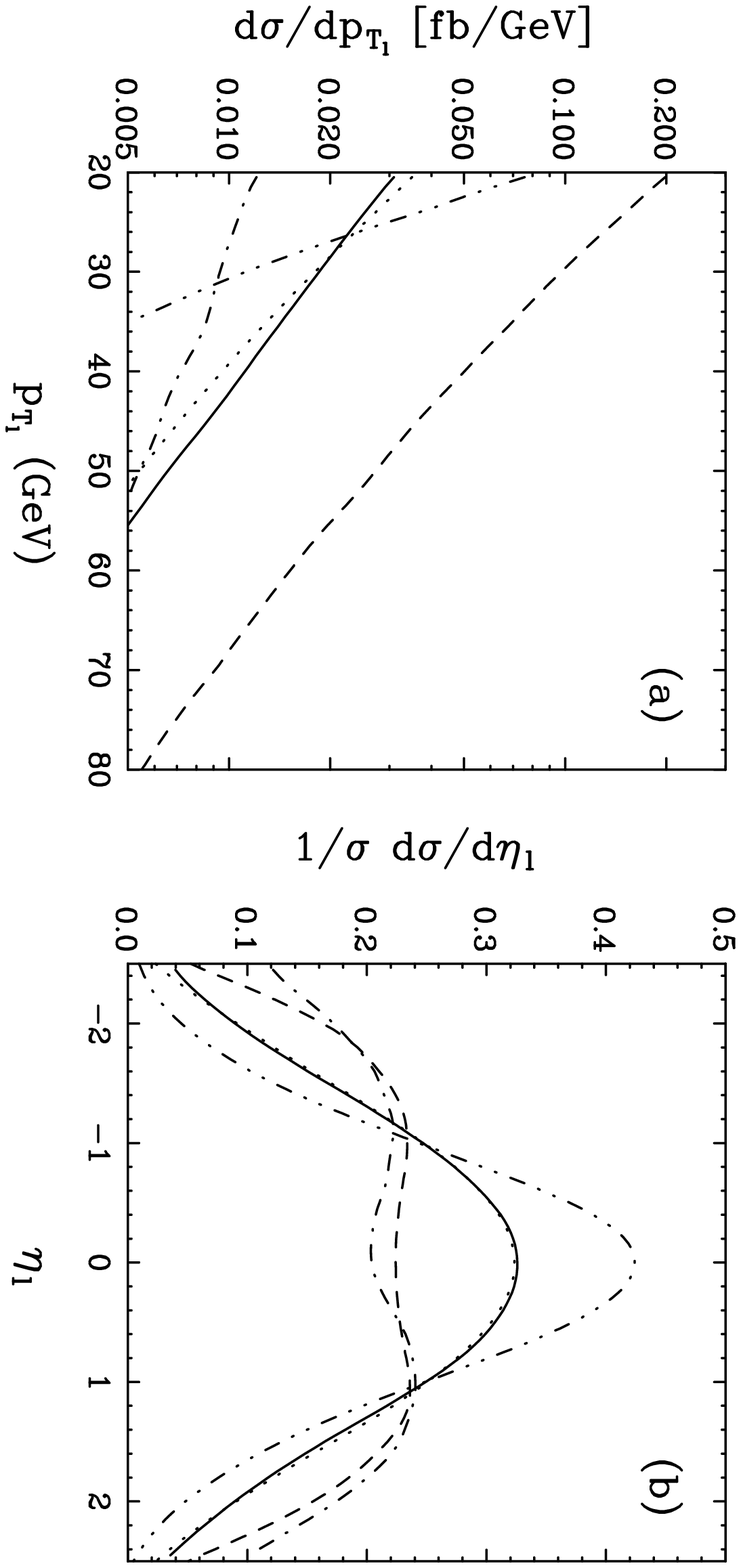}
\end{picture}
\vspace{5.2cm}
\caption{(a) Transverse momentum and (b) pseudorapidity distributions of the 
charged ``$\tau$'' decay lepton after the cuts of 
Eqs.~(\protect\ref{eq:basic}-\protect\ref{eq:gap},
\protect\ref{eq:pT_j}-\protect\ref{eq:x1x2}), for the 
$m_H = 120$~GeV signal (solid line), and backgrounds: QCD $Zjj$ production 
(dashed line), EW $Zjj$ events (dotted line), $Wj+jj$ events (dot-dashed line), 
and $b\bar{b}jj$ production (dash-double dotted line).}
\label{fig:lepton}
\end{figure}

Additional cuts beyond forward tagging as discussed in Section~\ref{sec:WBF} are 
specific to the $H\to\tau\tau$ channel, with one $\tau$ decaying leptonically 
and the other one decaying hadronically. Crucial are charged lepton isolation 
and efficient identification of the hadronically decaying $\tau$, which are 
needed for the suppression of heavy quark backgrounds and non-$\tau$ hadronic 
jets. This part of the analysis I have adapted from Ref.~\cite{Cavalli}, which, 
however, was performed for $A,H\to\tau\tau$ events from gluon fusion, {\em i.e.}, 
without requiring two additional forward tagging jets. A more detailed 
assessment of lepton isolation and hadronic $\tau$ identification in the present 
context is beyond the scope of the present work and should be performed with a 
full detector simulation.

The elimination of the $Wj+jj$ reducible background depends highly upon the 
Jacobian peak in the transverse mass distribution of the $W$ decay products. The 
other backgrounds and the Higgs signal typically produce rather small values of 
$m_T(\ell,\sla{p_T})$, below 30~GeV, and thus well below the peak in $m_T(W)$.

\begin{figure}[t]
\vspace*{0.5in}
\begin{picture}(0,0)(0,0)
\includegraphics{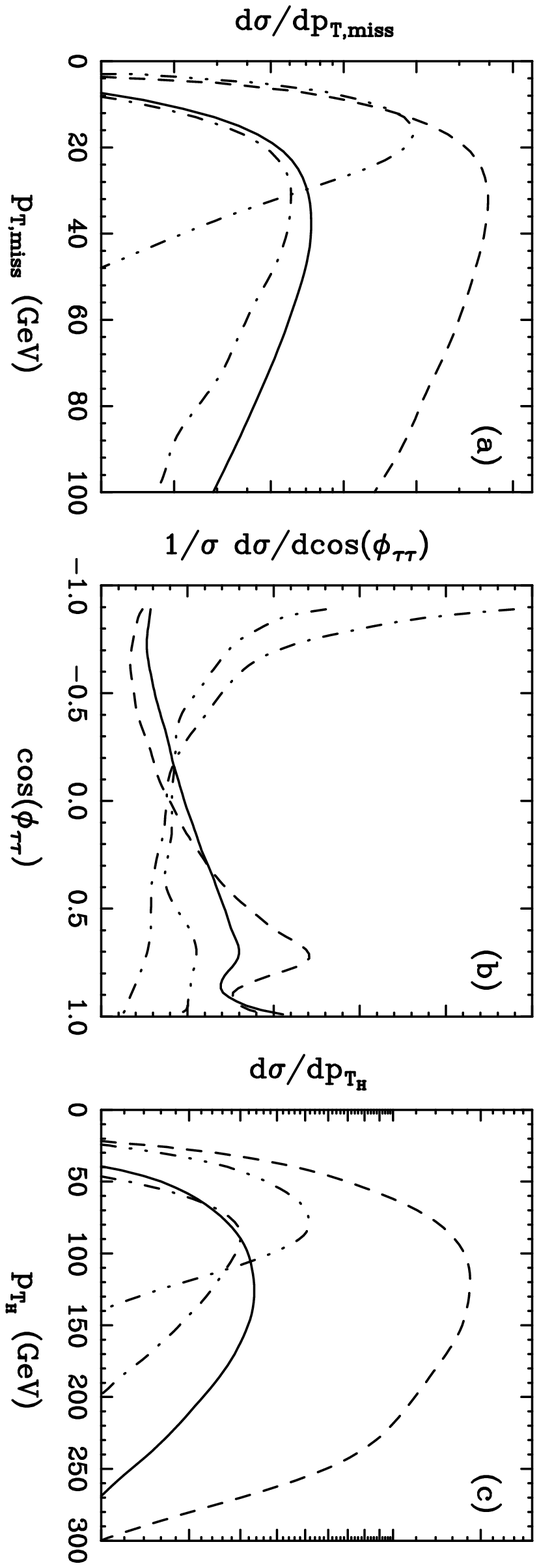}
\end{picture}
\vspace{4.0cm}
\caption{Shape comparison of various distributions for the Higgs signal (solid 
line) and the backgrounds: QCD $Zjj$ production (dashed line), $Wj+jj$ events 
(dot-dashed line), and $b\bar{b}jj$ production (dash-double dotted line). Shown 
are the (a) $\sla p_T$, (b) cos($\phi_{\tau\tau}$) and (c) transverse momentum 
distribution of the reconstructed $\tau\tau$ system, after the cuts of 
Eqs.~(\protect\ref{eq:basic}-\protect\ref{eq:gap},
\protect\ref{eq:pT_j}-\protect\ref{eq:x1x2}).}
\label{fig:taudata}
\end{figure}

Another distinguishing feature of real $\tau$ decays are the reconstructed 
momentum fractions $x_{\tau_l}$ and $x_{\tau_h}$ of the charged decay lepton and 
of the decay hadrons. Misidentified ``$\tau$'s'' tend to produce unphysically 
large values for these momentum fractions and can thereby be eliminated to a 
substantial degree (see Fig.~\ref{fig:x1x2}). The reconstruction of these $\tau$ 
momentum fractions is possible since the $\tau^+\tau^-$ pairs are typically 
being produced with sizable transverse momenta (see Fig.~\ref{fig:taudata}c). 
As a result back-to-back $\tau^+\tau^-$ decay products are rare (see 
Fig.~\ref{fig:taudata}b) and this in turns allows the mass reconstruction of 
the $\tau$-pair, which is crucial for the suppression of the main physics 
background, $Z\to\tau\tau$.

I have not made full use of the differences between the Higgs signal and the 
various backgrounds in some of these distributions. Additional examples are 
shown in Figures \ref{fig:lepton} and \ref{fig:taudata}. Fig.~\ref{fig:lepton} 
shows the $p_{T\ell}$ and $\eta_\ell$ distributions for the observable charged 
lepton, which will form an important part of the event trigger. As a result of 
the lepton isolation cut, the $p_{T\ell}$ falloff is considerably steeper for 
the $b\bar{b}jj$ background than for the signal and the other backgrounds. Not 
much leeway is present in applying more stringent cuts, however, without losing 
a substantial fraction of the signal. One can also take advantage of the 
$\eta_{\ell}$ distribution for the QCD $Zjj$ background, which, at the final 
level of cuts, remains important in particular for small values of the Higgs 
boson mass.

In addition to the lepton $p_T$, one may use the missing transverse momentum of 
the event, $\sla p_T$, Fig.~\ref{fig:taudata}(a), which is exceptionally small 
for the $b\bar{b}jj$ background. In combination with a more stringent cut on the 
$\tau$ pair opening angle, cos($\phi_{\tau\tau}$), shown in 
Fig.~\ref{fig:taudata}(b) (where an even more striking distinction between the 
physics and the reducible processes is found), both the $Wj+jj$ and $b\bar{b}jj$ 
backgrounds can be reduced even below the level discussed in 
Section~\ref{sec:tau_redbkg}. Such a strategy, however, may not increase the 
statistical significance of the signal. In fact I find that slightly looser 
cuts, for example on the dijet invariant mass, $m_{jj}$, can somewhat increase 
the significance of the signal while reducing the signal-to-background ratio. 
These points demonstrate that I have not yet optimized the search strategy for 
$H\to\tau\tau$ decays. This might be possible by combining the information from 
all the distributions mentioned above in a neural-net analysis. It is premature 
at this stage, however, to perform such an analysis since the issues of 
$\tau$-identification or of suppression of heavy quark decays in a realistic 
detector need to be addressed simultaneously, for the specific processes 
considered here.

Beyond the utility of confirmation of The Higgs' existence, independent 
measurement of a Higgs-fermion coupling will be another important reason to 
strive for observation of $H\to\tau\tau$ decays at the LHC. For such a 
measurement, via the analysis outlined in this paper, $\tau$-identification 
efficiencies, minijet veto probabilities etc. must be precisely known. For 
calibration purposes, the presence of the $Z\to\tau\tau$ peak in 
Fig.~\ref{fig:Mtautau} will be of enormous benefit. The production rates of the 
QCD and EW $Zjj$ events can be reliably predicted and, thus, the observation of 
the $Z\to\tau\tau$ peak allows for a direct experimental assessment of the 
needed efficiencies, in a kinematic configuration which is very similar to the 
Higgs signal.


%% file: body/concl.tex

The wonderfully developed and accurate theory of the electromagnetic, weak and 
strong forces of nature that is the Standard Model is nearly fully verified. It 
remains to be determined that the posited mechanism of spontaneous electroweak 
symmetry breaking and fermion mass generation, the Higgs mechanism, is correct. 
It is possible that some dynamical mechanism is instead responsible, and our 
precision electroweak data cannot determine this. It is also possible that the 
Higgs mechanism is {\it partially} correct: perhaps it provides for the 
symmetry breaking but not the fermion masses, or the other way around. 
Regardless, and barring any near-future discovery of a Higgs scalar at either 
LEP or the Tevatron, the LHC will have the capability to find a Higgs regardless 
of its possible mass.

Of course, we are most interested in the lower end of the mass range, since 
precision electroweak data suggest the Higgs lies there. I have presented in 
this dissertation three additional modes in which to search for a Standard Model
(or Standard Model-like) Higgs, via weak boson fusion in association with two 
hard tagging jets, providing for an experimental signature different from that 
of the inclusive (gluon fusion) searches. These modes completely cover the mass 
range from 100-200~GeV, safely overlapping both the LEP and Tevatron searches at 
the lower end and the ``gold-plated'' $H\to ZZ \to 4\ell$ mode at the upper end. 
Furthermore, and importantly, all modes I consider here allow for Higgs mass 
reconstruction.

\begin{figure}[t]
\begin{picture}(0,0)(0,0)
\includegraphics{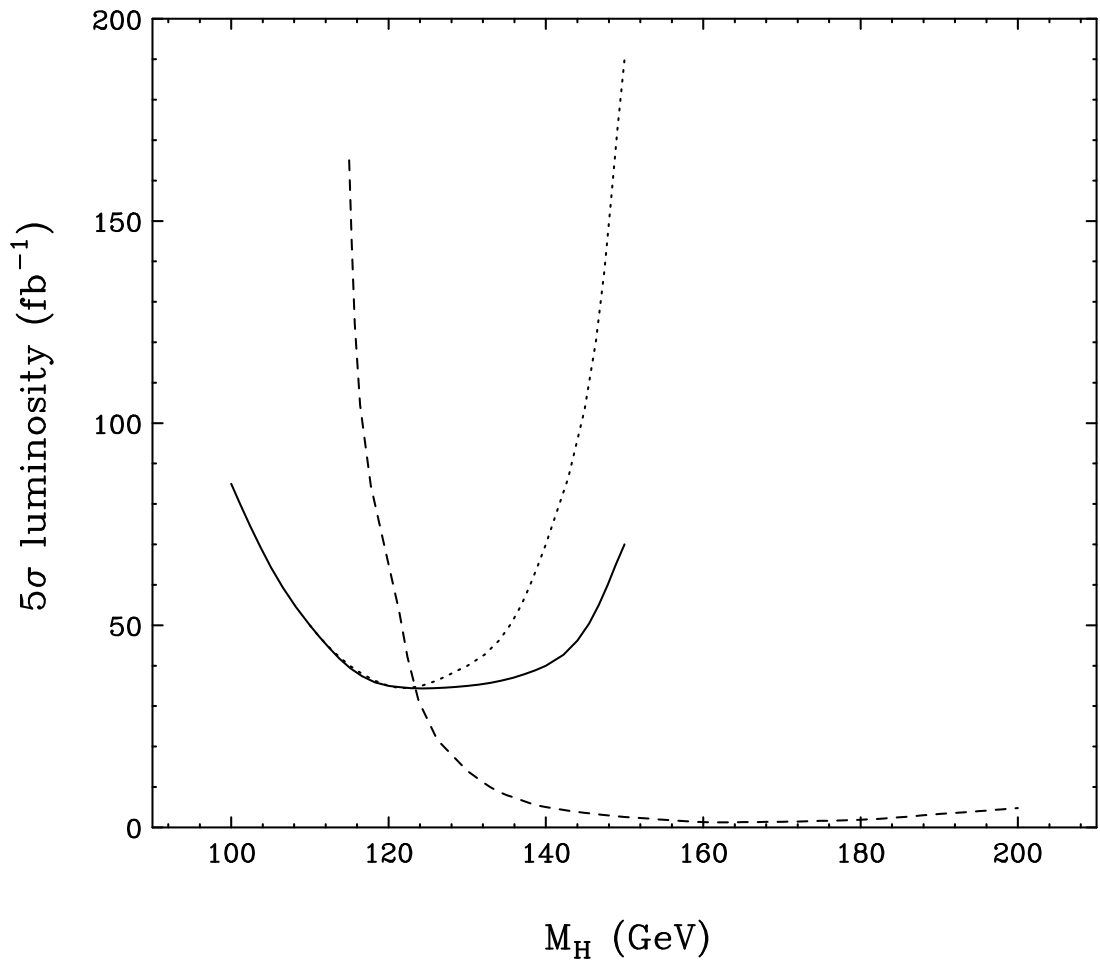}
\end{picture}
\vspace{8.5cm}
\caption{Minimum integrated luminosity (fb$^{-1}$) required to observe WBF Higgs 
production and subsequent decay in each of the three modes described in this 
dissertation: $H\to\gamma\gamma$ (solid); 
$H\to W^{(*)}W^{(*)} \to e^\pm\mu^\mp\sla{p}_T$ (dashed); and 
$H\to \tau^+\tau^- \to \ell^\pm h^\mp \sla{p}_T$ (dotted). Values quoted are for 
low machine luminosity ($10^{33}\; {\rm cm}^{-2}\; {\rm s}^{-1}$) and thus do 
not include additional efficiency factors due to minimum bias event rejection.}
\label{fig:lum}
\end{figure}

The first decay mode, $H\to\gamma\gamma$, covering the mass range of about 
110-145~GeV with 40-50~fb$^{-1}$ at low luminosity 
($10^{33}$~cm$^{-2}$~s$^{-1}$) for a $5\sigma$ statistical significance (the 
mass range can be expanded with additional luminosity, see Fig.~\ref{fig:lum}), 
nicely complements the same search in gluon fusion, which can be accomplished 
with about 20~fb$^{-1}$ with the CMS detector. At higher machine luminosities 
there will be an additional efficiency factor due to event rejection from 
minimum bias events. As the machine luminosity profile over time is not known, I 
do not attempt to include any factor in my estimates. In the event that nature 
doesn't generate fermion masses via the Higgs mechanism, this mode in weak boson 
fusion becomes extremely important to ensure that the Higgs does not go 
undetected. Mass resolution for this mode will be on the order of a GeV.

The second decay mode, $H\to W^{(*)}W^{(*)} \to e^\pm\mu^\mp\sla{p}_T$, is much 
more complicated a search channel but also highly unique in its event 
characteristics. Observation of this mode is possible for very low integrated 
luminosities, on the order of 2-10~fb$^{-1}$, if the Higgs boson lies in the 
mass range between about 130 and 200 GeV. $5\sigma$ observation can be pushed 
down to about 115~GeV with about 200~fb$^{-1}$ at low luminosity, see 
Fig.~\ref{fig:lum}. There is even room for improvement from my presentation, 
should a multivariate analysis with full detector simulation be performed on the 
observables I discussed.

The third decay mode, $H\to\tau^+\tau^-\to\ell^\pm h^\mp \sla{p_T}$, can cover 
the Higgs mass range 110-140~GeV with reasonable integrated luminosity, about 
60~fb$^{-1}$ ($\sim 30$~fb$^{-1}$ for $M_H \approx 115-130$~GeV), and the mass 
range can be extended to 150~GeV with order 200~fb$^{-1}$ of data, see 
Fig.~\ref{fig:lum}. This search is also challenging, but I have shown that it is 
feasible and desirable, as it would provide the first direct Higgs-fermion 
coupling measurement. It will also be an extremely important mode for 
investigating the possibility of a MSSM Higgs sector, as it (combined with the 
WBF $H\to\gamma\gamma$ mode) can cover the entire $\tan\beta - M_A$ parameter 
plane. If either of $h$ or $H$ is not observed, the MSSM is in serious jeopardy.

In all three cases, I advocate taking advantage of an additional fundamental 
characteristic of QCD and EW processes. Color-singlet exchange in the 
$t$-channel, as encountered in Higgs boson production by weak boson fusion (and 
in EW $Vjj$ backgrounds), leads to additional soft jet activity which differs 
strikingly from that expected for the QCD backgrounds in both geometry and 
hardness: gluon radiation in QCD processes is typically both more central 
and harder than in WBF processes. I exploit this radiation, via a veto on 
events with central minijets of $p_T > 20$~GeV, and estimate a typical $70\%$ 
reduction in QCD backgrounds, and a $20-25\%$ suppression in EW backgrounds, but 
only about a $10\%-15\%$ loss of the WBF Higgs signal.

Beyond the possibility of discovering the Higgs boson in the $H\to WW$ mode, 
or confirmation of its existence in the others, measuring the cross sections in 
both weak boson and gluon fusion will be important both as a test of the 
Standard Model and as a search for new physics. For such measurements, via the 
analyses described here, minijet veto probabilities must be precisely known. 
For calibration purposes, one can analyze $Zjj$ events at the LHC. The 
production rates of the QCD and EW $Zjj$ events can be reliably predicted 
and, thus, the observation of the $Z\to\ell\ell$ peak allows for a direct 
experimental assessment of the minijet veto efficiencies, in a kinematic 
configuration very similar to the Higgs signal.

Weak boson fusion at the LHC will be an exciting process to study, for a weakly 
coupled Higgs sector just as much as for strong interactions in the symmetry 
breaking sector of electroweak interactions.

%


%% file: body/parms.tex

I simulate $pp$ collisions at the CERN LHC, $\protect\sqrt{s} = 14$~TeV. 
For all our numerical results I have chosen ${\rm sin}^2\theta_W = 0.2315$,
$M_Z = 91.19$~GeV, and $G_F = 1.16639\cdot 10^{-5}\;{\rm GeV}^{-2}$, which
translates into $M_W = 79.94$~GeV and $\alpha(M_Z)^{-1} = 128.74$ when using
the tree-level relations between these input parameters. This value of $M_W$
is somewhat lower than the current world average of $\approx 80.35$ GeV.
However, this difference has negligible effects on all cross sections, e.g.
the $qq\to qqH$ signal cross section varies by about $0.5\%$ between these two
$W$ mass values. The tree level relations between the input parameters are
kept in order to guarantee electroweak gauge invariance of all amplitudes.
For all QCD effects, the running of the strong coupling constant is
evaluated at one-loop order, with $\alpha_s(M_Z) = 0.118$. We employ
CTEQ4L parton distribution functions~\cite{CTEQ4_pdf} throughout. Unless
otherwise noted the factorization scale is chosen as $\mu_f =$ min($p_T$)
of the defined jets, and the renormalization scale is chosen as the $p_T$ of the 
final-state colored parton emitted from a QCD vertex. For processes with more 
than one QCD vertex, the overall $\alpha_s$ is taken to be the geometric mean of 
each vertex's individual $\alpha_s(p_T)$, thus taking into account both the 
relevant scale for the hard scattering process and the soft scale for additional 
gluon emission in the relevant 3-jet processes. Inclusive WBF Higgs production 
cross sections as a function of mass are shown in Table~\ref{table_BR}, as well 
as the {\sc hdecay}~\cite{Hdecay}-corrected branching ratios to $\gamma\gamma$, 
$WW$, and $\tau\tau$ that are used in this study.

\begin{table}
\caption{Signal inclusive cross sections (pb) for $Hjj$ events of various Higgs 
masses in $pp$ collisions at $\protect\sqrt{s}=14$~TeV, and the HDECAY-corrected 
branching ratios to photon, $W$ and tau pairs.}
\label{table_BR}
\begin{center}
\begin{tabular}{|p{1.35in}|p{0.7in}|p{1.0in}|p{1.15in}|p{1.0in}|}
\hline\hline
Higgs mass (GeV) & $\sigma_{incl}$ (pb) &
   BR($H\to\gamma\gamma)$ & BR($H\to WW)$ & BR($H\to\tau\tau)$ \\
\hline
100 & 4.8 & 0.00154 & -     & 0.0792 \\
110 & 4.4 & 0.00190 & 0.047 & 0.0763 \\
120 & 4.1 & 0.00218 & 0.139 & 0.0681 \\
130 & 3.8 & 0.00221 & 0.299 & 0.0532 \\
140 & 3.5 & 0.00191 & 0.499 & 0.0347 \\
150 & 3.3 & 0.00134 & 0.700 & 0.0175 \\
160 & 3.1 & 0.00051 & 0.924 & 0.0033 \\
170 & 2.9 & -       & 0.967 & -      \\
180 & 2.7 & -       & 0.936 & -      \\
190 & 2.5 & -       & 0.779 & -      \\
200 & 2.4 & -       & 0.737 & -      \\
\hline\hline
\end{tabular}
\end{center}
\end{table}


%% file: body/smear.tex

The QCD processes discussed in the previous Chapters lead to steeply falling jet 
transverse momentum distributions. As a result, finite detector resolution can 
have a sizable effect on cross sections. These resolution effects are taken into 
account via Gaussian smearing of the energies of jets/$b$'s and charged leptons. 
All components of $p^\mu$ are smeared by the same factor derived from the energy 
smearing, {\em i.e.}, $p^2_{sm} = 0 = p^2$. We use 
\bq
\label{eq:smear_had}
{\triangle{E} \over E} =
{5.2 \over E} \oplus {0.4 \over {\sqrt E}} \oplus .009 \: ,
\eq
for jets (with individual terms added in quadrature), based on 
ATLAS expectations~\cite{CMS-ATLAS}. For charged leptons and 
photons I use
\bq
\label{eq:smear_em}
{\triangle{E} \over E} = 2\% \: ,
\eq
which is quite conservative compared to CMS expectations~\cite{CMS-ATLAS}.

In addition, finite detector resolution leads to fake 
missing-transverse-momentum in events with hard jets. An ATLAS 
analysis~\cite{ATLAS_tau} showed that these effects are well parameterized by a 
Gaussian distribution of the components of the fake missing transverse momentum 
vector, $\vec\sla p_T$, with resolution 
\bq
\sigma(\sla p_x,\sla p_y) = 0.46 \cdot \sqrt{\sum{E_{T,had}}} \: ,
\eq
for each component. The coefficient 0.46 is valid only in the low-luminosity 
scenario, which is sufficient as the searches in Chapters~\ref{ch:WW} 
and~\ref{ch:tautau} do not require more than 30~fb$^{-1}$. In my calculations, 
these fake missing transverse momentum vectors are added linearly to the 
neutrino momenta.


%% file: body/minijet.tex

\section{Introduction}

First introduced in Section~\ref{sec:minijet}, the minijet veto is a powerful 
suppression tool to extract small WBF process signatures from large QCD 
backgrounds. While the necessary information on angular distributions and 
hardness of additional radiation is available in the 3-jet and $t\bar t + jets$ 
processes discussed in Chapters~\ref{ch:gammagamma}-\ref{ch:tautau}, one must 
either regulate or reinterpret these divergent cross sections. Here I will 
represent cross sections for 2-jet processes by $\sigma_2$. These calculations 
are completely perturbative and well-behaved, and will be regarded as inclusive 
cross sections for the respective processes under consideration. Cross sections 
for 3-jet processes, which are in general the respective 2-jet hard scattering 
process with one additional gluon emission, and all crossing related diagrams, 
will be represented by
\bq
\sigma_3 = \int_{p_{T,soft}}^\infty {d\sigma_3 \over dp_{T3}} dp_{T3} \: .
\eq
These cross sections are divergent for low $p_T$ of the additional gluon 
emission. For WBF processes, $\sigma_3 = \sigma_2$ typically for 
$p_{T,soft} \approx 10$~GeV, whereas for QCD processes, this occurs typically 
around $p_{T,soft} \approx 40$~GeV for the LHC processes considered in this 
dissertation.

In this Appendix I explain the technical details of both the truncated shower 
approximation (TSA)~\cite{TSA}, used to regulate the 3-jet calculations, 
and the exponentiation model, which alternatively reinterprets the meaning of 
the divergent cross sections $\sigma_3$. 


\section{The Truncated Shower Approximation}

When several soft gluons are emitted in a hard scattering event their transverse 
momenta tend to cancel, leading to a regularization of the small-$p_T$ 
singularity (where the $p_T$ is the recoil of the hard scattering system) which 
is present when considering single-parton emission only. In the TSA these 
effects are simulated by replacing the tree-level three-jet differential cross 
section, $d\sigma_3^{\rm TL}$, with
\bq
\label{eq:TSA}
d\sigma_3^{\rm TSA}=d\sigma_3^{\rm TL}
\left(1-e^{-p_{T3}^2/p_{TSA}^2}\right)\;.
\eq
Here the parameter $p_{TSA}$ is chosen to correctly reproduce the tree-level 
two-jet cross section, $\sigma_2$, within the cuts specified for the comparison, 
typically the core forward tagging cuts of Eqs.~(\ref{eq:basic}-\ref{eq:gap}), 
\ba
\label{eq:basic.}
& p_{T_j} \geq 20~{\rm GeV} \, ,\qquad |\eta_j| \leq 5.0 \, ,\qquad 
\triangle R_{jj} \geq 0.7 \, , & \nonumber\\
& |\eta_X| \leq 2.5 \, , \qquad \triangle R_{jX} \geq 0.7 \, . &
\ea
\bq
\label{eq:Xcen.}
\eta_{j,min} + 0.7 < \eta_{X_{1,2}} < \eta_{j,max} - 0.7 \, , \qquad
\eta_{j_1} \cdot \eta_{j_2} < 0 \: .
\eq
\bq
\label{eq:gap.}
\triangle \eta_{tags} = |\eta_{j_1}-\eta_{j_2}| \geq 4.4 \, ,
\eq
where $X$ is any observable Higgs decay product; plus any additional cuts on 
the tagging jets that alter the hardness of the underlying event, such as 
increased $p_{T_j}$ or $M_{jj}$ cuts. {\it I.e.}, $p_{TSA}$ is fixed by the 
matching condition
\bq
\label{eq:match}
\sigma_2 = \int_0^\infty {d\sigma_3^{\rm TSA}\over dp_{T3}} dp_{T3}\; .
\eq
$p_{TSA}$ values are typically $\lesssim 10$~GeV for WBF processes such as the 
Higgs signal, but much higher for QCD processes, characteristically 40-60~GeV.

Using $d\sigma_3^{\rm TSA}$ as a model for additional jet activity one can make 
an estimate of the probability of observing at least one additional jet with 
$p_{T_j} > p_T^{veto}$ in the central region. These events can then be vetoed, 
as their angular distribution is determined correctly. However, in the TSA only 
one soft parton is generated, with a finite probability to be produced outside 
the veto region of Eq.~(\ref{eq:etaveto}). Therefore the veto probability will 
never reach 1, no matter how low $p_{T,\rm veto}$ may be. 

At small values of $p_{T,\rm veto}$ one would expect to underestimate the veto 
probability for QCD processes because the TSA does not take into account 
multiple parton emission. Because of this our TSA veto probability estimates 
may be regarded as conservative.


\section{The Exponentiation Model}

In the soft region gluon emission dominates, and one may assume that this 
soft-gluon radiation approximately exponentiates, {\em i.e.}, the probability 
$P_n$ for observing $n$ soft jets in the veto region is given by a Poisson 
distribution,
\bq\label{eq:expon}
P_n = {\bar n^n\over n!}\; e^{-\bar n} \;,
\eq
with
\bq
\label{eq:nbar}
\bar n = \bar n(p_{T,\rm veto}) = {1 \over \sigma_2}\; 
\int_{p_{T,\rm veto}}^{\infty} dp_{T3}\; 
{d\sigma_3 \over dp_{T3}}\; ,
\eq
where the unregularized three-parton cross section is integrated over the veto 
region of Eq.~(\ref{eq:mjveto}) and then normalized to the two-jet cross 
section, $\sigma_2$. A rough estimate of multiple emission effects is thus 
provided by using
\begin{equation}\label{Pvetoexp}
P_{exp}(p_{T,\rm veto}) \, = \, 1 - P_0 \, = \, 1 - e^{-\bar n(p_{T,\rm veto})} 
\end{equation}
for the veto probability.

Within the exponentiation model, $\bar{n} = {\sigma_3\over\sigma_2}$ represents 
the average multiplicity of minijets in the central region, between the two 
tagging jets. Even if the exponentiation model is of only limited accuracy, the 
ratio of three- to two-jet tree-level cross sections gives the best perturbative 
estimate available of the minijet activity in hard scattering events. One finds 
that the average minijet multiplicity depends strongly on the hardness of the 
underlying event. This is why I make estimates of veto probability only at the 
most stringent level of cuts on the hard scattering process. 


\section{Application}

To employ a minijet veto in this study, I must first establish an algorithm for 
selecting tagging jets in the three-jet simulations such that events are 
selected that reflect the phase space region of the hard scattering of the 
analogous two-jet events. Only when this is ensured will the matching condition 
of Eq.~(\ref{eq:match}) or the multiplicity definition, Eq.~(\ref{eq:nbar}), 
make sense. In our previous studies of the minijet veto 
technique~\cite{RZ_tautau,RSZ_vnj}, the matching condition (or calculation of 
$\bar{n}$) were performed without enforcement of the forward tagging cuts of 
Eqs.~(\ref{eq:Xcen.},\ref{eq:gap.}), even though tagging jet candidates were 
chosen for the purpose of identifying the veto candidate; tagging jet candidates 
were selected as the two most energetic~\cite{RSZ_vnj} or two 
highest-pT~\cite{RZ_tautau} defined jets ($p_T > 20$~GeV), in opposite detector 
hemispheres. However, without the additional forward tagging cuts, this will 
overestimate the effectiveness of the veto, especially for QCD processes. Taking 
QCD $Zjjj$ production as an example, with $Z\to\ell^+\ell^-$, the tagging 
candidate selection will frequently allow low-invariant-mass hard scattering 
events with gluon radiation forward of the quark jets to contribute to 
$\sigma_3$, where one of the tagging jets is the gluon instead of the hard 
quarks of the LO process. Because the $p_T$ spectrum of the central quark jets, 
one of which is then identified as a veto candidate, can be very hard, on the 
order of $40-60$~GeV, the minijet ``gluon radiation'' spectrum is shifted 
toward higher $p_T$ values.

A more realistic estimate of the minijet $p_T$ spectrum is obtained by applying 
the matching condition (or calculating $\bar{n}$) only in the phase space 
region where a comparison of signal and background will take place: after all 
acceptance cuts, determined at the two-jet level, have been imposed; especially 
the forward tagging cuts of Eqs.~(\ref{eq:Xcen.},\ref{eq:gap.}), which 
significantly alter the hardness of the event selection.

Once the full level of cuts for a given search scenario are imposed, one may 
examine different tagging jet selection algorithms to optimize the veto. 
Ideally, the outgoing quarks would always be selected, so that the additional 
gluon radiation is always the veto candidate. In practice, this is impossible, 
but for the Higgs signal various algorithms can achieve ``proper'' quark tagging 
with about $75\%$ efficiency, a high success rate. Briefly, these might be the 
two highest-$p_T$ jets, or the two jets closest to the reconstructed Higgs. 
Most algorithms have very little difference from each other in the case of the 
WBF signature. Thus, I choose an algorithm that allows more suppression of the 
QCD backgrounds. The final algorithm I chose is to select the highest-$p_T$ jet 
as the first tagging jet, since it will almost always be part of the hard 
scattering, and then choose the other tagging jet such that the event is more 
likely to pass the forward tagging cuts: look for jets with $p_T > 20$~GeV in 
the opposite hemisphere, such that the candidate Higgs decay products are 
between the tagging jets, satisfying Eq.~(\ref{eq:Xcen.}). This performs 
somewhat superior to merely choosing the two highest-$p_T$ jets.

Also in contrast to our previous studies~\cite{RZ_tautau,RSZ_vnj}, the veto 
candidates are defined jets ($p_T > 20$~GeV) anywhere between the tagging jets,
\bq
\label{eq:vetocand}
\eta_{tag,min} < \eta_{j,veto} < \eta_{tag,max} \: .
\eq
Previously, the veto candidate also had to be at least 0.7 units of rapidity 
away from the tagging jets, but the choice of Eq.~(\ref{eq:vetocand}) allows 
for more suppression of the backgrounds than the more restrictive selection.

Once the tagging selection algorithm is established, I need an estimate of the 
veto survival probability, $P_{surv}$, for the Higgs signal. The WBF Higgs 
processes including additional gluon emission were first discussed in 
Section~\ref{sec:minijet}, and the determination of $p_{TSA}$ and estimation of 
$P_{surv}$ with the TSA method were calculated via $pp\to Hjjj$ production with 
decays to both $H\to\gamma\gamma$ and $H\to\tau\tau$ with the full level of 
acceptance cuts for each decay mode as discussed in their respective Chapters. 
This includes the forward tagging acceptance cuts of 
Eqs.~(\ref{eq:basic.}-\ref{eq:gap.}) in both cases. For $H\to\gamma\gamma$ 
decays, the additional cuts are those of Eq.~(\ref{eq:phofinal}):
\ba
\label{eq:phofinal.}
p_{T_{j(1,2)}} \geq 40, 20~{\rm GeV}\; ,
\nonumber \\ 
p_{T_{\gamma (1,2)}} \geq 50, 25~{\rm GeV}\; .
\ea
In this case, I calculate $\bar{n} = 0.067$ and $p_{TSA} = 5.4$~GeV. This 
translates to $P_{surv} = 0.94$ for the exponentiation model and 
$P_{surv} = 0.88$ for the TSA. Once $p_{TSA}$ is set, I may switch to a tagging 
algorithm that selects the two defined jets closest in rapidity to the 
reconstructed Higgs, and find instead $P_{surv} = 0.86$. The small differences 
there are reassuring and due to slight changes in the allowable phase space of 
events. Considering the nature of the approximations in the exponentiation model 
and the TSA, it is also highly reassuring that their respective results agree to 
such a degree. For the purposes of this study I take the average of the results, 
$P_{surv} = 0.89$. These results are independent of the Higgs mass within Monte 
Carlo errors.

For $H\to\tau\tau$ decays, the additional cuts beyond forward tagging are those 
of Eqs.~(\ref{eq:pT_j},\ref{eq:tauID}-\ref{eq:x1x2}):
\bq
\label{eq:pT_j.}
p_{T_{j(1,2)}} \geq 40, 20~{\rm GeV} \: ,
\eq
\bq
\label{eq:tauID.}
p_{T_{\tau,lep}} > 20~{\rm GeV} \, , \qquad
p_{T_{\tau,had}} > 40~{\rm GeV} \, ,
\eq
\bq
\label{eq:tauphys.}
\cos\theta_{\tau\tau} > -0.9 \, , \qquad  x_{\tau_{l,h}} > 0 \, ,
\eq
\bq
\label{eq:mjj_tau.}
m_{jj} > 1~{\rm TeV} \, ,
\eq
\bq
\label{eq:mTlnu.}
m_T(\ell,\sla p_T) < 30~{\rm GeV} \, ,
\eq
\bq
\label{eq:x1x2.}
x_{\tau_l} < 0.75 \, , \qquad   x_{\tau_h} < 1 \, ,
\eq
Here I find $\bar{n} = 0.128$ and $p_{TSA} = 5.5$~GeV, which gives 
$P_{surv} = 0.88$ for the exponentiation model, $P_{surv} = 0.87$ for the TSA 
with $p_T$-tagging as described above, and $P_{surv} = 0.84$ for the TSA with 
``$\eta$-tagging'' as before. The average is $P_{surv} = 0.87$. That the 
survival probability is slightly smaller here than for the $H\to\gamma\gamma$ 
case is probably due to the increased hardness of the event via the cut 
$M_{jj,tags} > 1$~TeV: the average maximum tagging jet $p_T$ is 
$\approx 100$~GeV for $H\to\gamma\gamma$ but $\approx 140$~GeV for 
$H\to\tau\tau$; and similarly for the minimum tagging jet $p_T$. By $p_T$ 
balancing, the additional gluon radiation in the $M_{jj,tags} > 1$~TeV events 
will be somewhat harder on average, leading to a slightly increased chance of 
being vetoed.

Because these results are so similar, and the $M_{jj}$ cut is softer for the 
case $H\to WW$, there I apply the average value of $P_{surv} = 0.89$ as found 
for the case of decays to photons.

I also need to establish veto survival probabilities for the backgrounds. It is 
not necessary to do this for all backgrounds, only for characteristic classes, 
{\em e.g.} QCD and EW production of weak bosons in association with two tagging 
jets, $t\bar{t}+jets$, etc.

For QCD and EW production of weak bosons in association with jets I examined QCD 
and EW $Zjj(j)$ production with subsequent decay $Z\to\tau\tau$, as discussed in 
Sections~\ref{sec:QCD_tau} \& \ref{sec:EW_tau}, with subsequent tau decay to 
$e^\pm \mu^\mp$ or $\ell^\pm h^\mp$, reflecting the backgrounds to $H\to WW$ and 
$H\to\tau\tau$, respectively. For tau decays to $e^\pm \mu^\mp$ all the cuts of 
Eqs.~(\ref{eq:Wlepmin}-\ref{eq:tau}) were used in addition to forward tagging:
\bq
\label{eq:Wlepmin.}
p_{T_l} \, > \, 20 \: {\rm GeV} \, ,
\eq
\bq
\label{eq:mjj_W.}
m_{jj} > 650~{\rm GeV} \; ,
\eq
\bq
\label{eq:ang.}
\phi_{e\mu} < 105^{\circ} \, , \, \, \, 
{\rm cos} \; \theta_{e\mu} > 0.2 \, , \, \, \, 
\triangle R_{e\mu} < 2.2 \, ,
\eq
\bq
\label{eq:adv.}
m_{e\mu} < 110 \; {\rm GeV} \, , \, \, \, 
p_{T_{e,\mu}} < 120 \; {\rm GeV}  \, ,
\eq
\bq
\label{eq:tau.}
{\rm veto \: if} \qquad x_{\tau_1} , \; x_{\tau_2} > 0 \qquad{\rm and} \qquad
m_Z - 25\ {\rm GeV}\; < m_{\tau\tau} < \; m_Z + 25\ {\rm GeV} \, .
\eq
For the QCD processes I find $\bar{n} = 1.46$ and $p_{TSA} = 43.5$~GeV, yielding 
the estimates $P_{surv,\bar{n}} = 0.23$, $P_{surv,TSA(p_T)} = 0.34$ and 
$P_{surv,TSA(\eta)} = 0.30$. The average is $P_{surv} = 0.29$. For the EW 
processes I find $\bar{n} = 0.25$ and $p_{TSA} = 11.2$, translating to 
$P_{surv,\bar{n}} = 0.78$, $P_{surv,TSA(p_T)} = 0.76$ and 
$P_{surv,TSA(\eta)} = 0.71$. The average is $P_{surv} = 0.75$. The intermediate 
value of $P_{surv}$ for the EW $Zjj$ case reflects the partial bremsstrahlung 
nature of the hard scattering process, which can allow radiation back into the 
central region.

For the decays $Z\to\tau\tau\to \ell^\pm h^\mp$, I use the forward tagging cuts 
and those of Eqs~(\ref{eq:pT_j.}-\ref{eq:x1x2.}). For the QCD processes I find 
$\bar{n} = 1.43$ and $p_{TSA} = 41.8$, yielding the 
estimates $P_{surv,\bar{n}} = 0.24$, $P_{surv,TSA(p_T)} = 0.32$ and 
$P_{surv,TSA(\eta)} = 0.28$. The average is $P_{surv} = 0.28$, essentially the 
same as that found for the cuts of the $H\to WW$ scenario above. For the EW 
processes I find $\bar{n} = 0.16$ and $p_{TSA} = 6.7$, translating to 
$P_{surv,\bar{n}} = 0.85$, $P_{surv,TSA(p_T)} = 0.82$ and 
$P_{surv,TSA(\eta)} = 0.78$. The average is $P_{surv} = 0.82$. This survival 
probability is slightly higher than that for $Z\to\tau\tau\to e^\pm \mu^\mp$, 
which is due to the fact that while the average $p_T$ distributions of the jets 
is the same for both cases, the higher $M_{jj,tags}$ requirement has moved 
event selection to the region where the bremsstrahlung component of $Z$ 
production is much less, further reducing the fraction of minijets expected in 
the central region.

I use the above values for all QCD \& EW backgrounds consisting of one or two 
weak bosons plus two tagging jets, for the respective cuts considered, and 
also the QCD backgrounds $Wj+jj$ and $b\bar{b}+jj$. The last is not quite the 
same, but does share hardness characteristics; a separate determination is not 
possible since the matrix elements for $b\bar{b}+jjj$ do not yet exist. For the 
QCD $\gamma\gamma jj$ backgrounds (including DPS) I use $P_{surv} = 0.30$, 
extrapolating the other results for a lower value of $M_{jj,tags}$; and for the 
EW $\gamma\gamma jj$ background I use $P_{surv} = 0.75$. The latter value 
probably underestimates the effectiveness of the veto as it is a yet lower 
$M_{jj,tags}$ case. Thus, I am conservative in estimating the total background 
rates in the analysis of Chapter~\ref{ch:gammagamma}.

Finally, I must consider $t\bar{t}+jets$ processes separately, as the typical 
hardness of the underlying event is not necessarily similar to any of the above 
processes. For these processes it is simpler in practice to reinterpret the 
divergent higher-order cross sections in the context of the exponentiation 
model, rather than use the TSA, which would need to be redefined in light of the 
additional jets. The survival probability for $t\bar{t}$ events was determined 
using the $t\bar{t}j$ processes where the jet is allowed to be soft and the two 
$b$ jets from top decay were identified as the tagging jets. Similarly, the 
survival probability for $t\bar{t}j$ events was determined using the 
$t\bar{t}jj$ processes, where one $b$ jets is identified as a tagging jet and 
one light quark or gluon jet is identified as the other tagging jet; one light 
quark or gluon jet is then allowed to be soft. I found a veto survival 
probability of $P_{surv} = 46\%$ for $t\bar{t}$ events and $P_{surv} = 12\%$ for 
$t\bar{t}j$ events. Both of these results disagree with our other estimates of 
$P_{surv}$ for QCD processes. This may be understandable for $t\bar{t}$ events, 
as at tree level this component does no contain any t-channel gluon exchange 
processes, which all of the other QCD backgrounds do. I also observed that the 
additional radiation in $t\bar{t}$ events typically falls outside the central 
gap. I did not explore this any further as the $t\bar{t}$ component is 
negligible. That the value of $P_{surv}$ found for $t\bar{t}j$ events with the 
exponentiation model is so much smaller than that for other QCD backgrounds 
within the TSA may be understood for two reasons: the exponentiation method 
always gives a lower value for $P_{surv}$ than the TSA for QCD processes; and 
Ref.~\cite{DittDrein} have shown that the off-shell top contribution to the 
$t\bar{t} + jets$ background is not negligible, but I do not include it here. 
As these two issues are not yet fully explored I prefer to remain conservative 
and apply the value $P_{surv} = 0.29$ for the $t\bar{t} + jets$ backgrounds. 
Thus, it is possible that I am overestimating their contribution.

\begin{table}
\caption{Summary of veto survival probabilities for $p_T^{veto} = 20$~GeV used 
in Chapters~\ref{ch:gammagamma}-\ref{ch:tautau}.}
\label{vetosum}
\begin{center}
\begin{tabular}
{|p{0.85in}|p{0.35in}|p{0.35in}|p{0.35in}|p{0.55in}|p{0.55in}|p{0.45in}|p{0.4in}
 |p{0.4in}|p{0.4in}|p{0.35in}|}
\hline\hline
search & $Hjj$ & $t\bar{t}$ & $t\bar{t}j,$ & QCD      & EW       & QCD    
& QCD          & DPS               \\
       &       &            & $t\bar{t}jj$ & $V(V)jj$ & $V(V)jj$ & $Wjjj$ 
& $b\bar{b}jj$ & $\gamma\gamma jj$ \\
\hline
$\gamma\gamma jj$  & 0.89 &  -   &  -   & 0.30 & 0.75 &  -   &  -   & 0.30 \\
$W^{(*)}W^{(*)}jj$ & 0.89 & 0.46 & 0.29 & 0.29 & 0.75 &  -   &  -   &  -   \\
$\tau\tau jj$      & 0.87 &  -   &  -   & 0.28 & 0.80 & 0.28 & 0.28 &  -   \\
\hline\hline
\end{tabular}
\end{center}
\end{table}
%


%% file: body/taudecay.tex

\section{Tau Decays}

This analysis critically employs transverse momentum cuts on the charged 
$\tau$-decay products and, hence, some care must be taken to ensure realistic 
momentum distributions. Because of its small mass, I simulate the $\tau$ decays 
in the collinear approximation. The energy fraction $z$ of the charged decay 
lepton in $\tau^\pm \to \ell^\pm\nu_\ell\nu_\tau$ is generated according to the
decay distribution
\bq
{1\over \Gamma_\ell} {d\Gamma_\ell\over dz} =
{1\over 3} (1-z) \left[ (5+5z -4z^2) + \chi_\tau(1+z-8z^2) \right]\; .
\eq
Here $\chi_\tau$ denotes the chirality of the decaying $\tau$ (which, for a 
negative helicity $\tau^-$ or positive helicity $\tau^+$, is given by 
$\chi_\tau=-1$ in the relativistic limit). Similarly the pion spectrum for 
$\tau^\pm \to \pi^\pm\nu_\tau$ decays is given by
\bq
{1\over \Gamma_\pi } {d\Gamma_\pi\over dz} \simeq 1 + \chi_\tau(2z-1)\; .
\eq
Decay distributions for $\tau\to\rho\nu_\tau$ and $\tau\to a_1\nu_\tau$ are 
taken from Ref.~\cite{HMZ}. I add the decay distributions from the various 
hadronic decay modes according to their branching ratios. The vector meson 
decays are simulated in the narrow width approximation, which is adequate for 
my purposes, where the energy fraction is that of the vector meson. The decay 
of the Higgs scalar produces $\tau$'s of opposite chirality, 
$\chi_{\tau^+}=-\chi_{\tau^-}$ and this anti-correlation of the $\tau^\pm$ 
polarizations is taken into account in our study. For the Higgs signal, there is 
no correlation of this final state with Higgs production, as the scalar carries 
no spin information. For vector boson production and decay there is correlation, 
however, which I simulate exactly for the case of the QCD $Z\to\tau\tau$ 
background and approximate in the EW $Z\to\tau\tau$ simulation.


\section{Tau Reconstruction}

A collider experiment will measure the momenta of massless final-state particles 
in an event, as well as the missing transverse momentum. I may denote these 
quantities for two massless final-state particles as $\vec{k}_1,\vec{k}_2$, and 
$\vec\sla{p}_T$, respectively. For a hadron collider, the longitudinal component 
of missing momentum cannot be determined. For dual tau decays 
$\vec{k}_1,\vec{k}_2$ are the momenta of the charged particles in the decay, 
leptons or hadrons. I will ignore all other final-state particles in an event 
here and assume that the missing momentum $\vec\sla{p}_T$ comes entirely from 
the escaping neutrinos in the two tau decays. I also neglect the $\tau$ mass, 
as the collinearity condition already requires a $\tau$ transverse momentum much 
larger than the mass. 

Once the conditions for a collinear $\tau$ decay approximation are satisfied, 
the only unknowns are the two fractions of parent $\tau$ energy which each 
observable decay particle carries, which I denote by $x_{\tau_i}$ in the text 
and abbreviate as $x_1,x_2$ here. If $\vec{p}_1,\vec{p}_2$ are the tau momenta 
before decay, then we may write conservation of momentum in the transverse 
plane:
\bq
\label{eq:pTcons}
(\vec{p}_{1_T} + \vec{p}_{2_T}) \; = \; 
{\vec{k}_{1_T} \over x_1} + {\vec{k}_{2_T} \over x_2} \; = \; 
\vec{k}_{1_T} + \vec{k}_{2_T} + \vec\sla{p}_T \, .
\eq
The transverse momentum vectors are then related by
\bq
\label{eq:taurecon}
\vec\sla{p}_T = \biggl({1 \over x_1} - 1 \biggr) \; \vec{k}_{1_T} +
\biggl({1 \over x_2} - 1 \biggr) \; \vec{k}_{2_T} \, .
\eq
As long as the the decay products are not back-to-back, Eq.~(\ref{eq:taurecon}) 
gives two conditions for $x_{\tau_i}$ and provides  the $\tau$ momenta as 
$\vec{p}_i/x_i$. 
Splitting into $x$ and $y$ components, this may be rewrittien in matrix form,
\bq
\label{eq:taurecon_matrix}
\left( 
\begin{array}{cc} k_{1_x} & k_{2_x} \\ k_{1_y} & k_{2_y} \end{array} 
\right) \quad 
\left( 
\begin{array}{c} {1\over x_1} - 1 \\ {1\over x_2} - 1 \end{array} 
\right) \quad = \quad
\left( 
\begin{array}{c} \sla{p}_{T_x} \\ \sla{p}_{T_y} \end{array} 
\right) \, ,
\eq
which may be inverted to the form
\bq
\label{eq:x1x2inv_matrix}
\left( 
\begin{array}{c} {1\over x_1} - 1 \\ {1\over x_2} - 1 \end{array} 
\right) \quad = \, 
{1 \over k_{1_x}k_{2_y} - k_{1_y}k_{2_x}} \; 
\left( 
\begin{array}{c} k_{2_y}\sla{p}_{T_x} - k_{2_x}\sla{p}_{T_y} \\ 
                 k_{1_x}\sla{p}_{T_y} - k_{1_y}\sla{p}_{T_x}    \end{array} 
\right) \, .
\eq

Once $x_1,x_2$ have been solved, it is then simple to reconstruct the tau-pair 
invariant mass as follows:
\bq
\label{eq:tautaumass}
m^2_{\tau^+\tau^-} \; = \; (p_1+p_2)^2 \; = \; 2 (p_1 \cdot p_2 + m^2_\tau) 
\; \approx \; 2 \biggl({k_1 \cdot k_2 \over x_1 x_2} + m^2_\tau \biggr) 
\; \approx \; {2 k_1 \cdot k_2 \over x_1 x_2} \, .
\eq
This technique is discussed in more detail in Ref.~\cite{tautaumass}.


%% file: body/programs.tex

\section{Program Structure}

A typical Monte Carlo calculation begins with initialization of the phase space
integration limits and other constants, such as fundamental parameters. A loop 
is then performed for a large number of iterations, until the sampling error 
becomes small, wherein a set of random numbers is generated which is translated 
into four-momenta for the initial- and final-state particles. If the phase space 
point is valid and the configuration passes the cuts imposed on the final-state 
configuration desired, then the matrix element is calculated and squared for 
each subprocess. Each subprocess is folded with the structure functions for the 
incoming hadrons before they are summed together to be multiplied by the phase 
space weight. At the end of the calculation, collected statistics may be output 
to histograms for various phase space variables.

Our general Monte Carlo package structure is a suite of {\sc fortran} (77 or 90) 
programs, where each program performs a specific task in the process described 
above. The main program which drives the integration is generally named after 
the process being calculated. For example, to calculate the Higgs signal, the 
suite consists of the following program files:
\begin{verbatim}
          qqhmain.f
          koppln.f
          qqhqq.f
          tbv.f
          monaco.f
          ps.f
          smear.f
          cuts.f
          tautau.f
          faketaus.f
          calcqsqr.f
          m2sZH.f
          hist.f
          func.f
\end{verbatim}
All parameters that need to be easily changed are put in {\tt .inc} files, 
which centralizes the location so a change has to be made only once.

qqhmain begins by initializing all the necessary variables and routines, from 
couplings ({\tt koppln.f, qqhqq.f, tbv.f}) to histograms ({\tt hist.f}). A loop 
over adaptive Monte Carlo iterations ({\tt monaco.f} (a modified version of 
{\sc vegas}~\cite{vegas})) and phase space points ({\tt ps.f}) involves smearing 
the momenta ({\tt smear.f}) for the acceptance cuts algorithm ({\tt cuts.f}), 
and any particle reconstruction algorithms may follow ({\tt tautau.f, 
faketaus.f}). If the phase space point selected survives the acceptance cuts 
(some of which may be in the reconstruction procedures), then the true momenta 
are passed to the matrix elements routine ({\tt m2sZH.f}), which first 
calculates ({\tt calcqsqr.f}) the $Q^2$ values necessary for structure function 
calls (CERN pdflib). The driver then submits the event information to the 
histogramming subroutine ({\tt hist.f}), which compiles results with CERN 
{\sc hbook} routines. All extra useful subroutines, such as lego plot 
calculations, may go in an extra file ({\tt func.f}).

Historically we kept the routines in separate files to keep recompiling times 
short, but the advent of much faster machines may mean that the ultimate 
optimization achieved is slightly less than optimal. All computer codes used in 
this research, both signal and background, are available from the author upon 
request.


\section{Matrix Element Generation}

Many present-day hadron collider signal searches require the calculation of 
extremely complicated background containing many suproccesses with hundreds of 
Feynman diagrams; the number of diagrams grows roughly with $N!$, the number of 
final-state particles in the configuration. Many calculations would be 
intractable or simply take years to prepare without the aid of Feynman 
configuration generators such as {\sc grace}~\cite{grace}, 
{\sc madgraph}~\cite{Madgraph}, or {\sc comphep}~\cite{comphep}. I have found 
{\sc madgraph} to be the most powerful program; it has made possible the 
calculation of backgrounds with up to 8 final-state particles in a reasonable 
amount of time. In using {\sc madgraph}, the user inputs a given subprocess and 
the program outputs {\sc fortran} code which calls the helicity amplitude 
calculation subroutines of the {\sc helas} package to find an amplitude for each 
graph, and then performs $\sum |{\cal M}|^2$ for the entire subprocess, with 
appropriate averaging. The user must determine each subprocess that is relevant 
and generate the code for it separately.


%% file: body/mssm.tex

\section{The Higgs Sector in the MSSM}

In the minimal supersymmetric extension of the SM (MSSM) the search strategy for 
a Higgs is less clear~\cite{mssm_review} than in the SM. The Higgs sector in the 
MSSM consists of two complex scalar doublets, one with hypercharge $+1$ and the 
other $-1$, which after ESB similar to that in Section~\ref{sec:ESB} result in 
five physical Higgs states: two CP even mass eigenstates, $h$ and $H$, a CP odd 
$A$, and charged Higgs bosons $H^{\pm}$. In this Appendix I summarize the reach 
of WBF with subsequent decay to $\tau\tau$ for observing the neutral CP even 
Higgses in the MSSM. I show that the WBF channels are most likely to produce 
significant $h$ and/or $H$ signals in the regions of MSSM parameter space left 
uncovered by the MSSM Higgs searches at LEP.

Relevant features of the MSSM Higgs sector can be illustrated in a particularly 
simple approximation~\cite{easy}: including the leading contributions with 
respect to $G_F$ and the top flavor Yukawa coupling, $h_t =m_t/(vs_\beta)$. The 
qualitative features remain unchanged in a more detailed description. All our 
numerical evaluations make use of a renormalization group improved 
next-to-leading order calculation~\cite{Hdecay,one_loop}. The inclusion of two 
loop effects is not expected to change the results dramatically~\cite{two_loop}. 
Including the leading contributions with respect to $G_F$ and $h_t$, the mass 
matrix for the neutral CP even Higgs bosons is given by
\ba
\label{eq:delta}
{\cal M}^2 &=&
  m_A^2 
  \left( \begin{array}{cc}
      s_\beta^2         & -s_\beta c_\beta \\
      -s_\beta c_\beta  & c_\beta^2 
          \end{array}  \right)  
+ m_Z^2 
  \left( \begin{array}{cc}
      c_\beta^2         & -s_\beta c_\beta \\
      -s_\beta c_\beta  & s_\beta^2 
          \end{array}  \right)  
+ \epsilon
  \left( \begin{array}{cc}
      0  & 0 \\ 0 & 1 
          \end{array}  \right),  \nonumber \\
\epsilon &=& \frac{3 m_t^4 G_F}{\sqrt{2}\pi^2}
           \frac{1}{s_\beta^2}
 \left[  \log \frac{M^2_{SUSY}}{m_t^2}
       + \frac{A_t^2}{M^2_{SUSY}} \left( 1 - \frac{A_t^2}{12 M^2_{SUSY}} \right)
 \right].
\ea
Here $s_\beta,c_\beta$ denote $\sin\beta,\cos\beta$. The bottom Yukawa coupling 
as well as the higgsino mass parameter are neglected ($\mu\ll M^2_{SUSY}$). 
The orthogonal diagonalization of this mass matrix defines the CP even mixing 
angle $\alpha$. Only three parameters govern the Higgs sector: the pseudo-scalar 
Higgs mass, $m_A$, $\tan\beta$, and $\epsilon$, which describes the corrections 
arising from the supersymmetric top sector. For the scan of SUSY parameter space 
I will concentrate on two particular values of the trilinear mixing term, 
$A_t=0$ and $A_t=\sqrt{6}M^2_{SUSY}$, which commonly are referred to as no mixing 
and maximal mixing.

\begin{figure}[htb] 
\centerline{\epsfxsize 3.5 truein \epsfbox{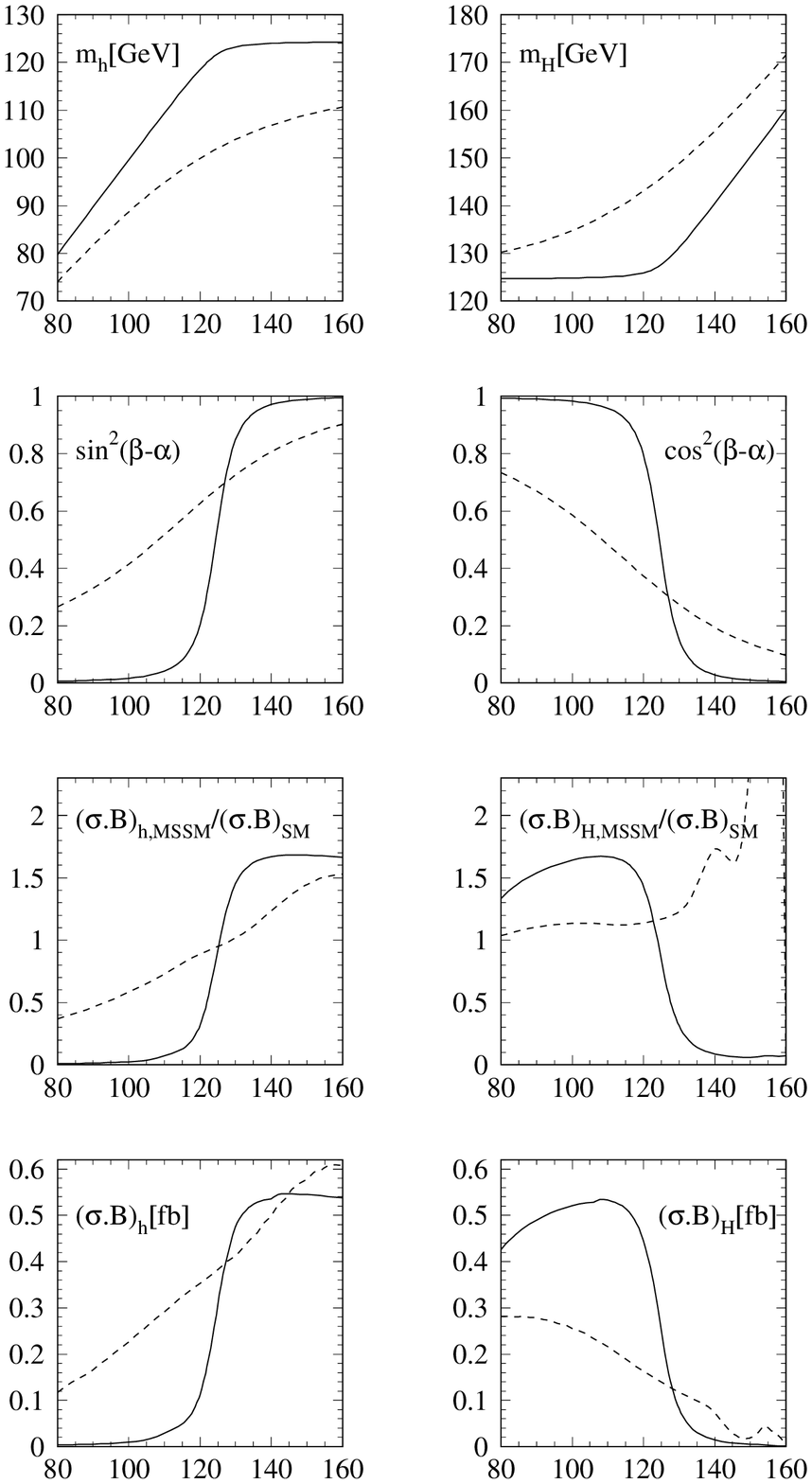}}
\caption{Variation of $h/H$ masses, couplings to $W/Z$, MSSM/SM strength ratio 
and total signal rate, for the CP even MSSM Higgs bosons as a function of the 
pseudoscalar Higgs mass. The complementarity of the search for the lighter $h$ 
(left column) and heavier $H$ (right column) is shown for $\tan\beta=4,30$ 
(dashed, solid lines). Other MSSM parameters are fixed to $\mu=200$~GeV, 
$M_{SUSY}=1$~TeV, and maximal mixing.}  
\label{fig:para} 
\end{figure}

Varying the pseudoscalar Higgs boson mass, one finds saturation for very large 
and very small values of $m_A$ -- either $m_h$ or $m_H$ approach a plateau:
\bq
\label{eq:limits}
\begin{array}{cccc}
m_h^2 & \simeq m_Z^2 (c_\beta^2-s_\beta^2)^2 + s_\beta^2 \epsilon 
& & \qquad {\rm for} \quad m_A \to \infty \, , \\ 
m_H^2 & \simeq m_Z^2 + s_\beta^2 \epsilon
& & \qquad {\rm for} \quad m_A \to 0 \, .
\end{array}
\eq
For large values of $\tan\beta$ these plateaus meet at 
$m_{h,H}^2 \approx m_Z^2+\epsilon$. Smaller $\tan\beta$ values decrease the 
asymptotic mass values and soften the transition region between the plateau 
behavior and the linear dependence of the scalar Higgs masses on $m_A$. These 
effects are shown in Fig.~\ref{fig:para}, where the variation of $m_h$ and $m_H$ 
with $m_A$ is shown for $\tan\beta=4,30$.  The small $\tan\beta$ region will be 
constrained by the LEP2 analysis of $Zh,ZH$ associated production, essentially 
imposing lower bounds on $\tan\beta$ if no signal is 
observed.~\footnote{Although the search for MSSM Higgs bosons at the Tevatron is 
promising~\cite{tev} I quote only the $Zh,ZH$ analysis of LEP2~\cite{lep} which 
is complementary to the LHC processes under consideration. The LEP2 reach is 
estimated by scaling the current limits for ${\cal L}=158$~pb$^{-1}$ and 
$\sqrt{s}=189$~GeV~\cite{lep} to ${\cal L}=100$~pb$^{-1}$ and 
$\sqrt{s}=200$~GeV.}

The theoretical upper limit on the light Higgs boson mass, to two loop order, 
depends predominantly on the mixing parameter $A_t$, the higgsino mass parameter 
$\mu$ and the soft-breaking stop mass parameters, which I treat as being 
identical to a supersymmetry breaking mass scale: 
$m_Q = m_U = M_{SUSY}$~\cite{one_loop}. As shown in Fig.~\ref{fig:para}, the 
plateau mass value hardly exceeds $\sim 130$~GeV, even for large values of 
$\tan\beta$, $M_{SUSY}=1$~TeV, and maximal mixing ~\cite{two_loop}. Theoretical 
limits arising from the current LEP and Tevatron squark search as well as the 
expected results from $Zh,ZH$ production at LEP2 assure that the lowest plateau 
masses are well separated from the $Z$ mass peak.

The production of the CP even Higgses in WBF is governed by the 
$hWW,HWW$ couplings, which, compared to the SM case, are suppressed by 
factors $\sin(\beta-\alpha),\cos(\beta-\alpha)$, respectively~\cite{mssm}. In 
the $m_h$ plateau region (large $m_A$), the mixing angle approaches 
$\alpha=\beta-\pi/2$, whereas in the $m_H$ plateau region (small $m_A$) one 
finds $\alpha\approx-\beta$.  This yields asymptotic MSSM coupling factors of 
unity for $h$ production and $|\cos(2\beta)|\gtrsim 0.8$ for the $H$ channel, 
assuming $\tan\beta \gtrsim 3$. As a result, the production cross section of 
the plateau states in WBF is essentially of SM strength. In 
Fig.~\ref{fig:para} the SUSY cross sections for $\sigma(qq \to qqh/H)$ are 
shown as a function of $m_A$; these may be compared to an expected rate for 
a SM Higgs of about 0.35~fb. The WBF cross section is sizable mainly in the 
plateau regions, and here the $h$ or $H$ masses are in the interesting range 
where decays into $b\bar{b}$ and $\tau^+\tau^-$ are expected to dominate.

The $h$ and $H$ couplings to $bb$ and $\tau\tau$ are also modified in the MSSM 
by trigonometric factors of $(\beta - \alpha)$, but the details are unimportant 
here. For effective production of $h$ or $H$ by WBF, we have 
$\sin^2(\beta-\alpha)\approx 1$ or $\cos^2(\beta-\alpha)\approx 1$, 
respectively. The coupling of the observable resonance to $bb$ and $\tau\tau$ 
is essentially of SM strength in these cases. The SUSY factors for the top and 
charm couplings are suppressed at large $\tan\beta\,$, however, which leads to 
$b\bar{b}$ and $\tau\tau$ branching ratios similar to or exceeding the SM 
values for a given mass.

Once additional off-diagonal contributions to the Higgs mass matrix are 
included, it is likely that the $h\tau\tau$ and $H\tau\tau$ couplings become 
highly suppressed for $\sin(2\alpha) = 0$, which can occur in a physical region 
of the MSSM parameter space, although where exactly this may occur is strongly 
dependent on approximations made in the perturbative expansion. If the observed 
Higgs sector turns out to be located in this parameter region, the vanishing 
coupling to $bb,\tau\tau$ would render the total widths small. This can 
dramatically increase the $h/H\to\gamma\gamma$ branching ratio, even though 
$\Gamma(h/H\to\gamma\gamma)$ may be suppressed compared to the SM case. In 
this region, the total $h/H\to\gamma\gamma$ should be sufficient for the search 
of Chapter~\ref{ch:gammagamma} to be applicable, thus covering any region 
where $h/H\to\tau\tau$ becomes impossible. Additional details may be found in 
Ref.~\cite{PRZ_mssm}.


\section{Higgs Search in Weak Boson Fusion}

Using the SUSY factors of the last section for production cross sections and 
decay rates, one can directly translate the SM results into a discovery reach 
for SUSY Higgs bosons. The expected signal rates, $\sigma B(h/H\to\tau\tau)$ are 
shown in Fig.~\ref{fig:para}. They can be compared to SM rates, within cuts, of 
$\sigma B(H\to\tau\tau) = 0.35$~fb and $\sigma B(H\to\gamma\gamma)=2$~fb for 
$m_H=120$~GeV. Except for the small parameter region where the $\tau\tau$ signal 
vanishes, and for very large values of $m_A$ (the decoupling limit), the 
$\gamma\gamma$ channel is not expected to be useful for the MSSM Higgs search in 
WBF. The $\tau\tau$ signal, on the other hand, compares favorably with the SM 
expectation over wide regions of parameter space. The SUSY factors for the 
production process determine the structure of $\sigma\cdot B(h/H\to\tau\tau)$. 
Apart from the typical flat behavior in the asymptotic plateau regions they 
strongly depend on $\beta$, in particular in the transition region, where all 
three neutral Higgs bosons have similar masses and where mixing effects are most 
pronounced.

\begin{figure}[htb] 
\centerline{\epsfxsize 4 truein \epsfbox{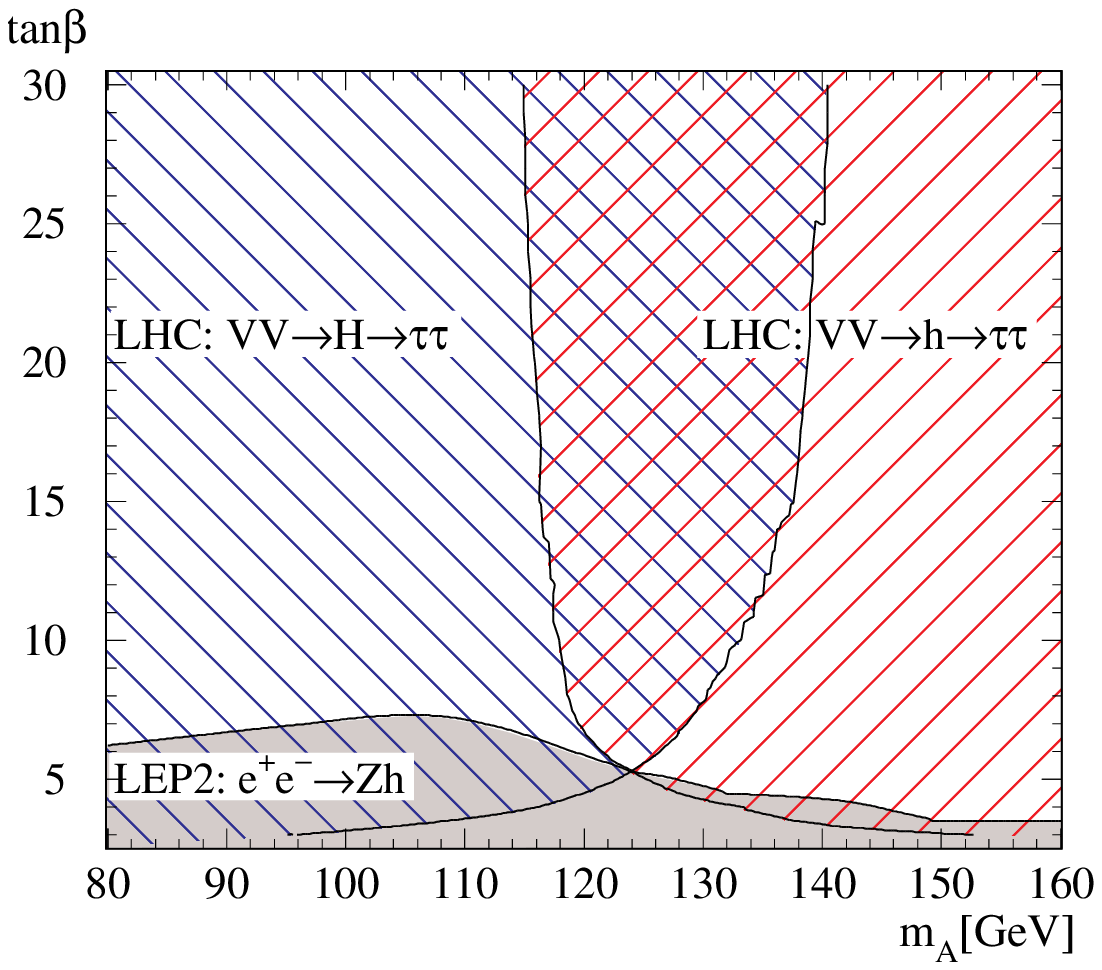}}
\caption{$5\sigma$ discovery contours for $h\to\tau\tau$ and $H\to\tau\tau$ in 
WBF at the LHC, with $70$~fb$^{-1}$. An additional efficiency factor of 0.8 
applied to the signal and all backgrounds due to pile-up is included beyond the 
first $30$~fb$^{-1}$. Also shown are the projected LEP2 exclusion limits (see 
text). Results are shown for SUSY parameters as in Fig.~\protect\ref{fig:para}, 
for maximal trilinear mixing, $A_t = \sqrt{6}M_{SUSY}$.}
\label{fig:maxmix}
\end{figure}
\begin{figure}[htb] 
\centerline{\epsfxsize 4 truein \epsfbox{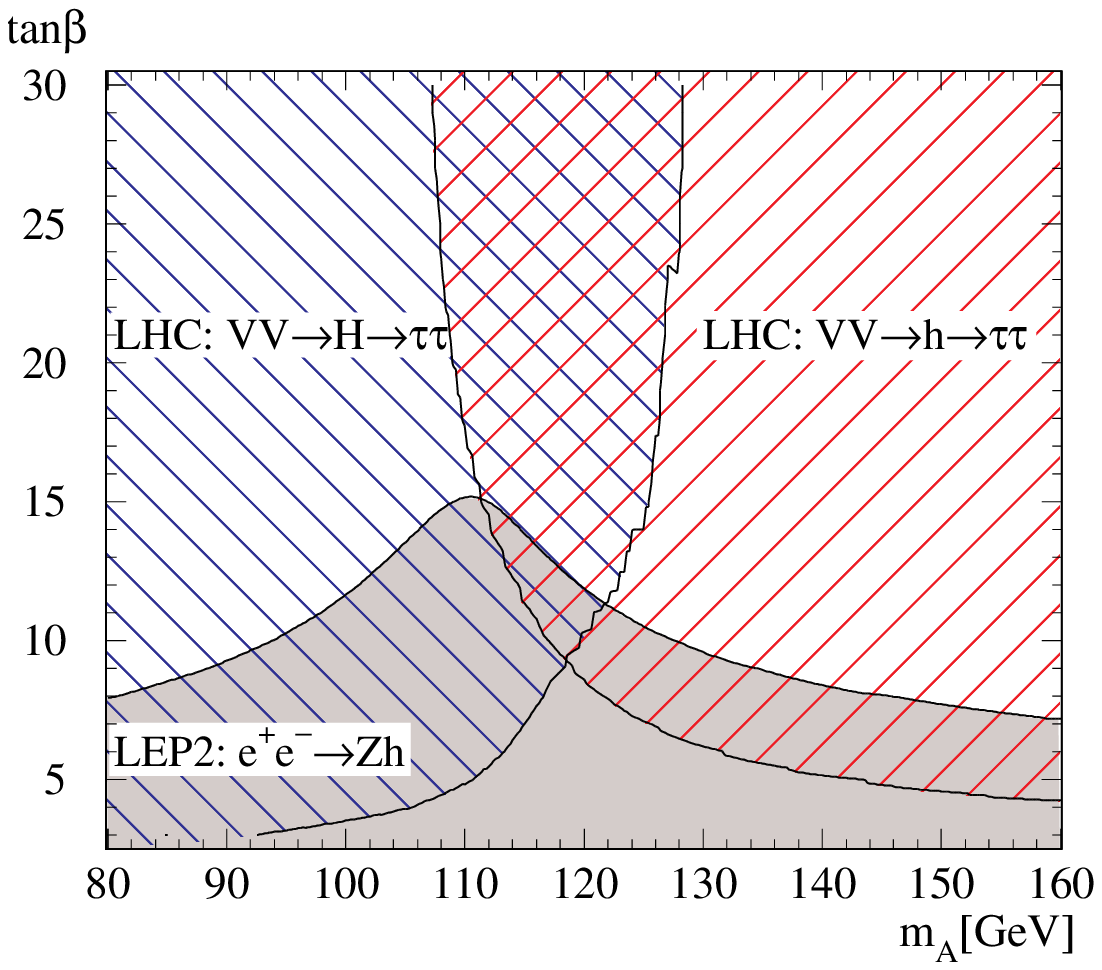}}
\caption{Same as Fig.~\ref{fig:maxmix}, but for the case of no trilinear mixing, 
$A_t = 0$.}
\label{fig:nomix}
\end{figure}

Given the background rates determined in Ch.~\ref{ch:tautau}, which are of order 
0.03~fb in a 20~GeV mass bin, except in the vicinity of the $Z$-peak, the 
expected significance of the $h/H\to \tau\tau$ signal can be determined. 
$5\sigma$ contours for an integrated luminosity of 70~fb$^{-1}$ are shown in 
Figs.~\ref{fig:maxmix},\ref{fig:nomix}, as a function of $\tan\beta$ and $m_A$. 
I include an additional efficiency factor of 0.8 for both the signal and all 
backgrounds due to pile-up for data taken beyond the first $30$~fb$^{-1}$.
Here the significances are determined from the Poisson probabilities of 
background fluctuations. Weak boson fusion, followed by decay to $\tau$-pairs, 
provides for a highly significant signal of at least one of the CP even Higgses. 
Even in the low $\tan\beta$ region, where LEP2 would discover the light Higgs, 
the WBF process at the LHC will give additional information. Most interesting 
is the transition region, where both $h$ and $H$ may be light enough to be 
observed via their $\tau\tau$ decay.

I have shown that the production of CP even MSSM Higgs bosons in WBF and 
subsequent decay to $\tau$ pairs gives a significant ($>5\sigma$) signal at the 
LHC. This search, with $\lesssim 100$fb~$^{-1}$ of integrated luminosity, and 
supplemented by the search for $h/H\to\gamma\gamma$ in weak boson fusion, 
should cover the entire MSSM parameter space left after an unsuccessful LEP2 
search, with a significant overlap of LEP2 and LHC search regions. The two CERN 
searches combined provide a no-lose strategy by themselves for seeing a MSSM 
Higgs. At the very least, the WBF measurements provide valuable 
additional information on Higgs couplings.

The present analysis relies only on the typical mixing behavior of the CP even 
mass eigenstates, and on the observability of a SM Higgs, of mass up to 
$\sim 150$~GeV, in WBF. This suggests that the search discussed here might also 
cover an extended Higgs sector as well as somewhat higher plateau masses, 
{\em e.g.} for very large squark soft-breaking mass parameters. Because decays 
into $\tau$ pairs are tied to the dominant decay channel of the intermediate 
mass range Higgs, $h/H\to \bar bb$, the search for a $\tau\tau$ signal in WBF is 
robust and expected to give a clear Higgs signal in a wide class of models.
